\documentclass[aps,pra,twocolumn,groupedaddress,floatfix,nofootinbib,longbibliography]{revtex4-2}
\usepackage{amsmath}
\usepackage{amssymb}
\usepackage{graphicx}
\usepackage{mathrsfs}
\usepackage{ntheorem}
\usepackage{mathtools}
\usepackage{enumerate}
\usepackage{enumitem}
\usepackage{physics}
\usepackage{subfigure}
\usepackage{overpic}
\usepackage{epstopdf}
\usepackage{array}
 \usepackage{multirow}
 \usepackage{array}

\usepackage{bm}
\usepackage{color,soul}
\usepackage[bookmarks=false]{hyperref}
\usepackage{xcolor}
\hypersetup{colorlinks=true,citecolor=blue,
linkcolor=blue,urlcolor=blue,pdfstartview=FitH,
bookmarksopen=true}
\usepackage{booktabs}

\def\blue#1{\textcolor{black}{#1}}
\def\red#1{\textcolor{black}{#1}}
\def\green#1{\textcolor{black}{#1}}

\newcommand{\ot}{\otimes}
\newcommand{\hjhket}[1]{|{#1}\rangle}

\newcommand{\bkt}[2]{\langle{#1}|{#2}\rangle}
\begin{document}

\title{Auto-correlative weak-value amplification under strong noise background}

\author{Jing-Hui Huang$^{1}$}\email{Email:jinghuihuang@cug.edu.cn}
\author{Xiang-Yun Hu$^{1}$} \email{Email:xyhu@cug.edu.cn}
\author{Adetunmise C. Dada$^{2}$}
\author{Jeff S. Lundeen$^{3}$}
\author{Kyle M. Jordan$^{3}$}
\author{Huan Chen$^{4}$}
\author{Jian-Qi An$^{5}$}

\address{$^{1}$Institute of Geophysics and Geomatics, China University of Geosciences, Lumo Road 388, 430074 Wuhan, China. }
\address{$^{2}$School of Physics and Astronomy, University of Glasgow, Glasgow G12 8QQ, UK }
\address{$^{3}$Department of Physics and Centre for Research in Photonics, University of Ottawa, 25 Templeton Street, Ottawa, Ontario, Canada K1N 6N5 }
\address{$^{4}$ School of Mathematics and Physics, China University of Geosciences, Lumo Road 388, 430074 Wuhan, China.}
\address{$^{5}$ School of Automation, China University of Geosciences, Lumo Road 388, 430074 Wuhan, China.}

\begin{abstract}
\blue{By choosing more orthogonality between pre-selection and post-selection states,} one can significantly improve the sensitivity in the general optical quantum metrology based on the weak-value amplification (WVA) approach. However, \blue{increasing the orthogonality} decreases the probability of detecting photons and makes the weak measurement difficult, especially when the weak measurement is disturbed by strong noise and the pointer is drowned in noise with a \blue{negative-dB} signal-to-noise ratio (SNR).
In this article, we investigate a modified weak measurement protocol with a temporal pointer, namely, the auto-correlative weak-value amplification (AWVA) approach. Specifically, a small longitudinal time delay (tiny phase shift) $\tau$ of a Gaussian pulse is measured by implementing two simultaneous auto-correlative weak measurements under Gaussian white noise with different SNR. The small quantities $\tau$ are obtained by measuring the auto-correlation coefficient of the pulses instead of fitting the shift of the mean value of the probe in the standard WVA technique.
Simulation results show that the AWVA approach outperforms the standard WVA technique in the time domain with smaller statistical errors, remarkably increasing the precision of weak measurement under strong noise background.
\end{abstract}
\maketitle

\section{Introduction}
\label{intro}
Higher precision in the measurement of various quantities is a persistent goal of scientific communities. It is well known that the pre- and post-selection of the measured system have played a crucial role in amplifying detector signals in a weak measurement. The use of the pre- and post-selection~\cite{AAV} originated from the work of Aharonov, Albert, and Vaidman in 1988.
Theoretically, they found the \red{measurement} result of the component of a particle spin, called the \emph{weak} \emph{value}, can be amplified by a large number, which opened up a gate for the quantum metrology~\cite{RevModPhys.86.307,PhysRevA.102.042601,Yin2021,PhysRevLett.126.020502,PhysRevLett.126.220801,PhysRevLett.108.070402,PhysRevLett.125.080501} with the weak-value amplification (WVA).

Chronologically, a standard weak measurement includes an initial preparation of the measured \blue{system (pre-selection)}, weak coupling between the system and the pointer, a post-selection of the system, and a \blue{projective} measurement on the pointer to read out the results~\cite{MA2021127027}. For now, the widely used pointers in weak measurement are 
shifts of mean values, such as the temporal shifts~\cite{PhysRevLett.105.010405,2013Weak,PhysRevA.102.023701,PhysRevA.105.013718}, the momentum shifts~\cite{2009Ultrasensitive,doi:10.1126/science.1202218}, the frequency \red{shifts}~\cite{PhysRevLett.111.033604,10.1109/JPHOT.2019.2942718}, and even the angular rotation shifts~\cite{DELIMABERNARDO20142029,PhysRevLett.112.200401}. But even so, the WVA scheme entails an inherent dilemma: when the pre- and post-selection are nearly orthogonal to achieve higher sensitivity, the probability of a successful post-selection will be reduced greatly~\cite{PhysRevLett.112.040406,PhysRevX.4.011032}. To obtain \blue{an} observable distribution function, \blue{the measurement must be repeated  a large number of times} and the requirements on the apparatus are \blue{stringent}~\cite{PhysRevA.103.032212}. For example, the resolution of the apparatus should be high enough and the intensity of the light source ought to be sufficiently strong.

Meanwhile, the determination of the pointer shift would inevitably be influenced by \blue{technical} noise and the surrounding environment. Numerous studies~\cite{JPHOT.2021.3057671,PhysRevX.4.011032,PhysRevLett.112.040406} have been done to investigate the advantage of WVA in presence of noise. In particular, Knee et al.~\cite{PhysRevX.4.011032} argued that the amplified displacement offered no fundamental metrological advantage, due to the necessarily reduced probability of success.  Using statistically rigorous arguments, Christopher et al.~\cite{PhysRevLett.112.040406} showed that the technique of WVA does not perform better than standard statistical techniques for single parameter estimation or signal detection. Considering the measured system cannot be completely isolated from the surrounding environment and the instability of the element itself, the determination of the pointer shift will inevitably be influenced by many sources~\cite{JPHOT.2021.3057671}: instability of the light source's spectrum and intensity, interference of the beam with itself, thermal noise and shot noise of the detection, and other noise sources. A natural thought is to equate these \blue{types of noise with Gaussian white noise}. In addition, Gaussian white noise has been replaced by colored noise (non-Gaussian )~\cite{PhysRevA.38.5938,PhysRevD.45.2843,WU2020124253,PhysRevE.101.052205} in a variety of areas of physics, such as quantum Brownian motion in a general environment with nonlocal dissipation and colored noise~\cite{PhysRevD.45.2843}, non-Gaussian noise-enhanced stability of foraging colony system~\cite{WU2020124253}. Colored noise also appears in gravitational wave interferometer, where the specific noise distribution depends on the quantum state of the gravitational field~\cite{PhysRevD.104.046021}. However, in this paper, we assume that the contribution of all noise is Gaussian white noise and investigate the weak-value amplification technique under strong Gaussian white noise background.

It is encouraging that the WVA scheme based on the imaginary weak-value (in the frequency domain) has sound potential to outperform the standard measurement in presence of technical noise~\cite{PhysRevLett.105.010405,PhysRevLett.107.133603,PhysRevA.85.060102,PhysRevX.4.011031,PhysRevLett.118.070802,PhysRevA.85.062108}. \blue{Furthermore}, Brunner et al.~\cite{PhysRevLett.105.010405} proposed that the WVA scheme using the imaginary weak value amplification can result in 3 orders of magnitude higher precision than the traditional interference method. Therefore, the WVA scheme based on the imaginary weak value is currently used in the field of biosensors, such as a new chiral sensor based on weak measurement for estimation of a trace amount of chiral molecule~\cite{LI2018103}, a tunable and high-sensitivity temperature-sensing method via WVA of Goos-H{\"a}nchen (GH) shifts in a graphene-coated system~\cite{ZHOU2021126655}, \red{and} even an optical system based on optical rotation via weak measurement for detection of single- and double-strand state of DNA~\cite{Guan2019}. Furthermore, there have been several optimization schemes to improve the WVA scheme based on analysis of its Fisher information~\cite{PhysRevLett.118.070802,JPHOT.2021.3057671,Zhu2016,PhysRevA.96.052128}.

Note that so far\blue{,} how to achieve higher precision based on the WVA scheme with the temporal pointer (in the time domain) is seldom investigated, especially under strong noise background (corresponding to the \red{negative-dB} SNR). \red{Nevertheless, different effects in the environment will inevitably cause signal noise, which makes the measurement with WVA more difficult.} Thus, further study of the WVA under \red{strong noise} is imperative.
In this paper, we proposed a modified weak measurement protocol with a temporal pointer, namely, auto-correlative weak-value \blue{amplification (AWVA)}. It is motivated by the widely used auto-correlation technique for signal denoising in engineering~\cite{1701108,machines9060123,Takahashi2013}, where auto-correlation is a signal processing method describing the correlation of a signal with a delayed copy of itself~\cite{machines9060123}. The AWVA technique can realize the WVA scheme under strong noise background. In particular, the measurements with  Gaussian white noises are studied at a certain signal-to-noise ratio (SNR). By simulating these measurements on Simulink and Matlab, we show that the measurement with AWVA is superior to the measurement with WVA under strong noise background.

The paper is organized as follows. In \blue{Sec.} II.A, we briefly review the standard WVA technique for measuring a time delay $\tau$, which \blue{serves} as the coupling strength in WVA. In \blue{Sec.} II.B, we derive the AWVA technique for the time delay $\tau$ measurement and introduce the auto-correlative intensity $\rm \Theta$ (units of voltage) to evaluate the weak value.
In \blue{Sec.} III, we present both the WVA scheme and the AWVA scheme under the Gaussian white noises.
In \blue{Sec.} IV, we show the analytic results with various types of noises and various coupling strengths. And \blue{Sec.} V is devoted to a summary and discussions.

\section{Theory}
\subsection{The standard WVA technique}
\begin{figure}[t]
	\centering
\subfigure
{
	\vspace{-0.4cm}
	\centering
	\centerline{\includegraphics[scale=0.39,angle=0]{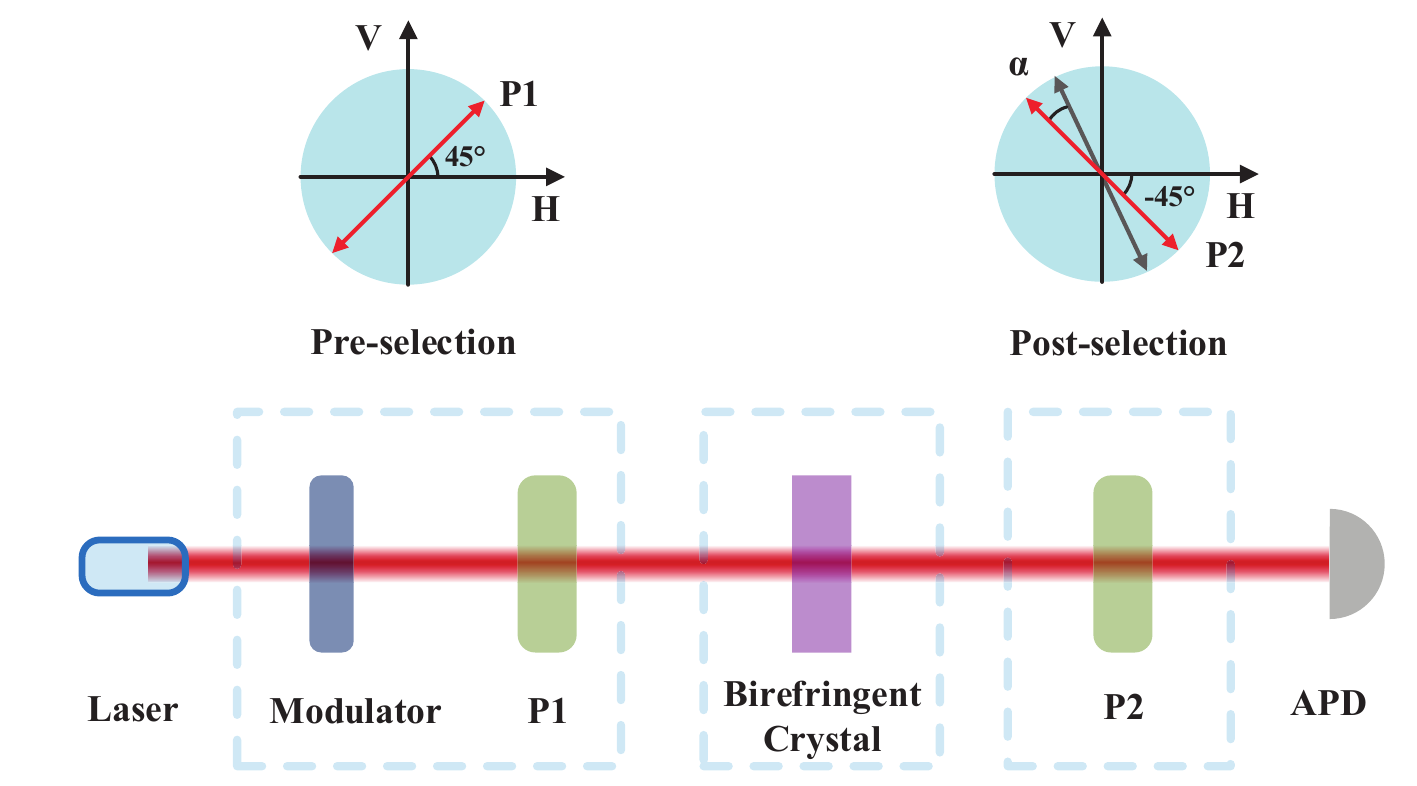}}
		\vspace{-0.4cm}
}
\vspace*{0mm} \caption{\label{Fig:Schemes_model1}Scheme of the standard WVA technique. The Gaussian beam is produced by the \blue{Laser} and Modulator. Then photons are pre-selected by the polarizer 1 (P1) with the optical axis set at $45^{0}$. A time delay $\tau$ (corresponding to a phase shift between the horizontal polarized state $\hjhket{H}$ and vertical polarized state $\hjhket{V}$) is induced by a \blue{birefringent crystal}. Finally, the photons are post-selected by the \blue{polarizer} 2 (P2) with an optical axis set at \blue{$\alpha - 45^{\circ}$}, and the arrival time of single photons is measured with an avalanche photodiode (APD).}
\end{figure}

Let us briefly review the standard WVA technique of Ref.~\cite{PhysRevLett.105.010405} with a two-level system (corresponding to the polarization state of the beam) in a quantum state $\ket{\Phi_{}}$ and a measurement device represented by a temporal pointer $\ket{\Psi_{}}$ to estimate a time delay. The scheme is shown in Fig.~\ref{Fig:Schemes_model1}. First, the system is pre-selected into a polarized state
\begin{equation}
\label{Eq:pre-sel-sys}
\ket{\Phi_{i}}= {\rm sin} (\frac{\pi}{4}) \ket{H}+ {\rm cos} (\frac{\pi}{4})\ket{V}=\frac{1}{\sqrt{2}}(\ket{H}+\ket{V})\, ,
\end{equation}
where $\ket{H}$ and $\ket{V}$ represent the horizontal and vertical polarized states\blue{,} respectively.
Thus, the initial joint state of the system and the pointer is given by
\begin{equation}
\label{Eq:pre-sel-sysAndprobe}
\ket{\Phi_{i}} \ot \ket{\Psi_{i}} \blue{\equiv \ket{\Phi_{i}} \ket{\Psi_{i}}} = \frac{1}{\sqrt{2}}(\ket{H}+\ket{V}) \ket{\Psi_{i}} ,
\end{equation}
\blue{where $\ot$ denotes tensor product. } Note that the Laser \red{prepares} a (bright) quantum coherent state of the field \red{for the} autocorrelation measurements. \red{A coherent state is what a laser prepares, but other types of light would work for our scheme as well. In particular, the state must be strong/bright so that we can ignore the effect of vacuum noise/partition noise in the following analysis in Sec~\ref{Sec:signal-to-noise}.}
Then, the system and the pointer are weakly coupled with the interaction Hamiltonian $\hat{H}=\tau \hat{A}\ot \hat{p}$, where the observable operator $\hat{A}=\ket{H} \bra{H}-\ket{V} \bra{V} $ and $\hat{p}$ is the momentum operator conjugated to the position operator $\hat{q}$. In the regime of weak measurement, the time shift $\tau$ is much smaller than the pointer spread $\omega$, and the final state of the pointer is given by:
\begin{eqnarray}
\label{inter_peobe_final}
\ket{\Psi_{f}}
&=&\bra{\Phi_{f}}e^{-i\tau\hat{A}\ot \hat{p}} \ket{\Psi_{i}}  \ket{\Phi_{i}} \nonumber \\
&\approx&\bra{\Phi_{f}}\left[ 1-i\tau\hat{A}\ot \hat{p}\right]\ket{\Psi_{i}}  \ket{\Phi_{i}} \nonumber \\
&=&\bkt{\Phi_{f}}{\Phi_{i}}\left[ 1-i\tau A_{w}\hat{p}\right]\ket{\Psi_{i}} \nonumber \\
&=&\bkt{\Phi_{f}}{\Phi_{i}}e^{-i\tau A_{w}\hat{p}}\ket{\Psi_{i}}\, ,
\end{eqnarray}
where  $A_{w}:={\bra{\Phi_{f}}\hat{A}\ket{\Phi_{i}}}/{\bkt{\Phi_{f}}{\Phi_{i}}}$\blue{---}the so-called weak value~\cite{AAV}\blue{---}represents the mean value of observable $\hat{A}$. Note that the time shift $\tau$ can be amplified by the weak value $A_{w}$. Normally, the weak value $A_{w}$ is a complex number~\cite{PhysRevA.76.044103}, thus the difference between the pointer in momentum space and that in position (temporal) space is used to perform the standard WVA. In particular, the imaginary part of $A_{w}$ is associated with a shift of the pointer in momentum space, while the time shift $\tau$ in the position of the pointer is amplified by the real part of $A_{w}$~\cite{PhysRevA.85.052110}:
\begin{equation}
\label{delta_q}
\Delta\langle\hat{q}\rangle=\frac{\int d q q\left|\left\langle q | \Psi_{f}\right\rangle\right|^{2}}{\int d q\left|\left\langle q | \Psi_{f}\right\rangle\right|^{2}}=\tau {\rm Re} \,[A_{\omega}] \,.
\end{equation}

In this paper, we design the weak measurement in the time domain and prepare the initial pointer with the Gaussian profile:
\begin{equation}
\label{Eq:initial-pointer}
I^{in}_{1}(t;\tau)=\left|\left\langle t | \Psi_{i}\right\rangle\right|^{2}=I_{0}\frac{1} {(2 \pi \omega^{2})^{1/4}}  e^{-(t-t_{0})^{2}/4\omega^{2}} \,.
\end{equation}
where $I_{0}$ represents the \blue{normalization} factor. In order to amplify the ultra-small time shift $\tau$, the system is post-selected into the state:
\begin{equation}
\label{Eq:post-sel-sys}
\ket{\Phi_{f}}= {\rm sin} (-\frac{\pi}{4}+\alpha) \ket{H}+ {\rm cos} (-\frac{\pi}{4}+\alpha)\ket{V},
\end{equation}
where $\alpha$ should not be chosen as $\alpha$=0 to obtain non-vanishing probability of post-selection. One then \blue{obtains} the weak value as:
\begin{equation}
\label{weak_value_1}
A_{w}=\frac{{\rm sin} (-\frac{\pi}{4}+\alpha) - {\rm cos} (\frac{\pi}{4}+\alpha)}{{\rm sin} (-\frac{\pi}{4}+\alpha) + {\rm cos} (\frac{\pi}{4}+\alpha)} =-{\rm cot} \alpha \, ,
\end{equation}
The corresponding time shift $\tau$ can be obtained from the peak shift $\delta t=|\tau {\rm Re}\,[ A_{w}]|= \tau  {\rm cot} \alpha$ of the signal detected by an avalanche photodiode (APD), with the detected signal $I_{1}^{out}$ calculated from Eq. (\ref{inter_peobe_final}) as:
\begin{eqnarray}
\label{Eq:schme1inter_peobe_final}
I_{1}^{out}(t;\tau)&=& |\bkt{\Phi_{f}}{\Phi_{i}}|^{2} e^{-2i\tau A_{w}\hat{p}} \left|\left\langle q | \Psi_{i}\right\rangle\right|^{2} \nonumber \\
&\approx &   I_{0}\frac{({\rm sin}\alpha)^{2}} {(2 \pi \omega^{2})^{1/4}}   e^{-(t-t_{0}-\delta t)^{2}/4\omega^{2}}
\end{eqnarray}

In principle, a larger peak shift $\delta t$ \red{as well as a larger weak value} are obtained by choosing a smaller $\alpha$, at a cost of decreasing the probability of post-selection $\mathcal{P}=|{\bkt{\Phi_{f}}{\Phi_{i}}}|^{2}=({\rm sin}\alpha)^{2}$ . Note that the low probability $\mathcal{P}$ leads to a weak signal and makes the weak measurement more difficult under strong noise background.
In the next subsection, we will improve it with an AWVA technique.

\subsection{AWVA technique with auto-correlative intensity}
\begin{figure}[htp!]
	\centering
\subfigure
{
	\vspace{-0.4cm}
	\centering
	\centerline{\includegraphics[scale=0.39,angle=0]{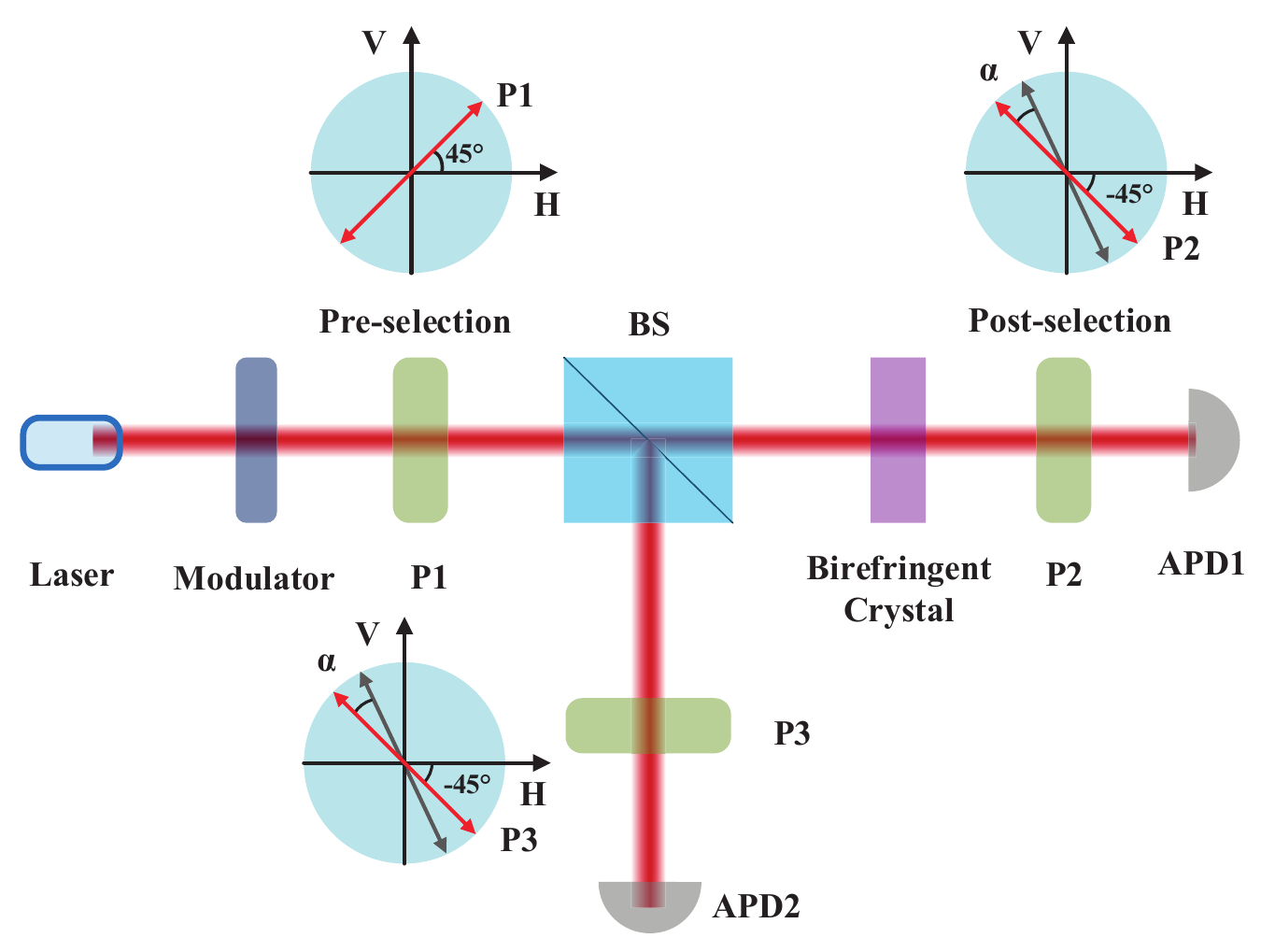}}
		\vspace{-0.4cm}

}
\vspace*{0mm} \caption{\label{Fig:Schemes_model2}Scheme of the AWVA technique. The light path is similar to that in the WVA scheme (Fig.~\ref{Fig:Schemes_model1}), except for a 50:50 beam splitter being inserted between \blue{polarizer} P1 and \blue{the birefringent crystal} to add a light path for an auto-correlative measurement. The optical axis of \blue{polarizer 3} (P3) is also set at $\alpha - 45^{0}$.
 }
\end{figure}

We display the scheme of the AWVA technique in Fig.~\ref{Fig:Schemes_model2}, in which we introduce an auto-correlative intensity $\rm \Theta$ for estimating the time shift $\tau$ introduced by the birefringent crystal. The main difference between the two schemes is that an additional light path is added in the AWVA scheme, by dividing the light after pre-selection into two light paths with a beam splitter (BS, splitting ratio 50:50). In one path, light passes through the birefringent crystal and P2 as in the WVA scheme, while in the other path light passes only through Polarizer 3 (P3) for an auto-correlative measurement.

The signal $I_{21}^{out}(t)$ detected at APD1 is similar to \blue{Eq.~}\ref{Eq:schme1inter_peobe_final} except the intensity being halved by the BS
\begin{eqnarray}
\label{Schem2:inter_peobe_final}
I_{21}^{out}(t;\tau)=\frac{I_{0}}{2} \frac{ ({\rm sin}\alpha)^{2}} {(2 \pi \omega^{2})^{1/4}}  e^{-(t-t_{0}-\delta t)^{2}/4\omega^{2}}
\end{eqnarray}

Considering the light only passing through P3, there is no shift of the mean value of the pointer and the signal $I_{22}^{out}(t;\tau)$ detected at APD2 is given as:
\begin{eqnarray}
\label{Schem2:inter_peobe_final22}
I_{22}^{out}(t;\tau)=\frac{I_{0}}{2} \frac{({\rm sin}\alpha)^{2}} {(2 \pi \omega^{2})^{1/4}}   e^{-(t-t_{0})^{2}/4\omega^{2}} \, .
\end{eqnarray}
\begin{figure}[htp!]
	\centering
\subfigure
{
	\vspace{-0.4cm}
	\begin{minipage}{6.5cm}
	\centering
	\centerline{\includegraphics[scale=0.35,angle=0]{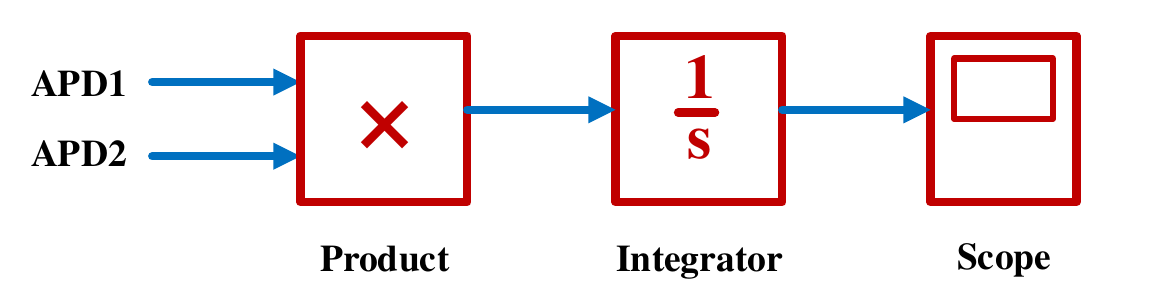}}
	\end{minipage}
}
	\vspace{-0.4cm}
\vspace*{0mm} \caption{\label{Fig:Schemes_model3}Scheme of the signal processing module with AWVA technique.  }
\end{figure}
\begin{figure*}[htp!]
	\centering
\subfigure
{
	\vspace{-0.4cm}
	\begin{minipage}{0.31\linewidth}
	\centering
	\centerline{\includegraphics[scale=0.135,angle=0]{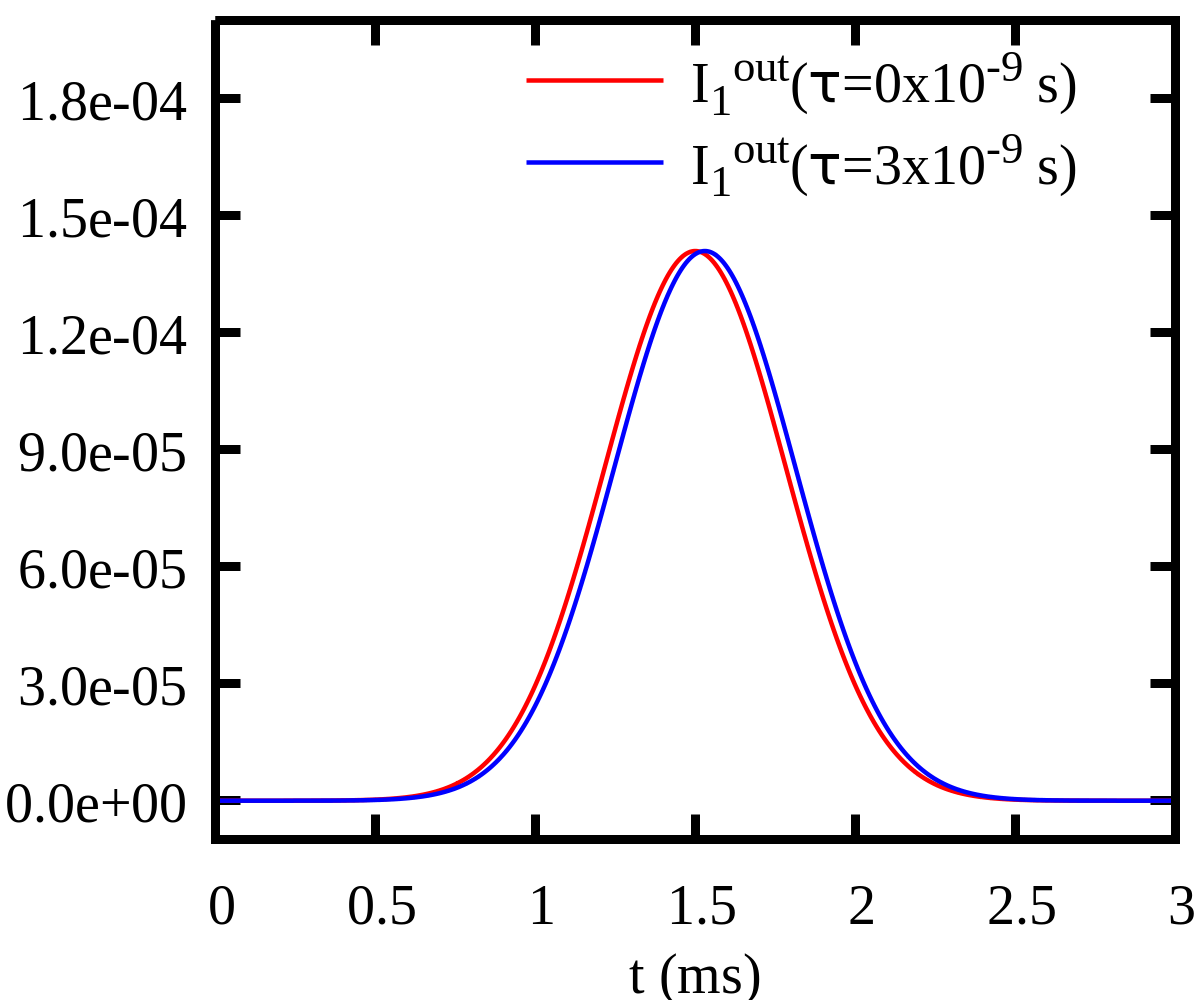}}
	\end{minipage}
}
\subfigure
{
	\vspace{-0.4cm}
	\begin{minipage}{0.31\linewidth}
	\centering
	\centerline{\includegraphics[scale=0.135,angle=0]{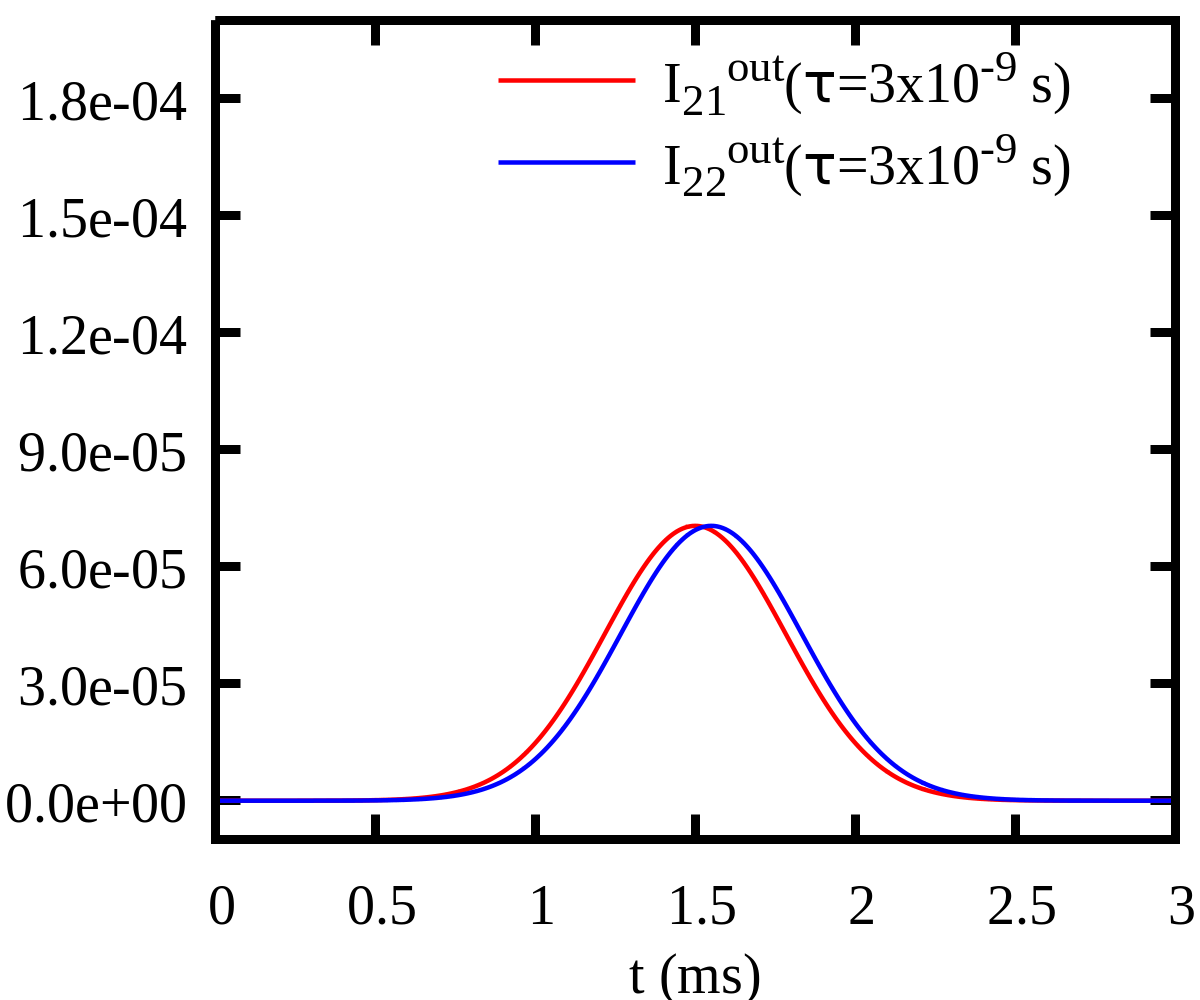}}
	\end{minipage}
}
\subfigure
{
	\vspace{-0.4cm}
	\begin{minipage}{0.31\linewidth}
	\centering
	\centerline{\includegraphics[scale=0.135,angle=0]{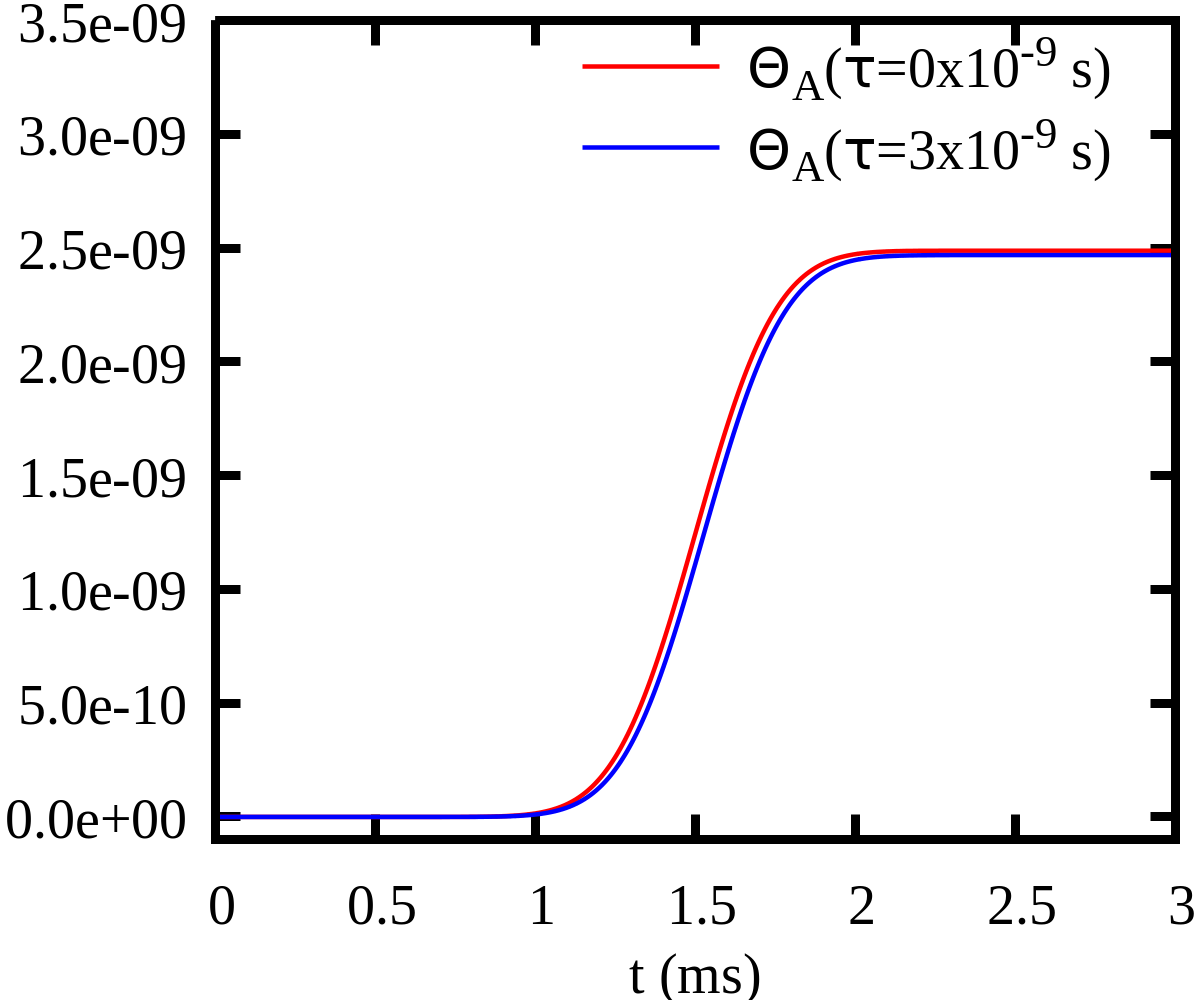}}
	\end{minipage}
}
\vspace*{0mm} \caption{\label{Fig:probeChangeInTwoSchemeNOnoise}Simulation results with WVA and AWVA schemes in absence of noises. Left panels: the signals $I_{1}^{out}(t;\tau)$ with $\tau =0 \times 10^{-9}$ s and $\tau =3 \times 10^{-9}$ s in the WVA scheme. Middle panels: the signals  $I_{21}^{out}(t;\tau)$ as well as $I_{22}^{out}(t;\tau)$ with  $\tau =3 \times 10^{-9}$ s in the AWVA scheme. Right panels: the signals ${\rm I}^{AC}_{A}$ with $\tau =0 \times 10^{-9}$ s and $\tau =3 \times 10^{-9}$ s in the AWVA scheme. The units of the \red{quantities} $I_{21}^{out}(t;\tau)$, $I_{21}^{out}(t;\tau)$ and $I_{22}^{out}(t;\tau)$ are units of $I_{0}$, the \red{quantity} ${\rm I}^{AC}_{A}$ is units of voltage.}
\end{figure*}

In this paper, we introduce the new \red{quantity} $\rm \Theta$ to estimate the time shift $\tau$ rather than the peak shift of the pointer. $\rm \Theta$ is measured with the scheme shown in Fig.~\ref{Fig:Schemes_model3}. The signals $I_{21}^{out}(t;\tau)$ and $I_{22}^{out}(t;\tau)$ detected at APD1 and APD2 pass through the \blue{``Product"} and the \blue{``Integrator"}. Finally, the \blue{``Scope''} detects the \blue{``$\rm \Theta$"} signal, which is mathematically given as:
\begin{eqnarray}
\label{Eq:ACIdefine}
{\rm \Theta}_{A}(t;\tau)&=&\int_{0}^{t} I_{21}^{out}(t^{\prime};\tau) \times  I_{22}^{out}(t^{\prime};\tau) dt^{\prime} \\
&=&  \frac{I_{0}^{2}}{4} \frac{({\rm sin}\alpha)^{4}} {(2 \pi \omega^{2})^{1/8} }   \int _{0}^{t} e^{-[(t^{\prime}-t_{0})^{2}+(t^{\prime}-t_{0}-\delta t)^{2}]/4\omega^{2}} dt^{\prime}\, . \nonumber
\end{eqnarray}
The numerical results in absence of noises in the WVA scheme and the AWVA scheme are displayed in Fig.~\ref{Fig:probeChangeInTwoSchemeNOnoise}. 
In the WVA scheme, the time shift $\tau$ can be estimated by fitting the Gaussian signals $I_{1}^{out}(\tau= 0 \times 10^{-9}s)$ and $I_{1}^{out}(\tau= 3 \times 10^{-9}s)$ by means of Least square method. In the AWVA scheme, the time shift $\tau$ is estimated with the values of ${\rm \Theta}$. Eq. (\ref{Eq:ACIdefine}) also indicates that the value of ${\rm \Theta}_{A}(t;\tau)$ depends strongly on the integral time $t$, which affects the sensitivity of the measurement with the AWVA technique. We will show and discuss the results in the next section.

The signal processing module shown in Fig.~\ref{Fig:Schemes_model3} can be implemented by both digital circuits and analog circuits. However, in this paper, we will only simulate the signal processing process in Simulink and MATLAB.
The tools in Simulink allow us to implement weak measurements under various types of noise, and these simulation results will be shown in the following part.

\section{WVA and AWVA under Gaussian white noises }
\label{Sec:WVAandAWVAunderNoises}
\begin{figure*}[htp!]
	\centering
\subfigure
{
	\vspace{-0.4cm}
	\begin{minipage}{0.31\linewidth}
	\centering
	\centerline{\includegraphics[scale=0.135,angle=0]{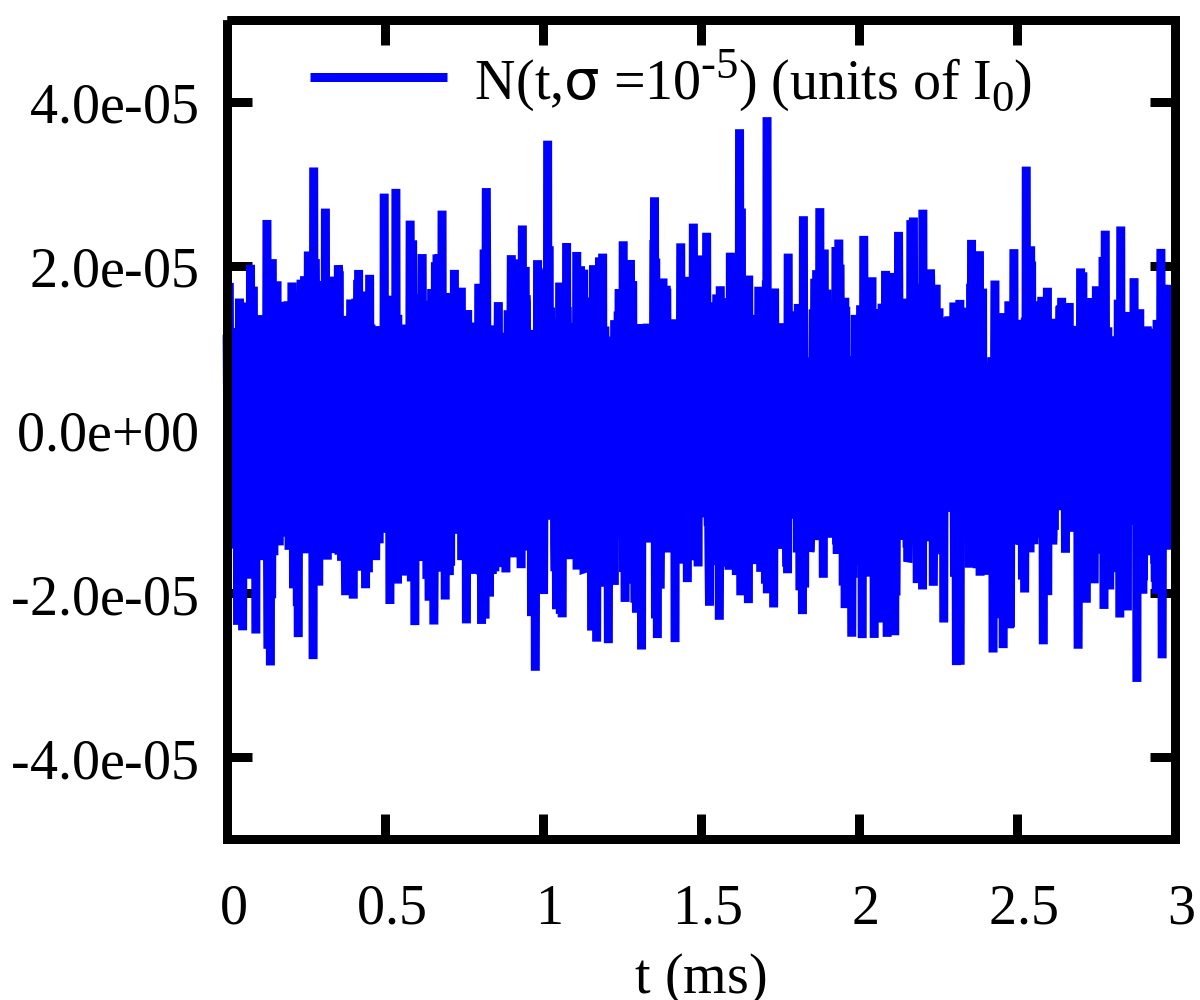}}
	\end{minipage}
}
\subfigure
{
	\vspace{-0.4cm}
	\begin{minipage}{0.31\linewidth}
	\centering
	\centerline{\includegraphics[scale=0.135,angle=0]{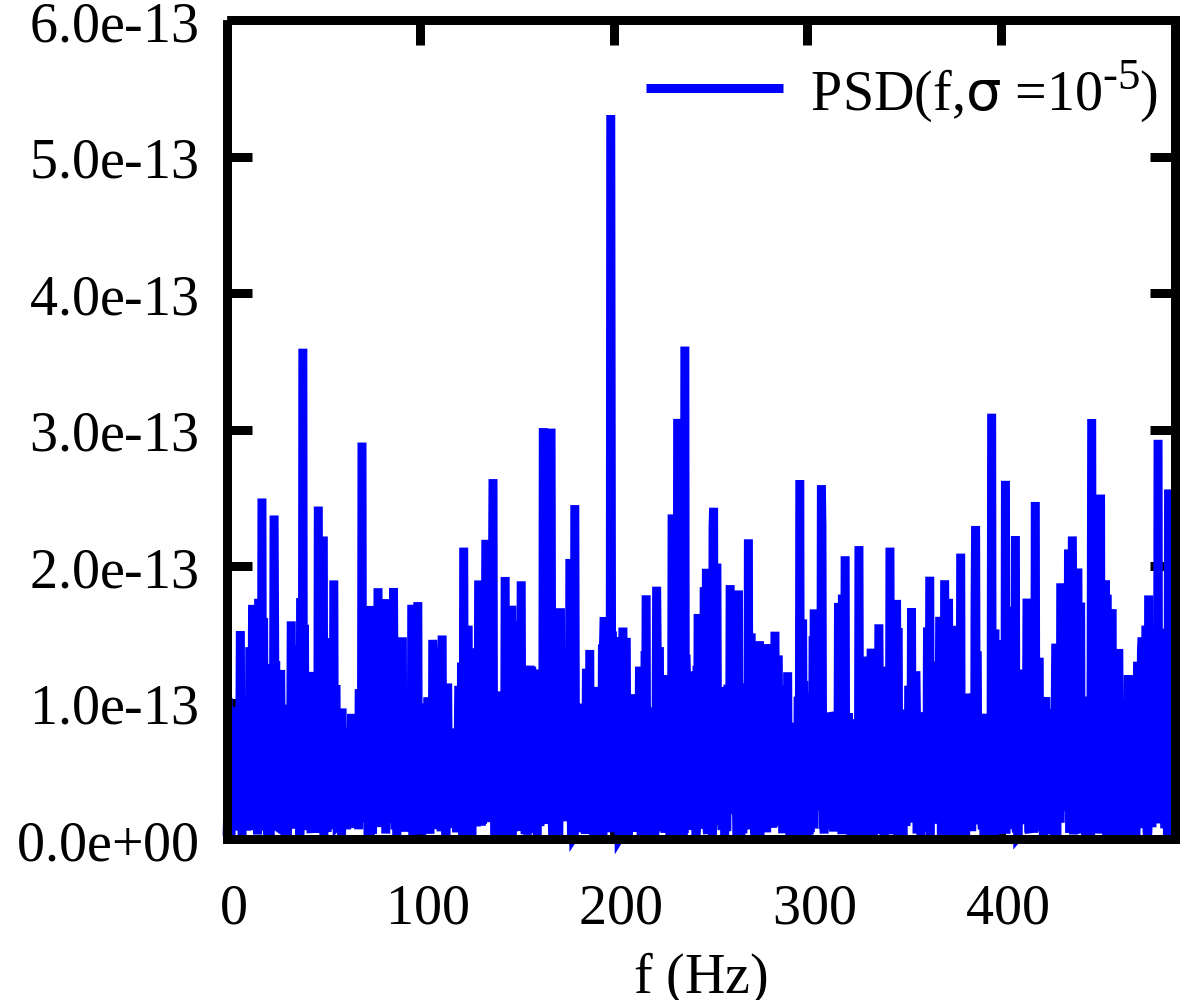}}
	\end{minipage}
}
\subfigure
{
	\vspace{-0.4cm}
	\begin{minipage}{0.31\linewidth}
	\centering
	\centerline{\includegraphics[scale=0.135,angle=0]{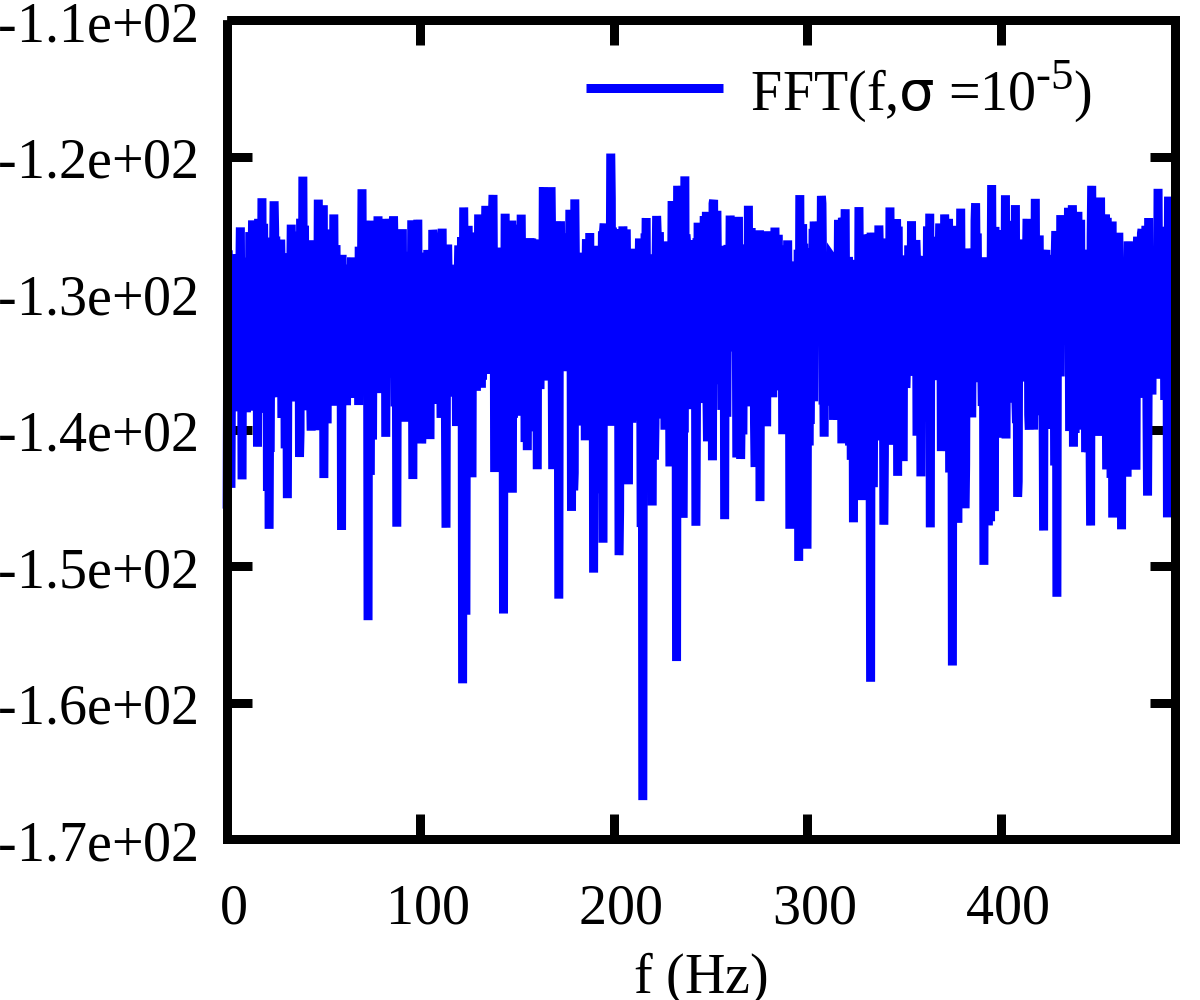}}
	\end{minipage}
}
\vspace*{0mm} \caption{\label{Fig:VariousNoise} The Gaussian noise signals $\textbf{N}(t,\sigma^{2}=1.0 \times 10^{-5},\xi=0)$ in the time domain (left panels), its power spectral densities $\textbf{PSD}(f,\sigma^{2}=1.0 \times 10^{-5},\xi=0)$ (middle panels) and its FFT result $\textbf{FFT}(f,\sigma^{2}=1.0 \times 10^{-5},\xi=0)$ (right panels). }
\end{figure*}
\begin{figure*}[htp!]
	\centering
\subfigure
{
	\vspace{-0.4cm}
	\begin{minipage}{0.31\linewidth}
	\centering
	\centerline{\includegraphics[scale=0.135,angle=0]{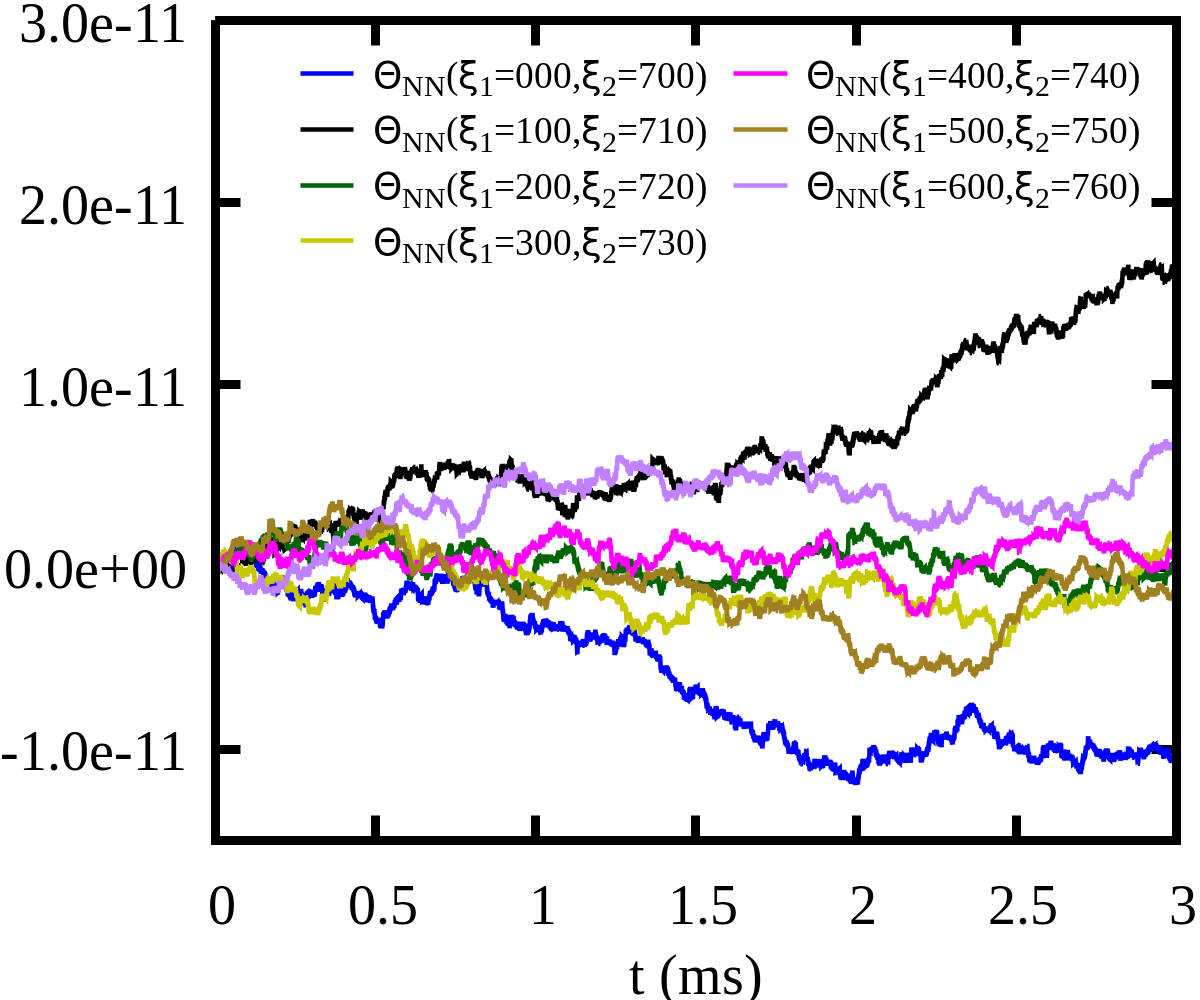}}
	\end{minipage}
}
\subfigure
{
	\vspace{-0.4cm}
	\begin{minipage}{0.31\linewidth}
	\centering
	\centerline{\includegraphics[scale=0.135,angle=0]{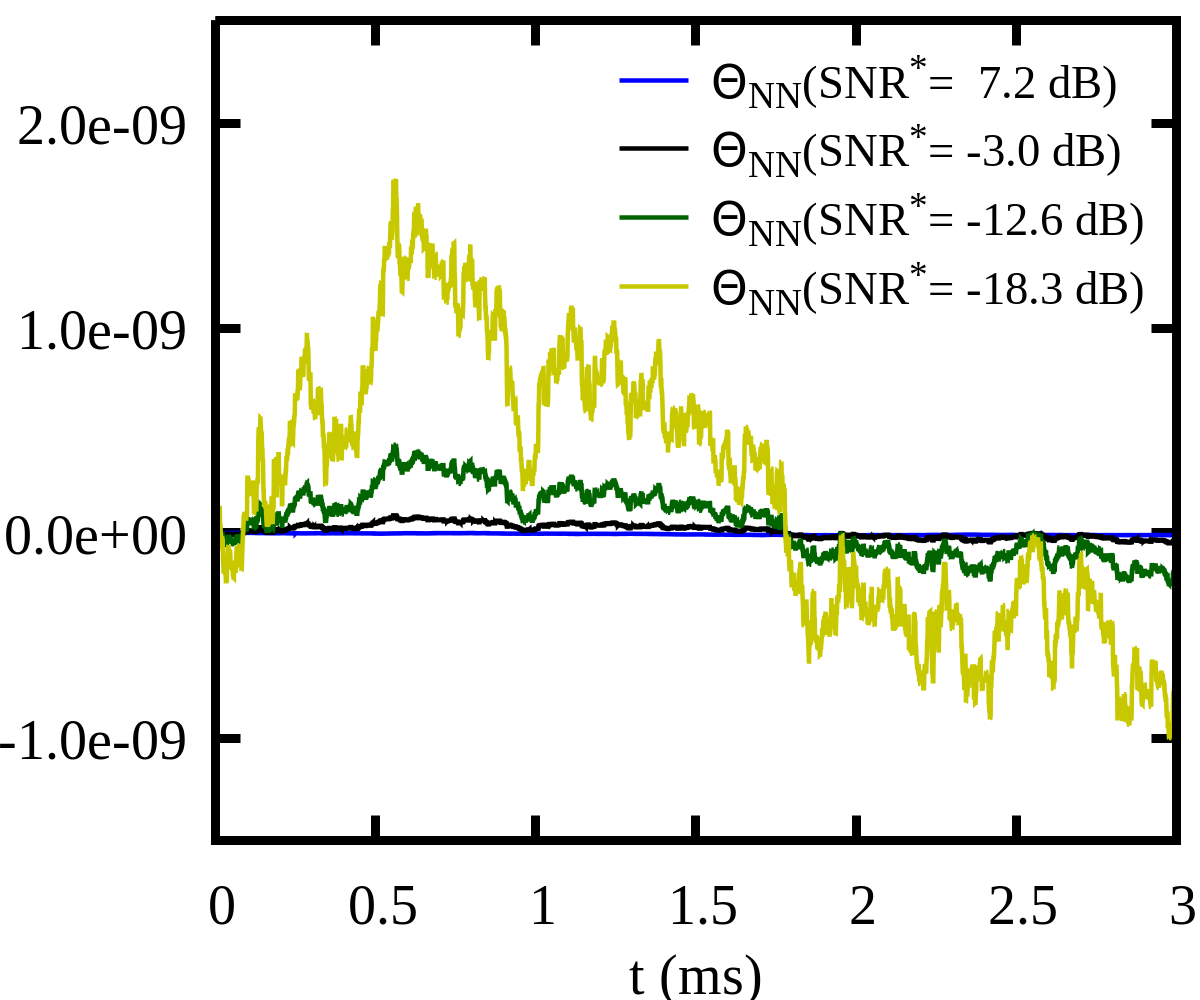}}
	\end{minipage}
}
\subfigure
{
	\vspace{-0.4cm}
	\begin{minipage}{0.31\linewidth}
	\centering
	\centerline{\includegraphics[scale=0.135,angle=0]{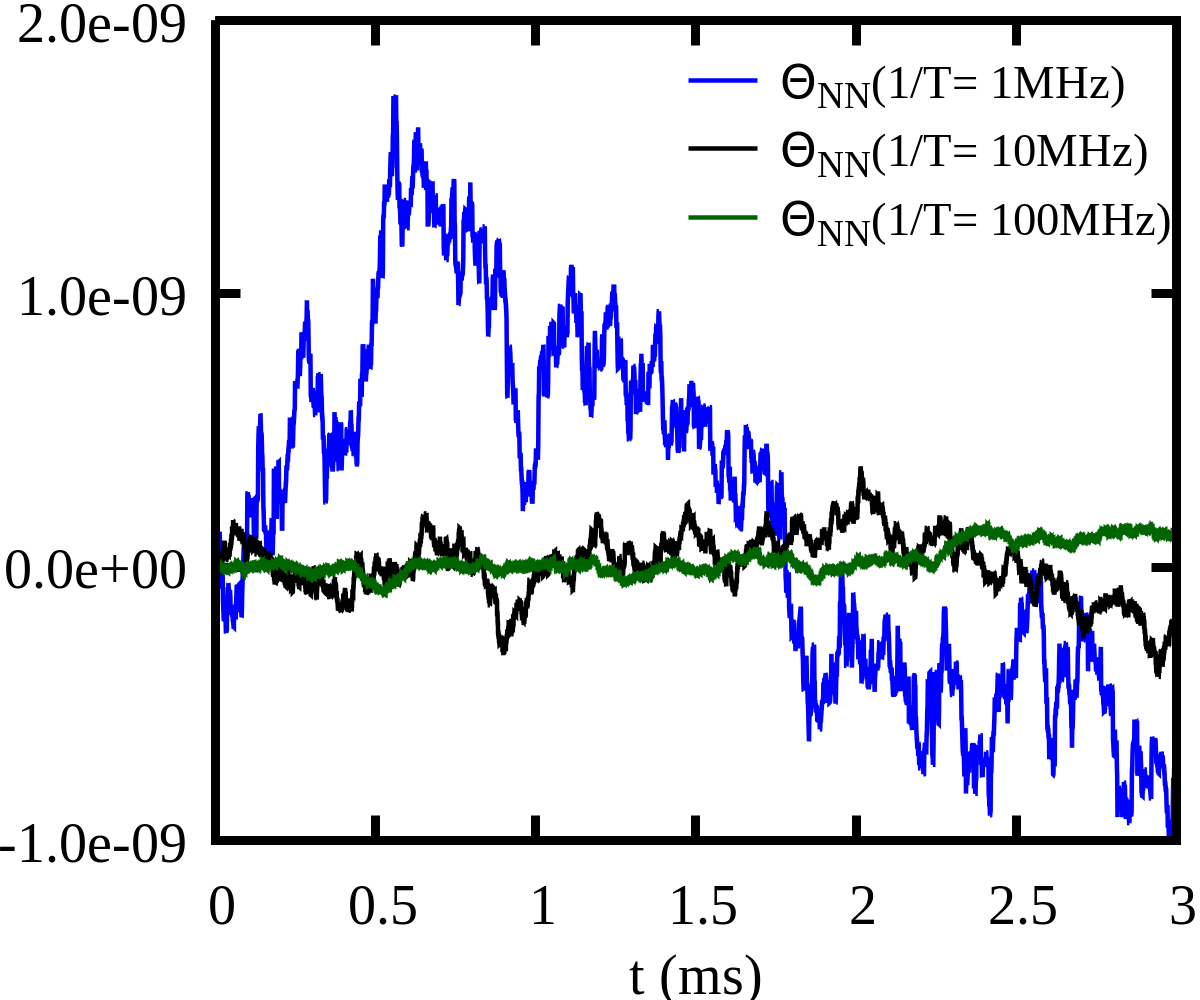}}
	\end{minipage}
}
\vspace*{0mm} \caption{\label{Fig:Variousauto-correlativeNoise}
{
The auto-correlative intensity ${\rm \Theta_{NN}}(t;\tau)$ of various noises with different sampling frequency 1/T of APD. Left panel: the auto-correlative results of noises $\textbf{N}(t,\sigma^{2},\xi_{1})$ and $\textbf{N}(t,\sigma^{2},\xi_{2})$ with different seeds $\xi_{1}$ and $\xi_{2}$ at $\rm SNR^*$=7.2 dB and 1/T= 1MHz. Middle panel: the auto-correlative results of noises $\textbf{N}(t,\sigma^{2},\xi_{1}\red{=0})$ and $\textbf{N}(t,\sigma^{2},\xi_{2}=700)$ at sampling frequency 1/T= 1MHz with different  $\rm SNR^*$s. Right panel: the auto-correlative results of noises $\textbf{N}(t,\sigma^{2},\xi_{1}\red{=0})$ and $\textbf{N}(t,\sigma^{2},\xi_{2}=700)$ at $\rm SNR^*$=-18.6 dB with different sampling frequency 1/T.
}}
\end{figure*}
It has been pointed out that \blue{technical noise}~\cite{JPHOT.2021.3057671,PhysRevX.4.011032,PhysRevLett.112.040406} has a great influence on \blue{weak measurements}. To simulate the WVA and AWVA schemes with the temporal pointer on Simulink in a realistic situation, the effects of Gaussian white noises with different SNR on the two schemes are investigated \blue{first}.

In this paper, \red{we approximate the optical noise due to} the instability of the light source, interference in the light path, thermal noise and shot noise of the detection, the partition noise (vacuum noise) due to the beamsplitter, and noises from other unknown sources as Gaussian noise. The characteristic of Gaussian white noise is that its power spectral density and the fast Fourier transform (FFT) result are uniformly distributed. \red{In our work}, the Gaussian normal distribution denoted as $\textbf{N}(t,\sigma^{2},\xi)$ is generated by the pseudo-random number generator in Simulink, where $\sigma^{2}$ is the variance of the random signal. $\xi$ is the random seed and represents the initial value used to generate a pseudo-random number in Simulink. The noise $\textbf{N}(t,\sigma^{2}, \xi)$ with different $\xi$ is corresponding to the results of multiple measurements (different times).
Note that thermal noise and shot noise of the detection may cause the different series of noises detected on APD1 and APD2. In addition, the influence of the noise of different time series is also investigated in Section~\ref{Sec:different_random_seed}.

The Gaussian white noise $\textbf{N}(t,\sigma^{2}=1.0 \times 10^{-5},\xi=0)$, the power spectral density $\textbf{PSD}(f,\sigma^{2}=1.0 \times 10^{-5},\xi=0)$ and the corresponding $\textbf{FFT}(f,\sigma^{2}=1.0 \times 10^{-5},\xi=0)$ results  are shown in Fig.~\ref{Fig:VariousNoise}.
Note that Gaussian white noise is not only uncorrelated but also statistically independent between random variables at two different moments. Thus, on the basis of auto-correlation technique for signal denoising in engineering~\cite{1701108,machines9060123,Takahashi2013},
the $\rm  {\Theta}_{NN}(t;\tau)$ of the Gaussian white noise $\textbf{N}(t,\sigma^{2})$ is defined as:
\begin{eqnarray}
\label{Eq:ACIdefineWihtnoise}
{\rm \Theta_{NN}}(t;\tau)&=&\int_{0}^{t} {\textbf{N}(t^{\prime},\sigma^{2},\xi_{1}) }\times  {\textbf{N}(t^{\prime},\sigma^{2},\xi_{2}) }dt^{\prime} \nonumber \\
&=& 0 (t \xrightarrow{} \infty)\, .
\end{eqnarray}
{where the $\xi_{1}$ and $\xi_{2}$ represent the different random seeds of the Gaussian white noise $\textbf{N}(t,\sigma^{2},\xi_{})$. In this paper, for the measurements with the AWVA technique, the corresponding time series of noises $\textbf{N}(t,\sigma^{2},\xi_{1})$ and $\textbf{N}(t,\sigma^{2},\xi_{2})$ detected at APD1 and APD2 respectively ought to be different. Even if the two noises only originate from the Laser (before the BS), the noise $\textbf{N}(t,\sigma^{2},\xi_{1})$ which passes through the Birefringent Crystal and the different lengths of the two optical paths in Fig.~\ref{Fig:Schemes_model2} will cause \red{a time} delay between the noises $\textbf{N}(t,\sigma^{2},\xi_{1})$ and $\textbf{N}(t,\sigma^{2},\xi_{2})$.
}

\begin{figure*}[htp!]
	\centering
	\subfigure
{
	\vspace{-0.4cm}
	\begin{minipage}{0.31\linewidth}
	\centering
	\centerline{\includegraphics[scale=0.135,angle=0]{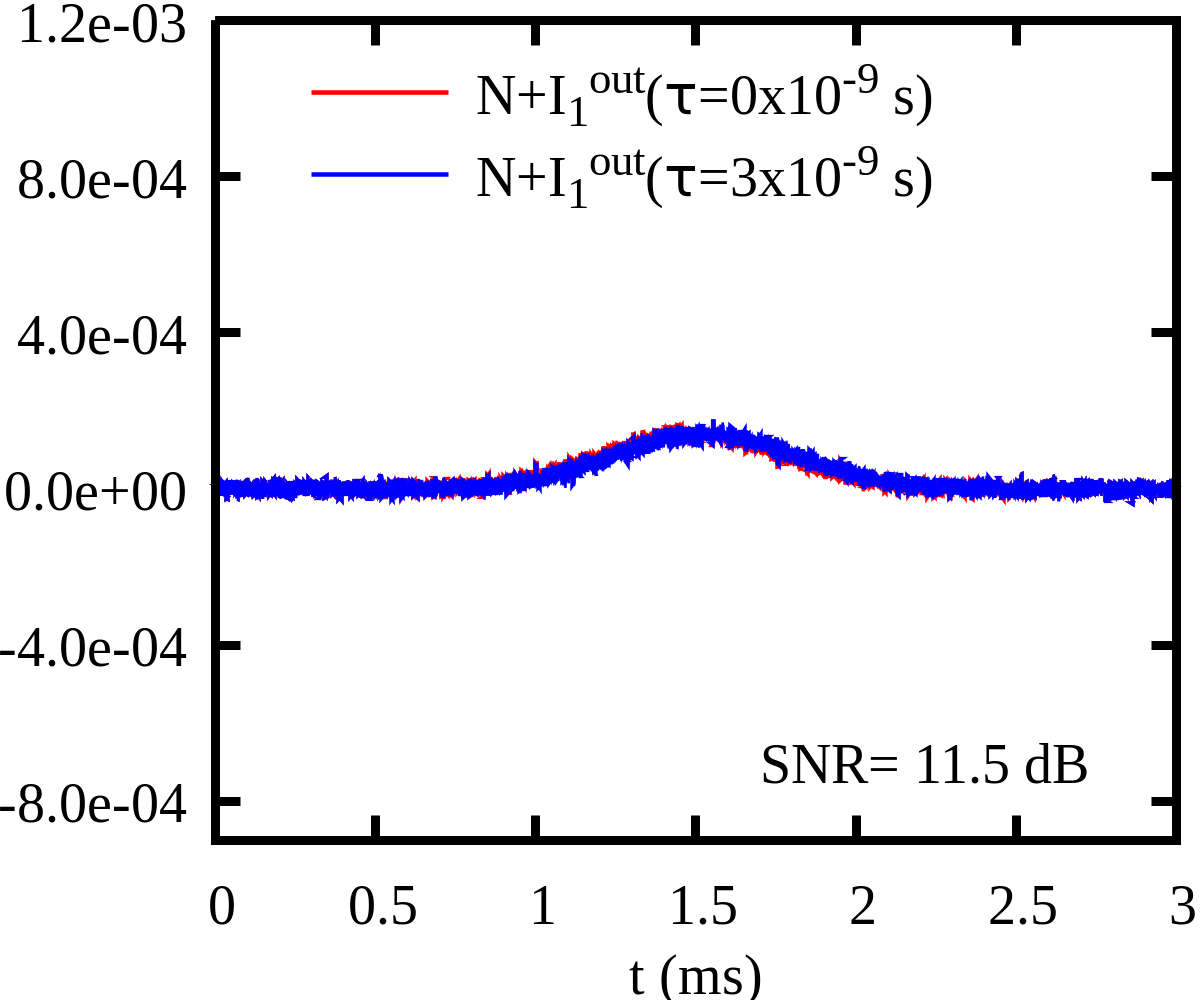}}
	\end{minipage}
}
\subfigure
{
	\vspace{-0.4cm}
	\begin{minipage}{0.31\linewidth}
	\centering
	\centerline{\includegraphics[scale=0.135,angle=0]{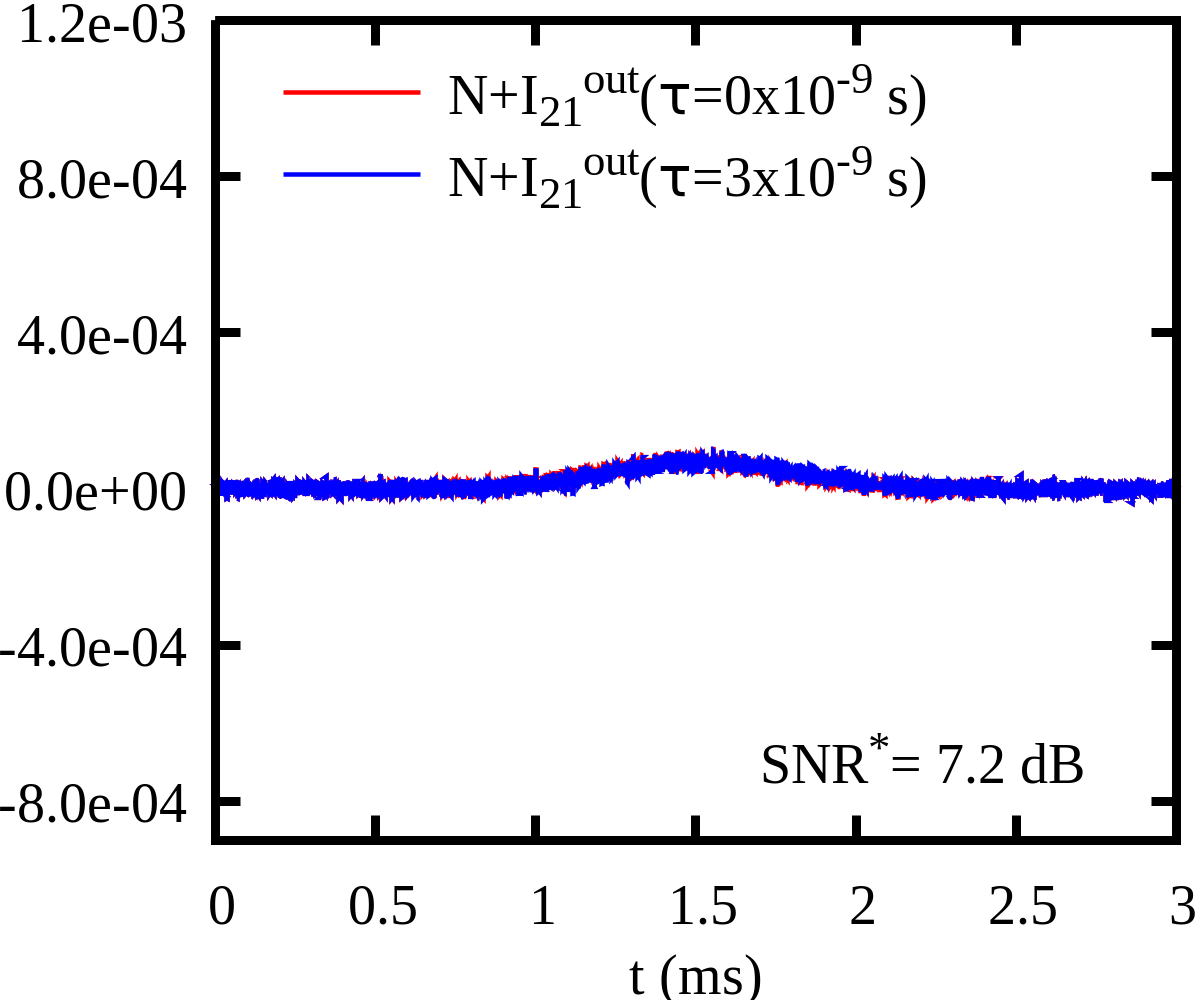}}
	\end{minipage}
}
\subfigure
{
	\vspace{-0.4cm}
	\begin{minipage}{0.31\linewidth}
	\centering
	\centerline{\includegraphics[scale=0.135,angle=0]{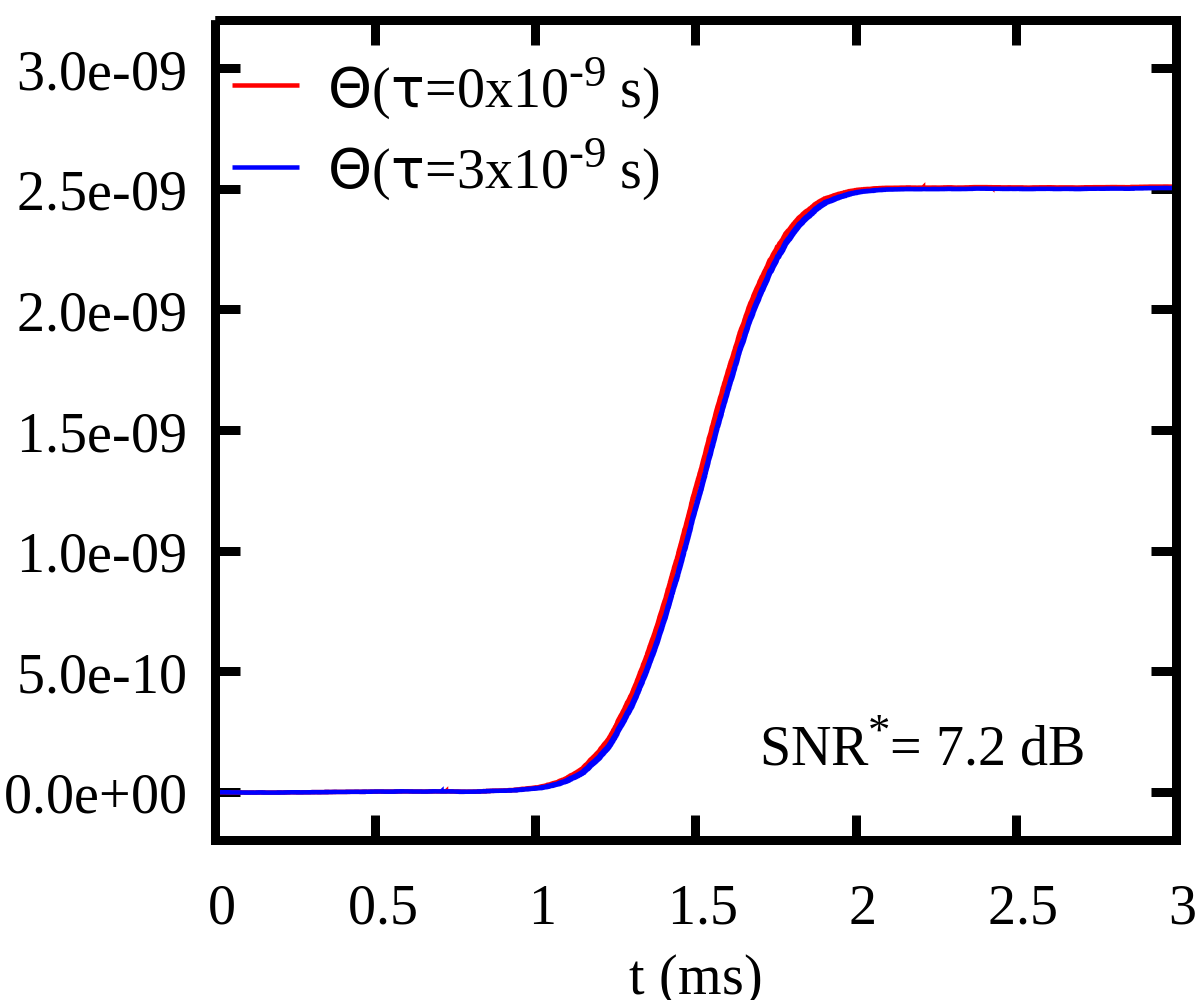}}
	\end{minipage}
}
	\vspace{-0.4cm}
	
\subfigure
{
	\vspace{-0.4cm}
	\begin{minipage}{0.31\linewidth}
	\centering
	\centerline{\includegraphics[scale=0.135,angle=0]{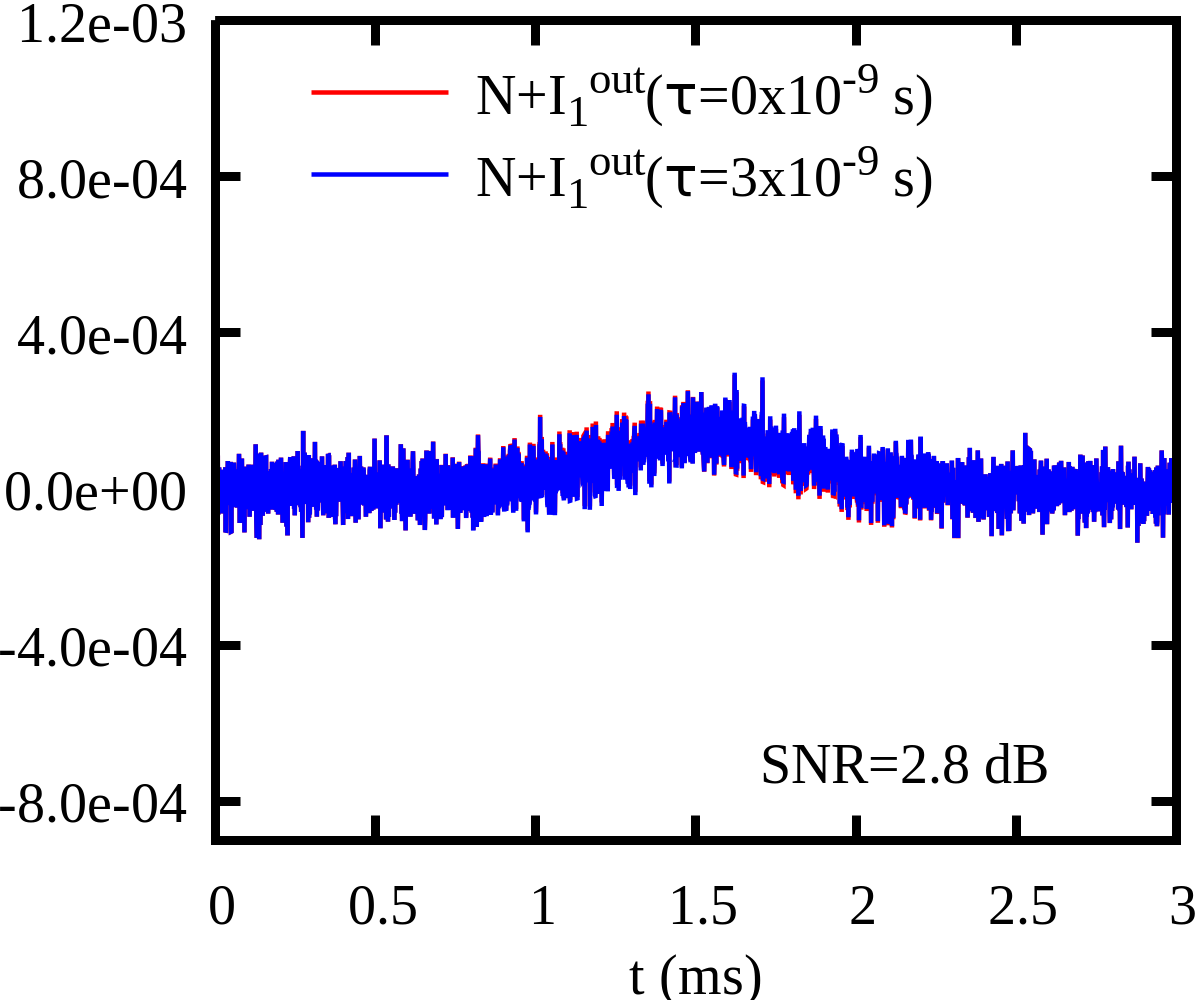}}
	\end{minipage}
}
\subfigure
{
	\vspace{-0.4cm}
	\begin{minipage}{0.31\linewidth}
	\centering
	\centerline{\includegraphics[scale=0.135,angle=0]{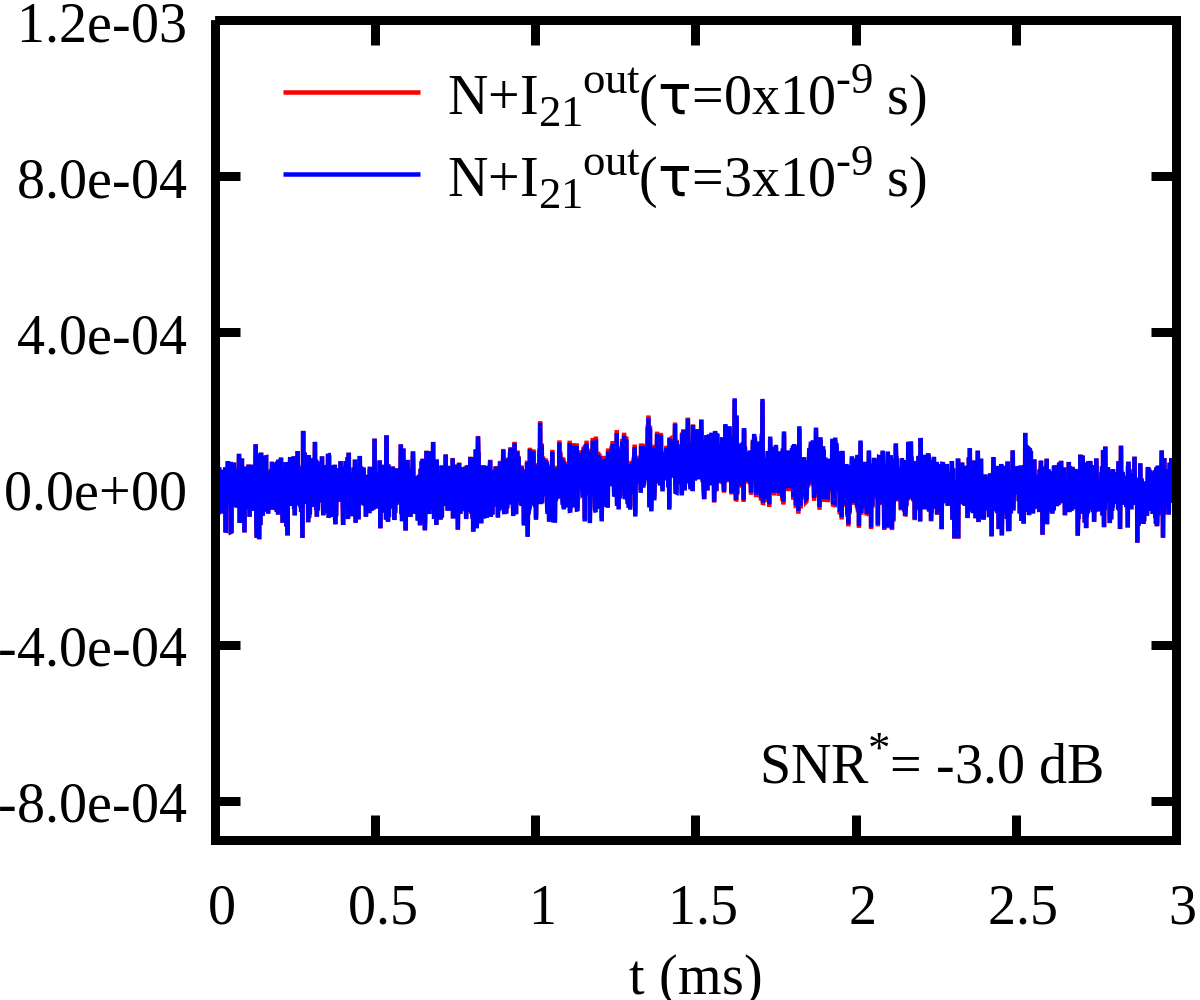}}
	\end{minipage}
}
\subfigure
{
	\vspace{-0.4cm}
	\begin{minipage}{0.31\linewidth}
	\centering
	\centerline{\includegraphics[scale=0.135,angle=0]{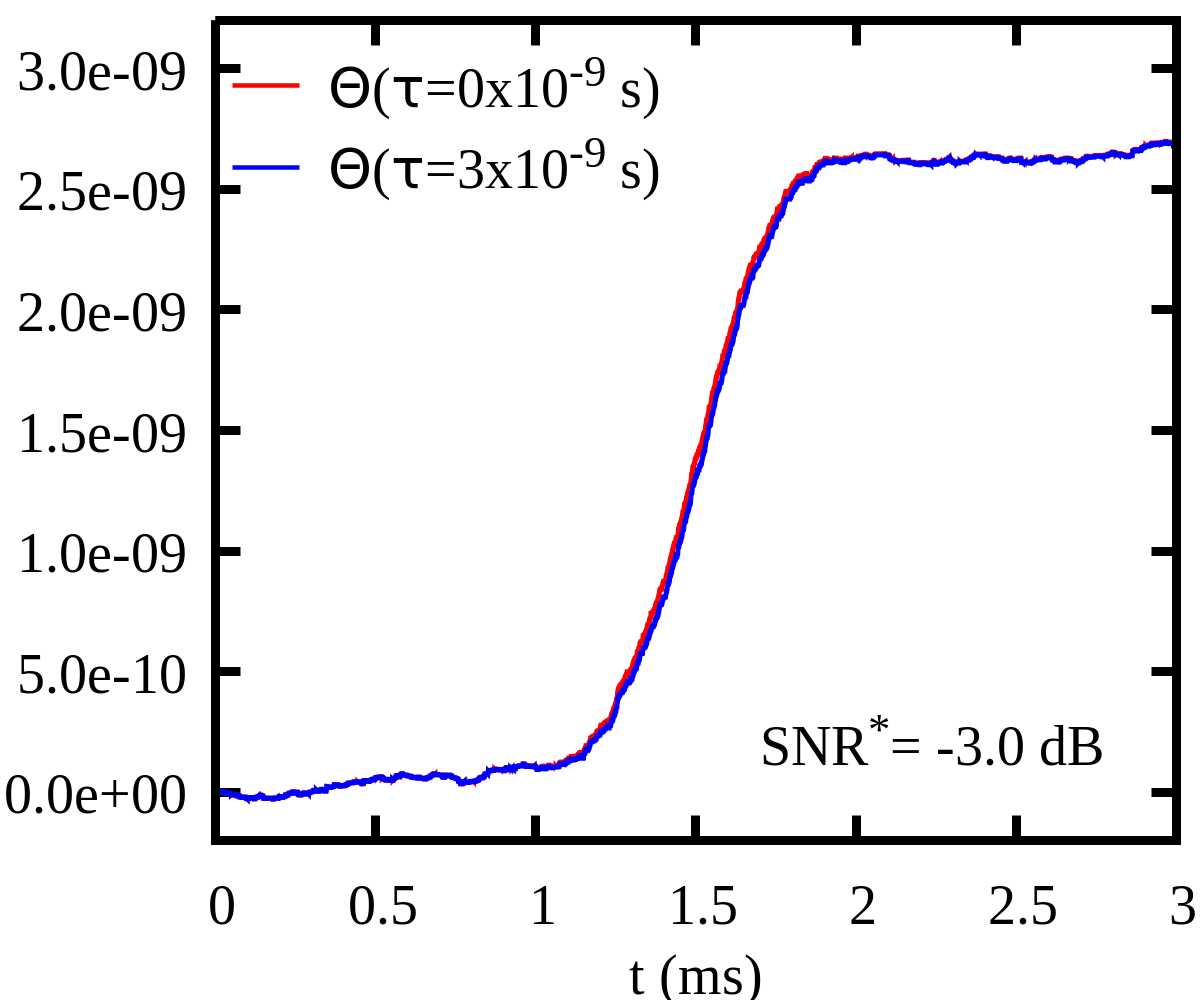}}
	\end{minipage}
}
	\vspace{-0.4cm}
	
\subfigure
{
	\vspace{-0.4cm}
	\begin{minipage}{0.31\linewidth}
	\centering
	\centerline{\includegraphics[scale=0.135,angle=0]{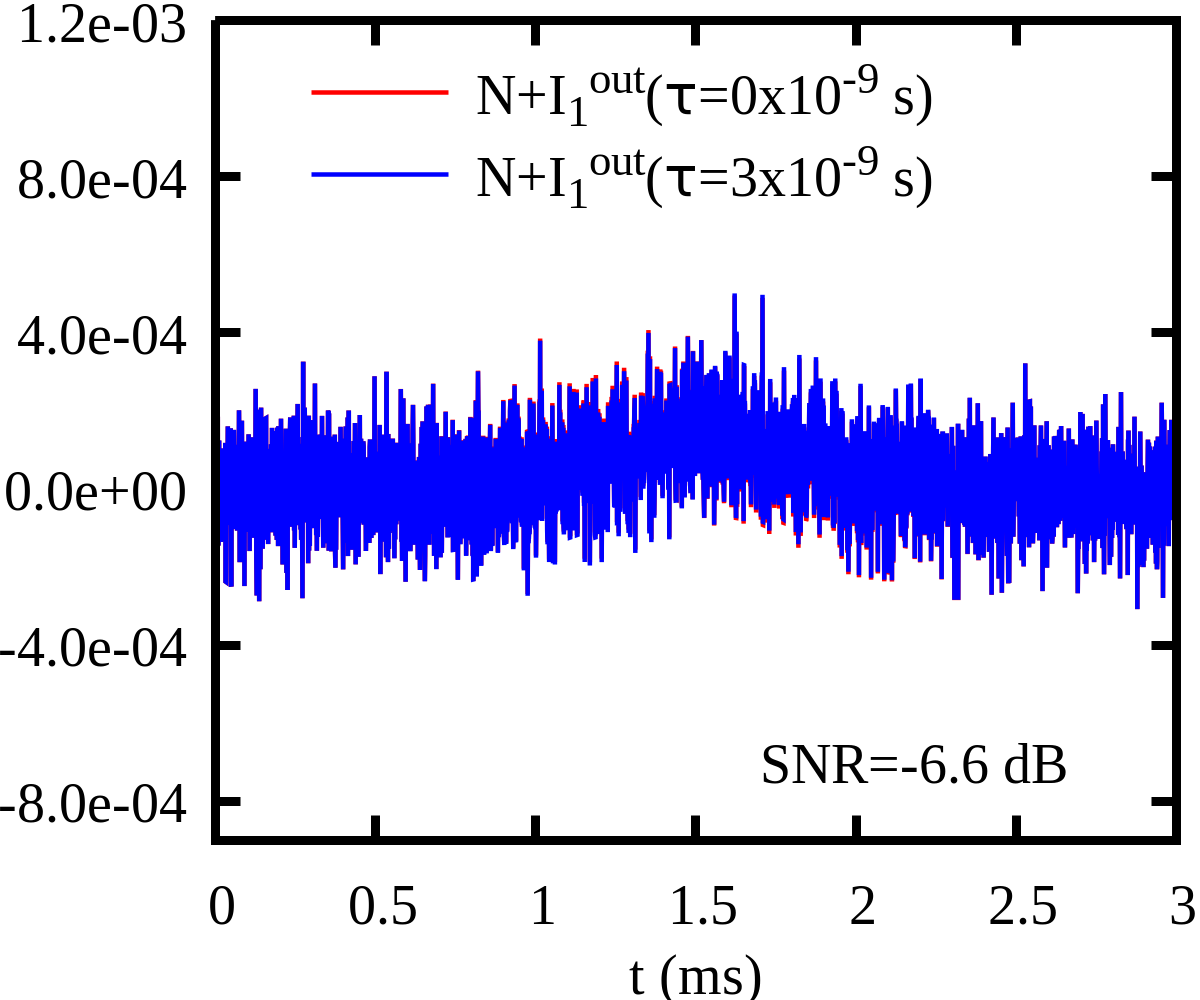}}
	\end{minipage}
}
\subfigure
{
	\vspace{-0.4cm}
	\begin{minipage}{0.31\linewidth}
	\centering
	\centerline{\includegraphics[scale=0.135,angle=0]{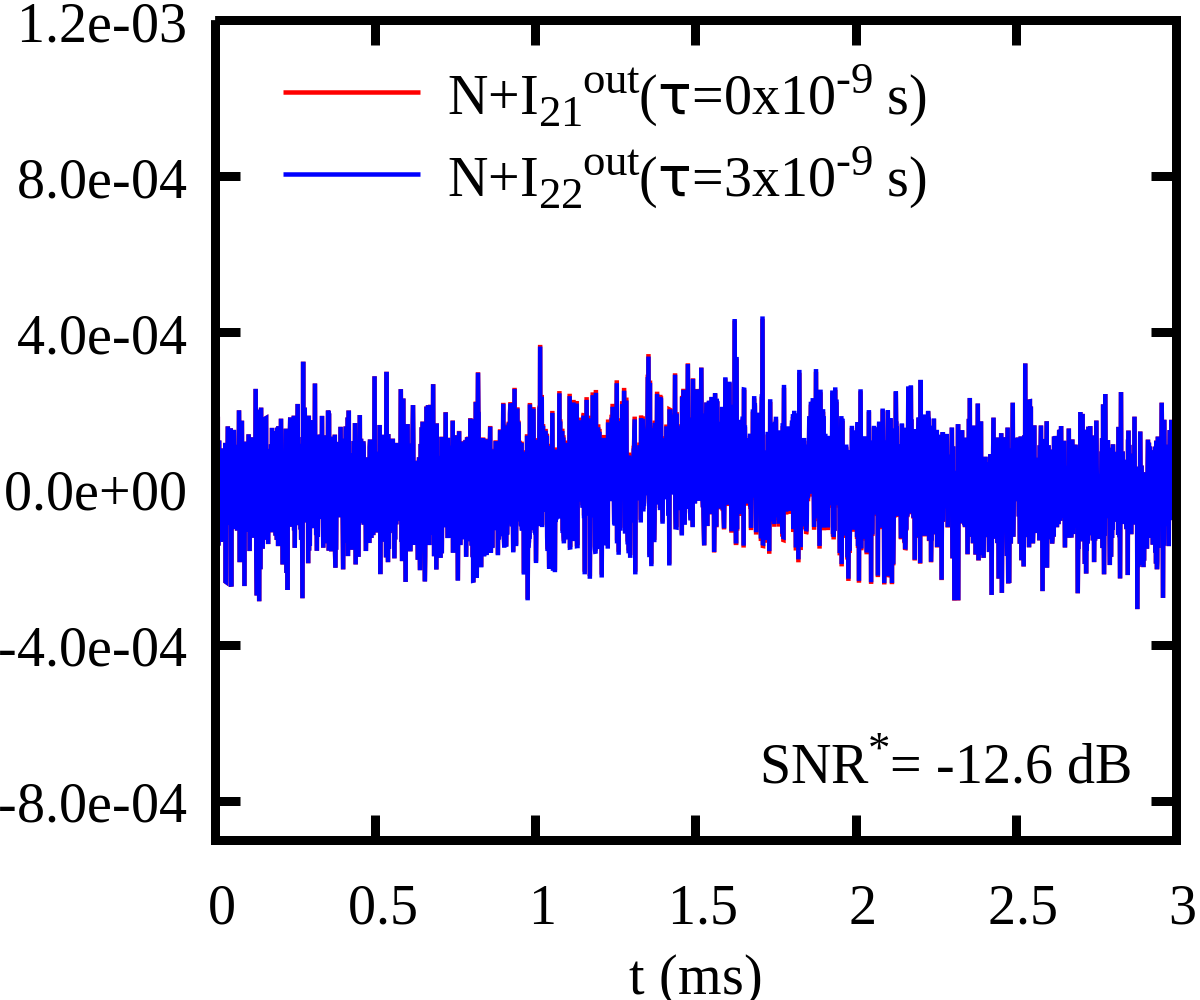}}
	\end{minipage}
}
\subfigure
{
	\vspace{-0.4cm}
	\begin{minipage}{0.31\linewidth}
	\centering
	\centerline{\includegraphics[scale=0.135,angle=0]{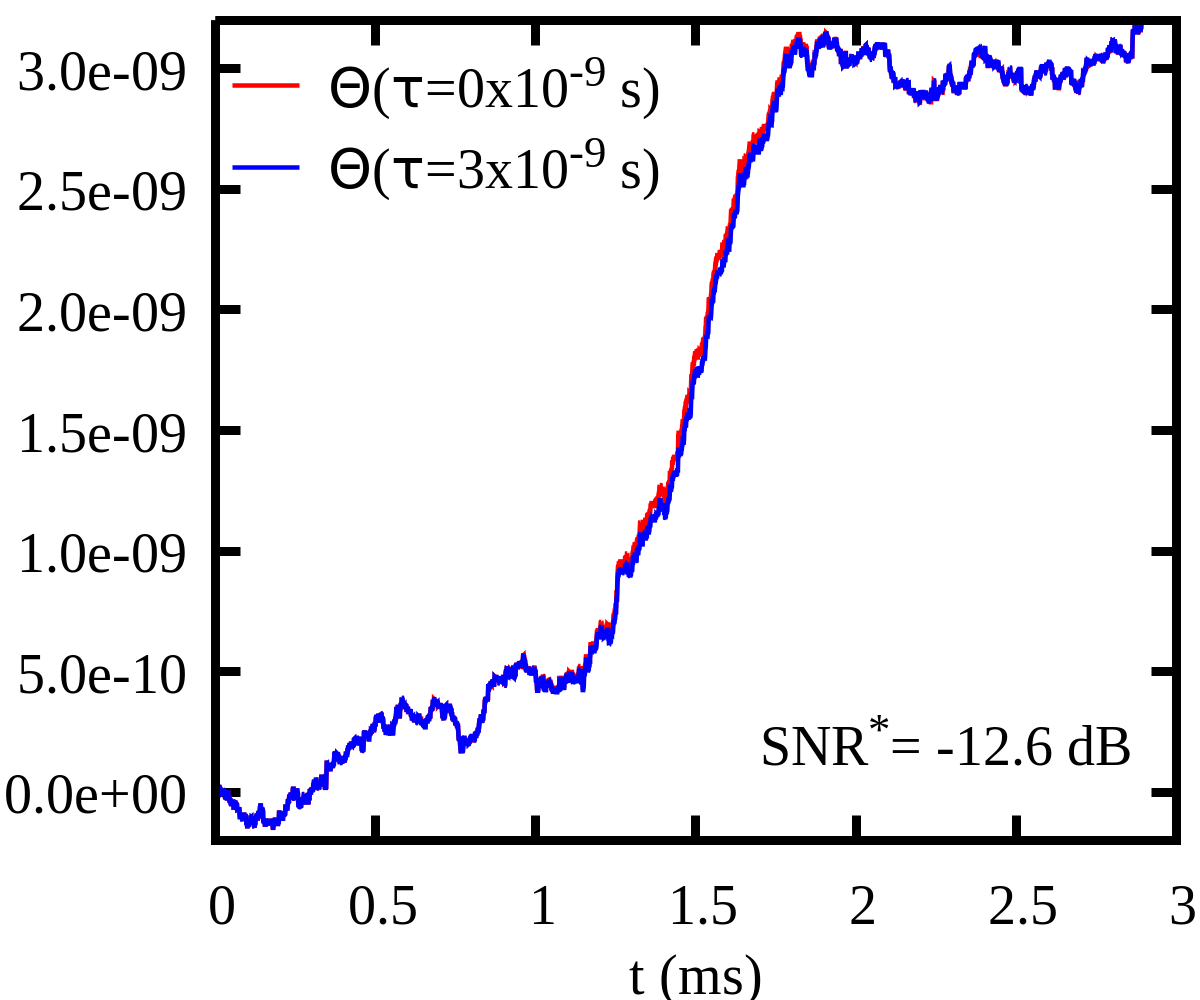}}
	\end{minipage}
}
	\vspace{-0.4cm}
	
\subfigure
{
	\vspace{-0.4cm}
	\begin{minipage}{0.31\linewidth}
	\centering
	\centerline{\includegraphics[scale=0.135,angle=0]{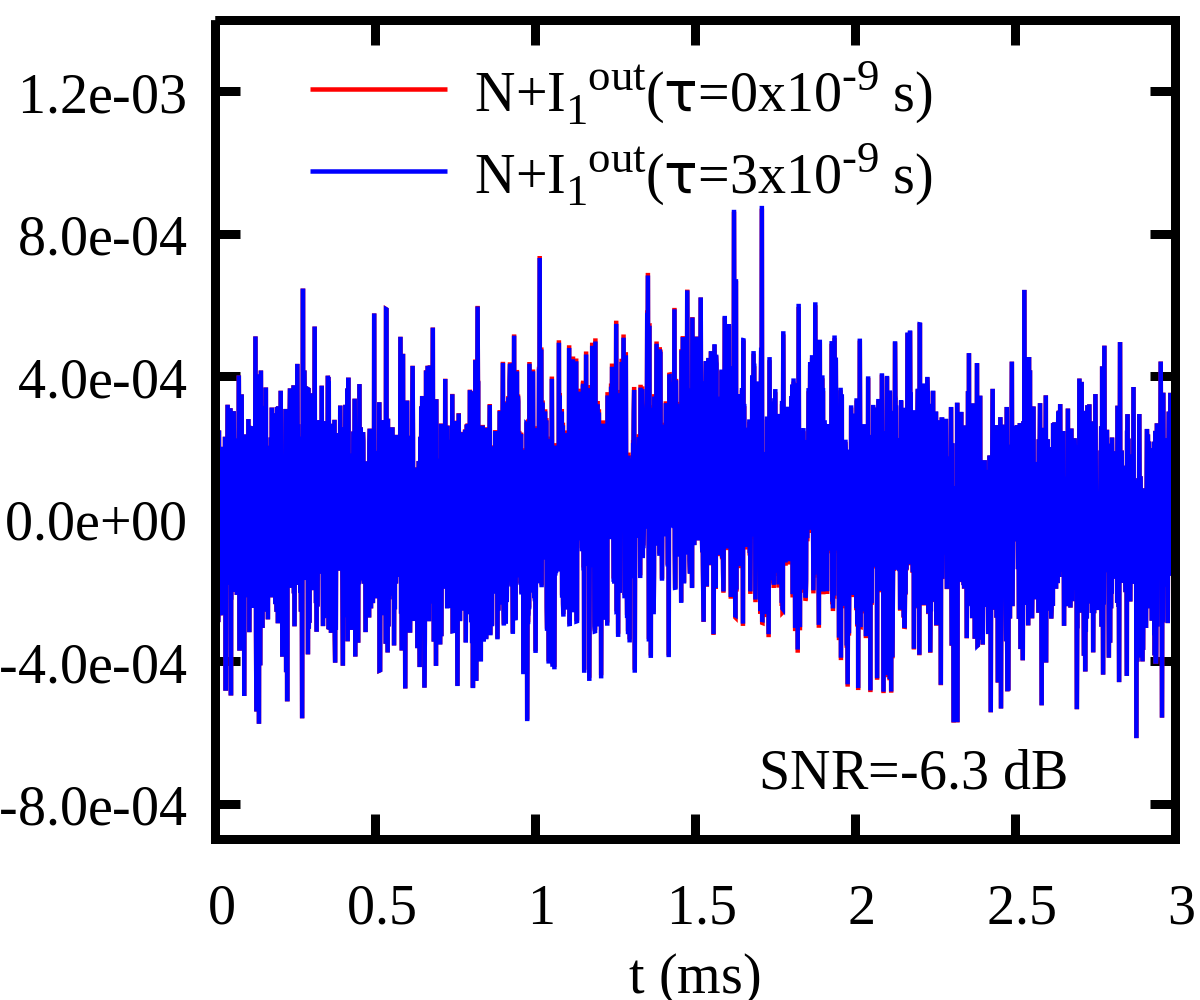}}
	\end{minipage}
}
\subfigure
{
	\vspace{-0.4cm}
	\begin{minipage}{0.31\linewidth}
	\centering
	\centerline{\includegraphics[scale=0.135,angle=0]{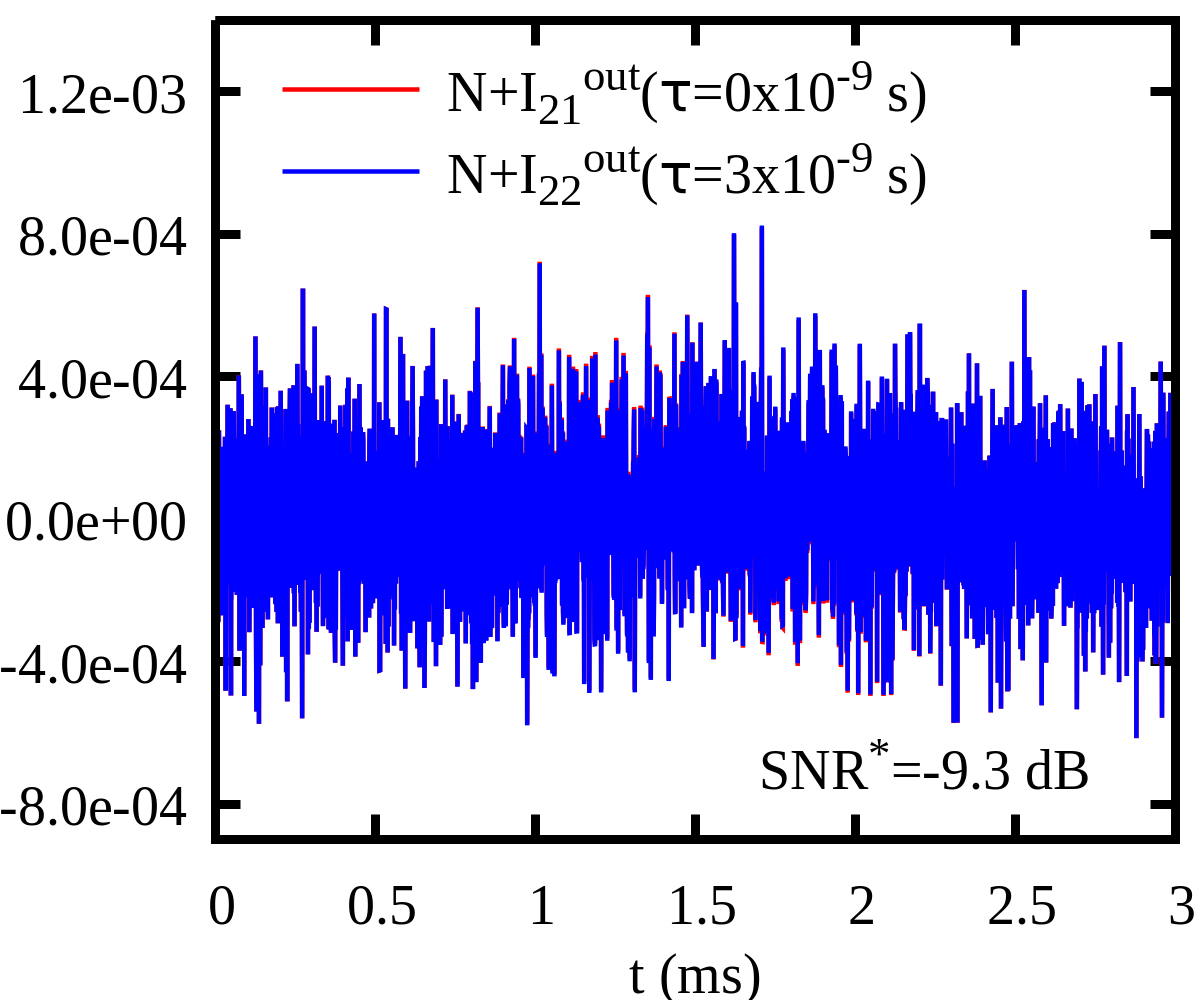}}
	\end{minipage}
}
\subfigure
{
	\vspace{-0.4cm}
	\begin{minipage}{0.31\linewidth}
	\centering
	\centerline{\includegraphics[scale=0.135,angle=0]{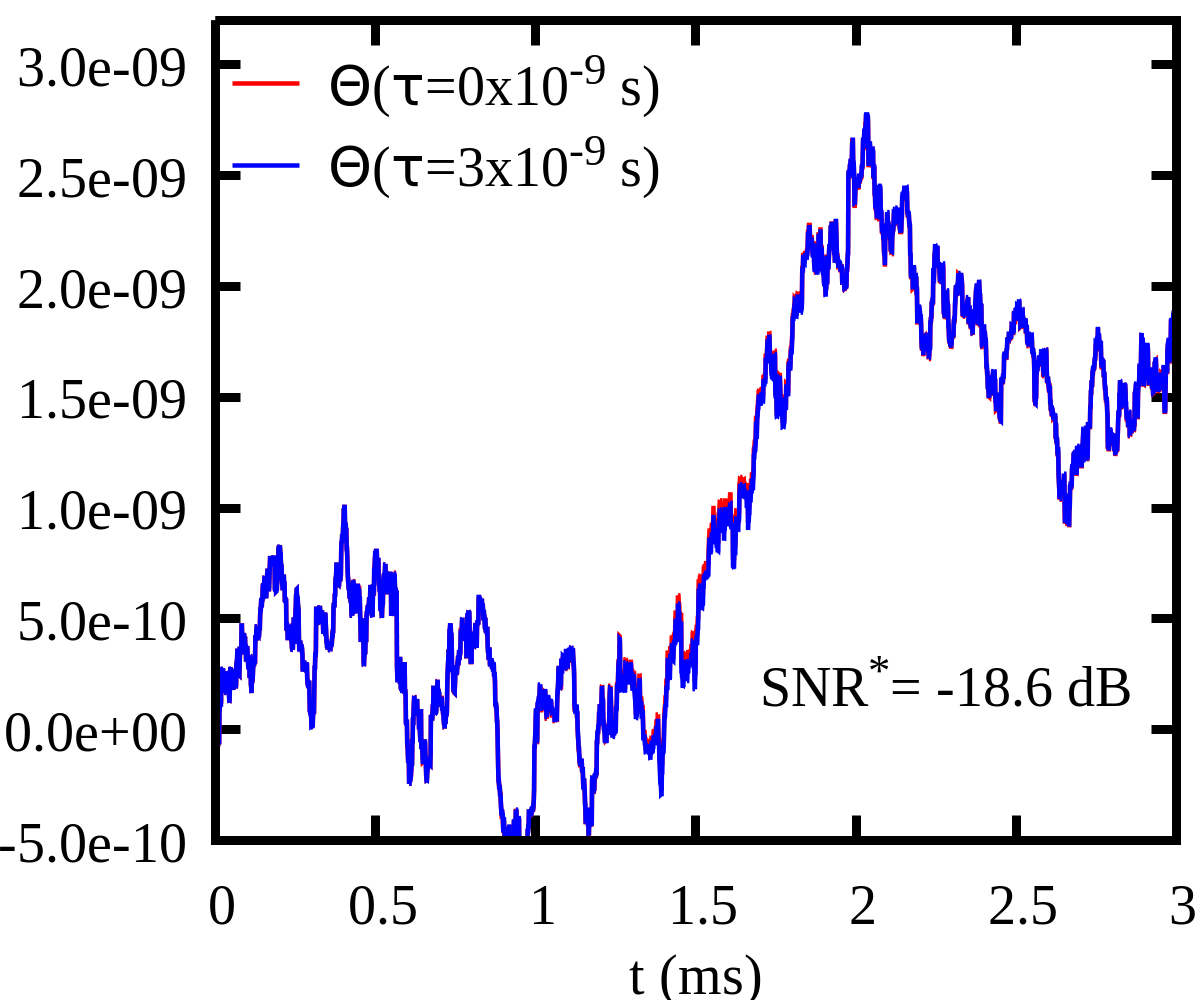}}
	\end{minipage}
}
\vspace{-0.4cm} \caption{\label{Fig:HighprobeChangeInTwoScheme}The simulation results in two schemes under the Gaussian white noise with different SNR.
{
Left panels: the signals $I_{1}^{out}(t;\tau)$ in the WVA scheme under noise $\textbf{N}(t,\sigma^{2},\xi\red{=0})$. Middle panels: the signals $I_{21}^{out}(t;\tau)$ and $I_{22}^{out}(t;\tau)$ in the AWVA scheme.
Right panels: the auto-correlative intensity $\rm \Theta$ in the AWVA scheme under noises $\textbf{N}(t,\sigma^{2},\xi_{1}\red{=0})$ and $\textbf{N}(t,\sigma^{2},\xi_{2}=700)$ with sampling frequency 1/T= 1MHz. The units of the quantities $I_{21}^{out}(t;\tau)$, $I_{21}^{out}(t;\tau)$ and $I_{22}^{out}(t;\tau)$ are units of $I_{0}$, the quality ${\rm I}^{AC}_{A}$ is units of voltage.}
}
\end{figure*}

Then, we add the noise $\textbf{N}(t,\sigma^{2};\xi)$ into both the WVA scheme and the AWVA scheme. For the measurement with the WVA technique, one \red{gets} the final signal $I_{1+\textbf{N}}^{out}(t;\tau)$ under $\textbf{N}(t,\sigma^{2},\xi_{})$ as:
\begin{eqnarray}
\label{schme1:inter_peobe_final+noise}
I_{1+\textbf{N}}^{out}(t;\tau)= I_{1}^{out}(t;\tau)   +\textbf{N}(t,\sigma^{2},\xi_{}) \, .
\end{eqnarray}
Now we need to evaluate the mean shift $\delta t$ by the Gaussian fitting the result of $I_{1+\textbf{N}}^{out}(t;\tau)$. Obviously, the noise $\textbf{N}(t,\sigma^{2},\xi)$ will lead to an uncertainty of estimating $\delta t$. The simulation results with different  simulation conditions are shown in Fig.~\ref{Fig:HighprobeChangeInTwoScheme} , Table.~\ref{Table:sensitivityOfWVA} and Table.~\ref{Table:sensitivityOfWVAdifferntTau}.
\begin{figure*}[htp!]
	\centering
\subfigure
{
	\vspace{-0.4cm}
	\begin{minipage}{0.31\linewidth}
	\centering
	\centerline{\includegraphics[scale=0.135,angle=0]{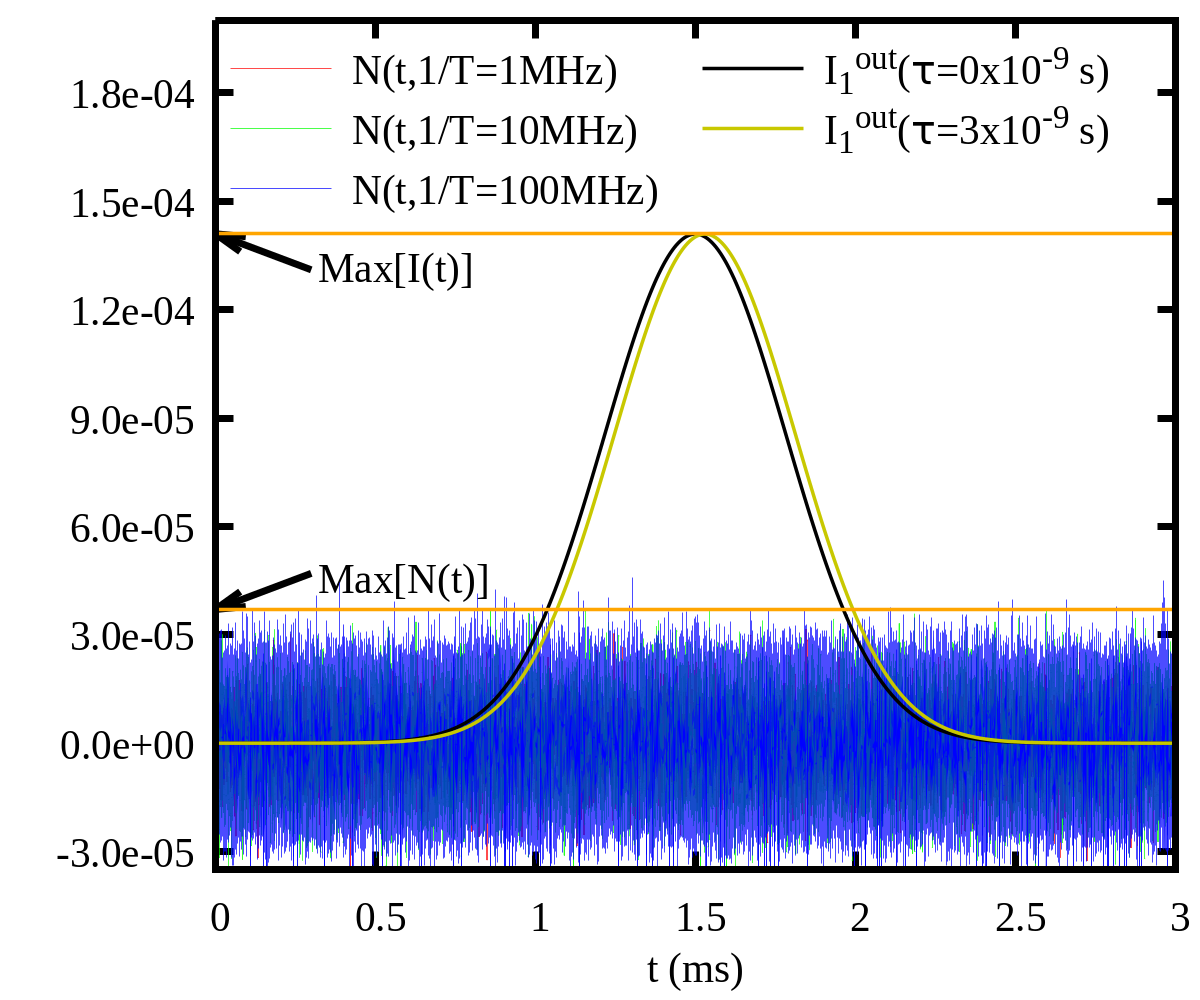}}
	\end{minipage}
}
\subfigure
{
	\vspace{-0.4cm}
	\begin{minipage}{0.31\linewidth}
	\centering
	\centerline{\includegraphics[scale=0.135,angle=0]{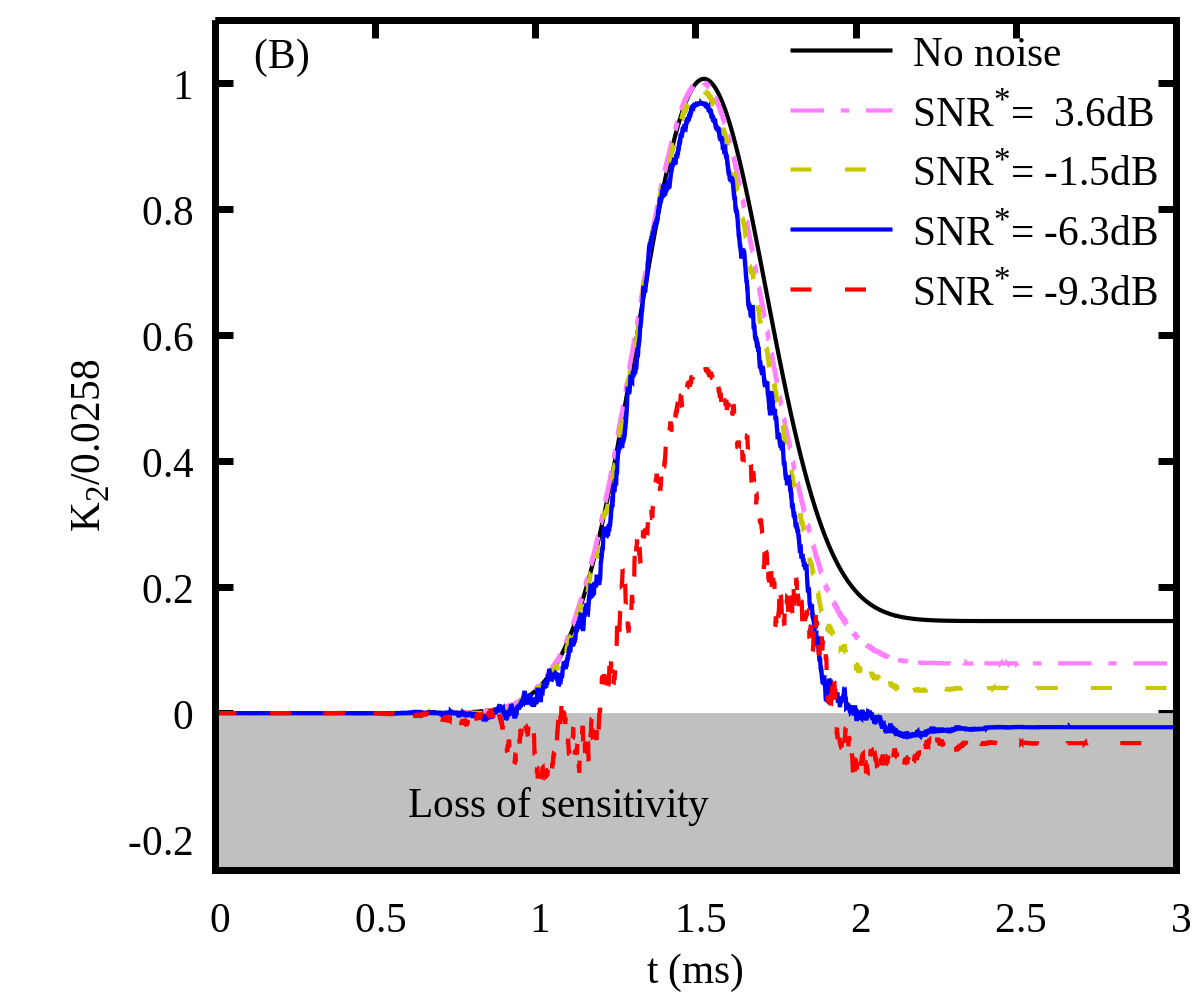}}
	\end{minipage}
}
\subfigure
{
	\vspace{-0.4cm}
	\begin{minipage}{0.31\linewidth}
	\centering
	\centerline{\includegraphics[scale=0.135,angle=0]{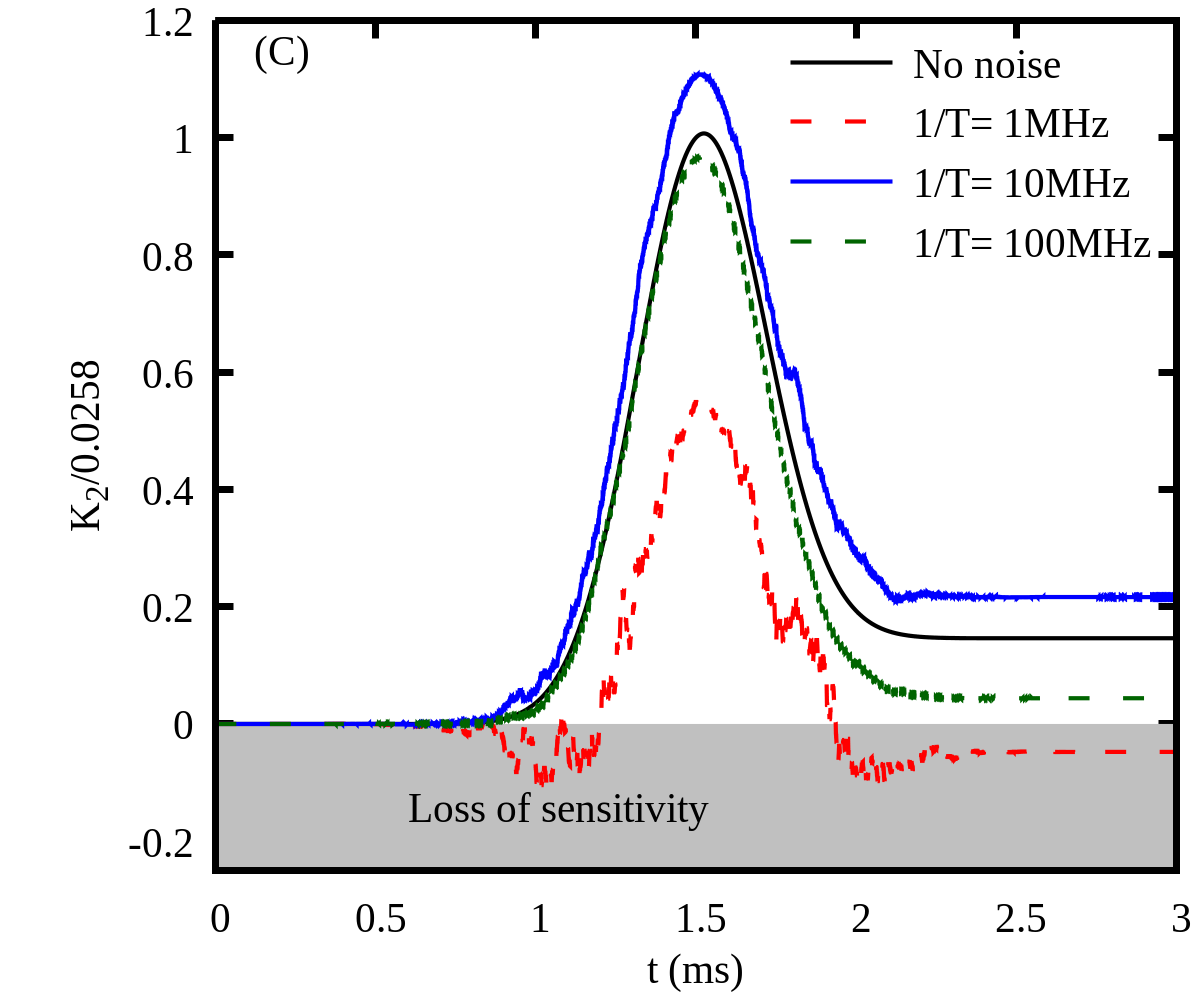}}
	\end{minipage}
}
\vspace*{0mm} \caption{\label{Fig:exzampelnoiseAndSensitivity}
{ Examples of estimating the SNR and the sensitivity the maximum sensitivity ${\rm K_{2}^{M}}$: (A). An example of estimating the SNR in WVA scheme under the Gaussian white noise $\textbf{N}(t,\sigma^{2}=10^{-11},\xi=0)$ with different sampling frequency; (B). The dependence of the sensitivity with different $\rm SNR^*$ on the integral time $t$ under the noises $\textbf{N}(t,\sigma^{2},\xi_{1}\red{=0})$ and $\textbf{N}(t,\sigma^{2},\xi_{2}=700)$ at sampling frequency 1/T= 1MHz; (C). The dependence of the sensitivity with different sampling frequency 1/T on the integral time $t$ under the noises $\textbf{N}(t,\sigma^{2},\xi_{1}\red{=0})$ and $\textbf{N}(t,\sigma^{2},\xi_{2}=700)$ at $\rm SNR^*$= -18.6 dB. The gray band represents the measurements failing to effectively detect the final signal.
}}
\end{figure*}

For the measurement with the AWVA technique, we obtain the quantity ${\rm \Theta}_{A+\textbf{N}}(\tau)$ with the noises $\textbf{N}(t,\sigma^{2},\xi_{1})$ and $\textbf{N}(t,\sigma^{2},\xi_{2})$. Then the ${\rm \Theta}_{A+\textbf{N}}(\tau)$ is defined as
\begin{eqnarray}
\label{Eq:ACIdefine+noise}
{\rm \Theta}_{A+\textbf{N}}(t;\tau)&=&\int_{0}^{t}  I_{21+\textbf{N}}^{out}(t^{\prime};\tau) \times  I_{22+\textbf{N}}^{out}(t^{\prime};\red{\tau}) dt^{\prime} \nonumber \\
&=& {\rm \Theta}_{A}(t;\tau)+{\rm \Theta}_{21\textbf{N}}(t;\tau) +  \\
&&{\rm \Theta}_{22\textbf{N}}(t;\tau)+{\rm \Theta_{NN}}(t;\tau)  \nonumber
\end{eqnarray}
with
\begin{eqnarray}
\label{Eq:ACIdefine+noise2}
{\rm \Theta}_{21\textbf{N}}(t;\tau)&=&\int_{0}^{t}  I_{21}^{out}(t^{\prime};\tau) \times  {\textbf{N}(t^{\prime},\sigma^{2},\xi_{1})} dt^{\prime}\, ,\\
{\rm \Theta}_{22\textbf{N}}(t;\tau)&=&\int_{0}^{t}  I_{22}^{out}(t^{\prime}) \times  {\textbf{N}(t^{\prime},\sigma^{2},\xi_{2})} dt^{\prime}\, .
\end{eqnarray}

However, due to the non-correlation between the signal and the random noise, we obtain that ${\rm \Theta}_{21\textbf{N}}(t;\tau)$ = ${\rm \Theta}_{22\textbf{N}}(t;\tau)$ = 0. Finally, we can get the relation ${\rm \Theta}_{A+\textbf{N}}(t;\tau)={\rm \Theta}_{A}(t;\tau)$ from the theoretical analysis when the integral time $t$ is infinite, which means that the noise has no influence on evaluating the values of $\rm \Theta$ by assuming that the noises $\textbf{N}(t,\sigma^{2},\xi_{1})$ and $\textbf{N}(t,\sigma^{2},\xi_{2})$ detected at APD1 and APD2 respectively are strictly time-independent.
{ In particular, we first
investigate the accuracy of the formula (\ref{Eq:ACIdefineWihtnoise}) using simulations on Simulink, based on Eq.~(\ref{Eq:ACIdefine+noise}).}

{
The left panel of Fig.~\ref{Fig:Variousauto-correlativeNoise} shows the auto-correlative intensity ${\rm \Theta_{NN}}(t;\tau)$ of various noises with different seeds $\xi_{1}$ and $\xi_{2}$ at $\rm SNR^*$=7.2 dB and 1/T= 1MHz. The values of ${\rm \Theta_{NN}}(t;\tau)$ with different initial time (seed) are on the order of 2.0$\times 10^{-11}$, which is smaller than the theoretical maximum value of ${\rm \Theta_{A}}(t;\tau)$ calculated from Eq.~(\ref{Eq:ACIdefine}) in Fig.~\ref{Fig:probeChangeInTwoSchemeNOnoise}. However, with the SNR of the Gaussian noise decreasing as shown in the middle panel of Fig.~\ref{Fig:Variousauto-correlativeNoise}, the values of ${\rm \Theta_{NN}}(t;\tau)$ \red{increase}. The larger values may have more negative effects on the AWVA technique. We will investigate the total auto-correlative intensity ${\rm \Theta_{A+N}}(t;\tau)$ in the next section.
Note that \blue{``}infinite" is a relative concept: {when the number of integral nodes (sample time) \red{M}$>>1$ in the integral time t}, the integral time can also be regarded as infinite. Therefore,  the auto-correlative results of noises $\textbf{N}(t,\sigma^{2},\xi_{1}\red{=0})$ and $\textbf{N}(t,\sigma^{2},\xi_{2}=700)$ at $\rm SNR^*$=-18.6 dB with different sampling frequency 1/T (T corresponding to sampling interval) are displayed in the right panel of Fig.~\ref{Fig:Variousauto-correlativeNoise}. The curves indicate that increasing the sampling rate (1/T) of APD can decrease the value of ${\rm \Theta_{NN}}(t;\tau)$ and improve the accuracy of the formula (\ref{Eq:ACIdefineWihtnoise}).
}

In conclusion, the equality ${\rm \Theta}_{A+\textbf{N}}(\tau)$ =${\rm \Theta}_{A}(\tau)$  represents that the AWVA scheme has strong robustness\footnote{The \blue{``}robustness" of the control system refers to the ability of the system to keep certain performance invariable under the disturbance(noise) of uncertainty} against \blue{noise}. The features of the realistic \blue{kinds of noise} are indeed more complicated than the features of the Gaussian noise~\cite{PhysRevA.38.5938,PhysRevD.45.2843,WU2020124253,PhysRevE.101.052205}.
However, it is necessary to first consider the effects of the strong Gaussian noise on the weak measurements \red{theoretically}, and the simulation results are shown in the next section.

\begin{figure*}[htp!]
	\centering
\subfigure
{
	\vspace{-0.4cm}
	\begin{minipage}{0.31\linewidth}
	\centering
	\centerline{\includegraphics[scale=0.135,angle=0]{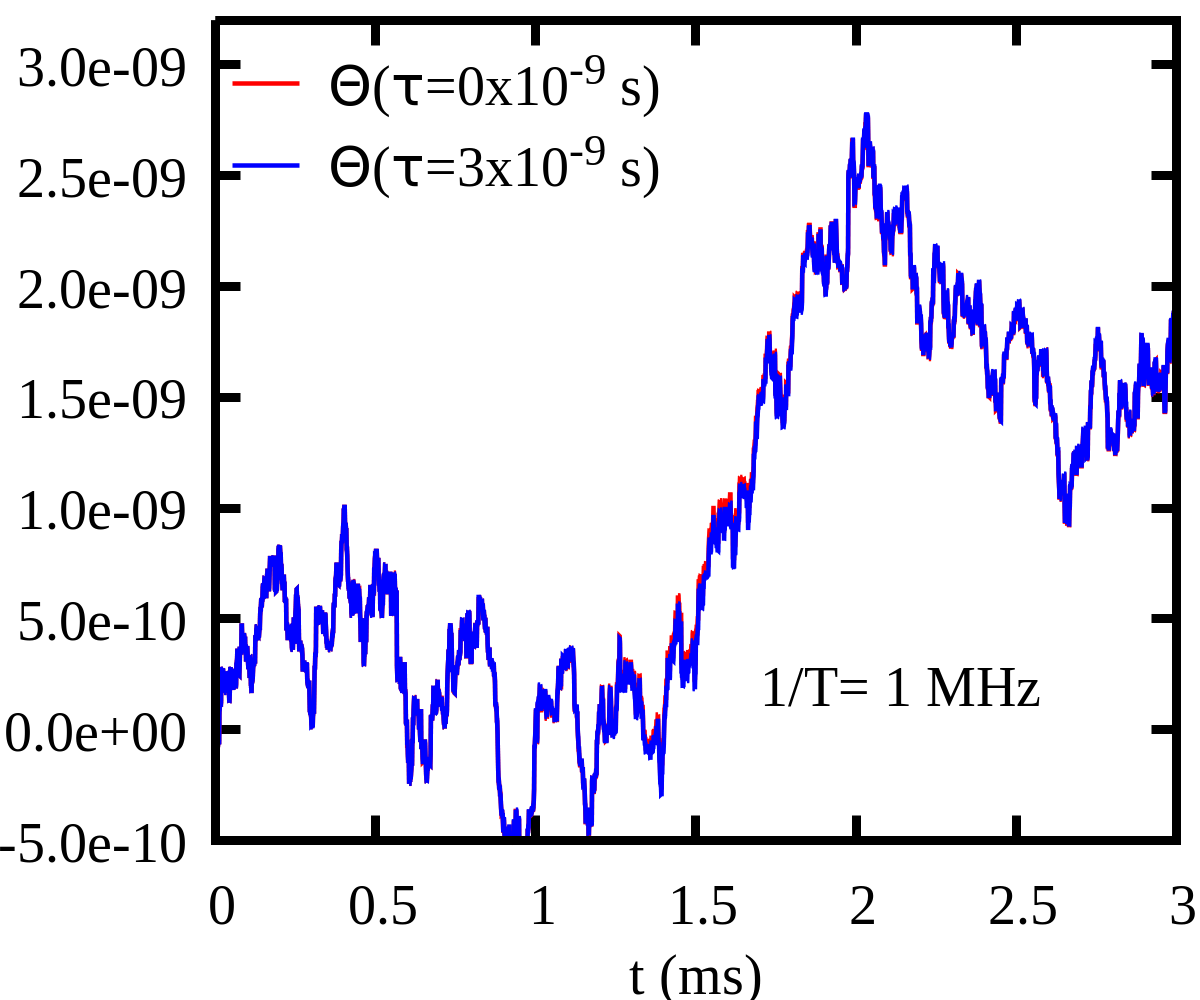}}
	\end{minipage}
}
\subfigure
{
	\vspace{-0.4cm}
	\begin{minipage}{0.31\linewidth}
	\centering
	\centerline{\includegraphics[scale=0.135,angle=0]{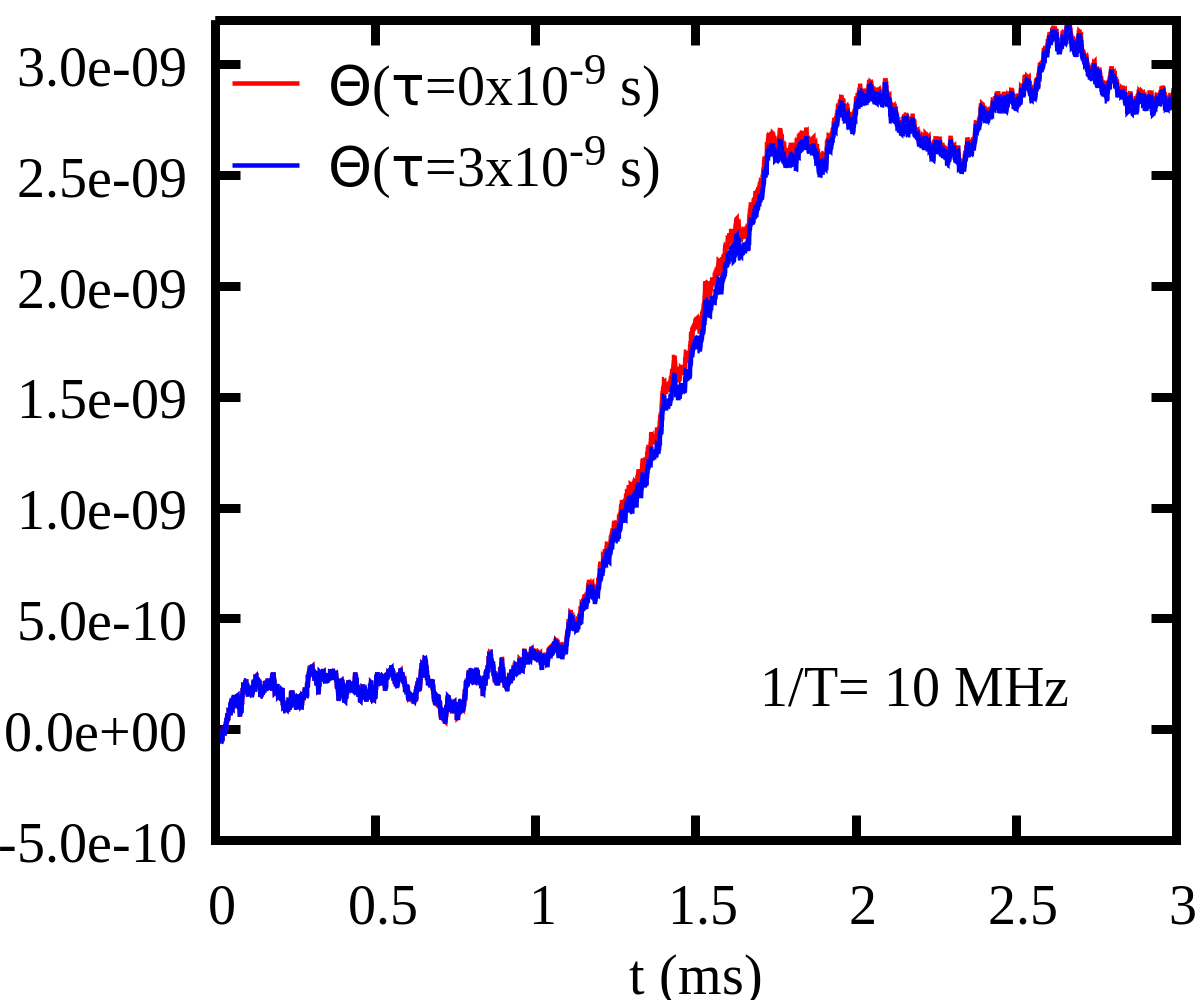}}
	\end{minipage}
}
\subfigure
{
	\vspace{-0.4cm}
	\begin{minipage}{0.31\linewidth}
	\centering
	\centerline{\includegraphics[scale=0.135,angle=0]{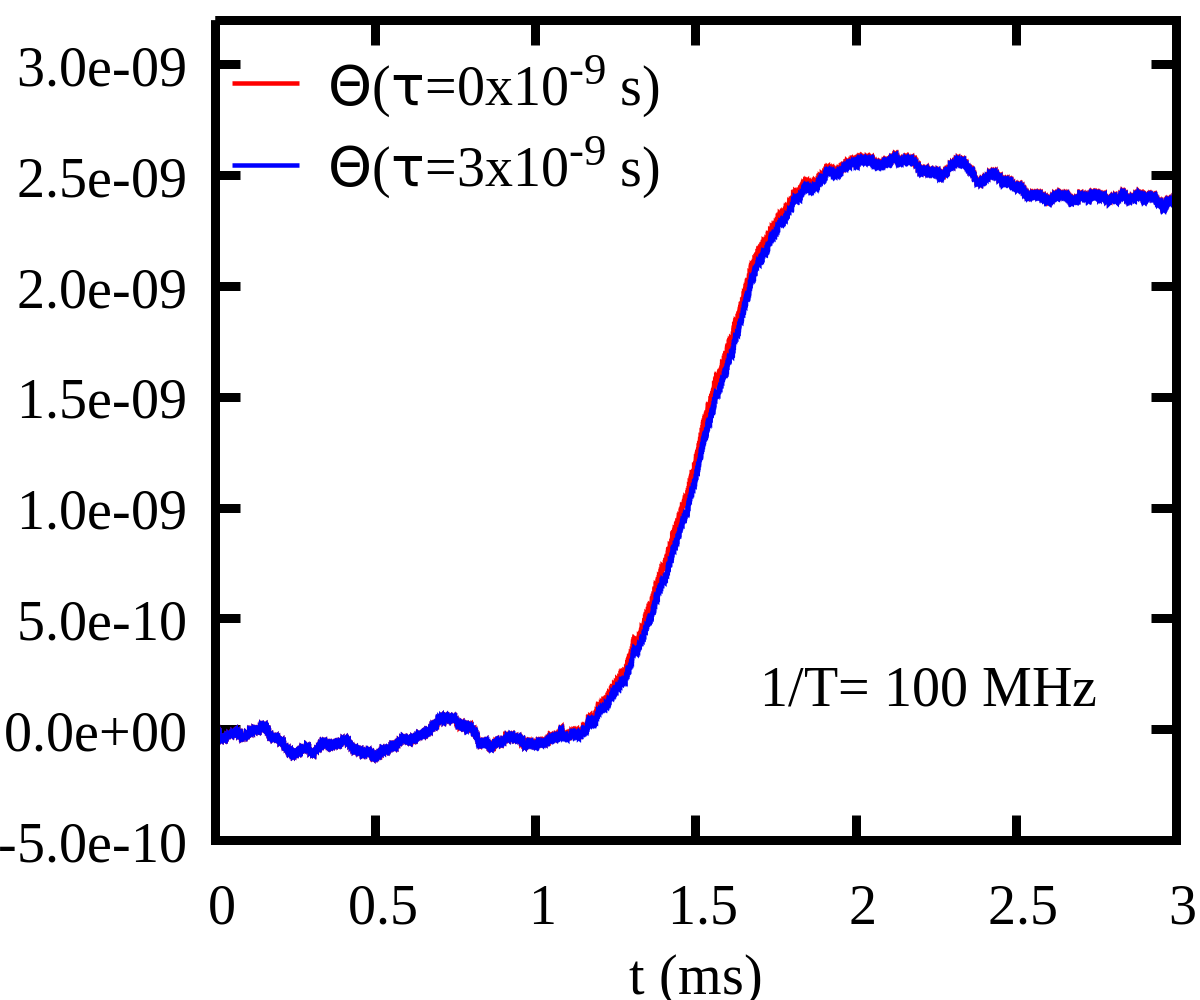}}
	\end{minipage}
}
\vspace{-0.4cm} \caption{\label{Fig:ACIschemeDifferentSamplingFrequnecy}
{
The simulation results the auto-correlative intensity $\rm \Theta$ in the AWVA scheme noises $\textbf{N}(t,\sigma^{2},\xi_{1}\red{=0})$ and $\textbf{N}(t,\sigma^{2},\xi_{2}=700)$ at $\rm SNR^*$=-18.6 dB with different sampling frequency 1/T.}
}
\end{figure*}

\begin{table*}[htp!]
\setlength{\tabcolsep}{2.8mm}{
\caption{\label{Table:sensitivityOfWVA} Parameters and some characteristic numerical results (dimensional quantities in unit of s) of measuring time shift $\tau=3.0 \times 10^{-9}$ under Gaussian noise with different SNR in the WVA scheme. The contents ${\rm E}_{t0}$ and ${\rm E}_{t \tau}$ in parentheses represent the standard error for estimating the time shifts $\delta t_{0} $ and $\delta t_{\tau} $ respectively. The data in red means the measurement is invalid. $\textbf{N}(\xi\red{=0})$, $\textbf{N}(\xi=100)$ ... $\textbf{N}(\xi=600)$ represent the multiple measurements with the different initial time.}
\begin{tabular}{rrrrr}
\toprule
\toprule
$\rm SNR$             & Noise       & $\delta t_{0} \, \rm (\pm E_{t 0})$                                                    & $\delta t_{\tau} \, \rm (\pm E_{t \tau})$                & $\rm K_{1}\, (\pm {\rm E}_{1})$                                                            \\ \hline
          -        & no-noise    & $0.00 \times 10^{-3}$ ($\pm 1.8 \times 10^{-12}$)                       & $3.00 \times 10^{-5}$ ($\pm 1.8 \times 10^{-12}$) & $10.0 \times 10^{3}$ ($\pm 1.21 \times 10^{-3}$)
          \\\cline{1-5}
                      & $\rm N(\xi= 000)$ & $9.20 \times 10^{-7}$ ($\pm 1.21 \times 10^{-6}$)                        & $3.07 \times 10^{-5}$ ($\pm 1.21 \times 10^{-6}$)  & $9.92 \times 10^{3}$ ($\pm 0.81 \times 10^{3}$)
                      \\
                      & $\rm N(\xi= 100)$ & $1.25 \times 10^{-6}$ ($\pm 1.24 \times 10^{-6}$)                        & $3.11 \times 10^{-5}$ ($\pm 1.24 \times 10^{-6}$)  & $9.95 \times 10^{3}$ ($\pm 0.82 \times 10^{3}$)
                      \\& $\rm N(\xi= 200)$ & $2.33 \times 10^{-6}$ ($\pm 1.20 \times 10^{-6}$)                        & $3.23 \times 10^{-5}$ ($\pm 1.20 \times 10^{-6}$)  & $9.99 \times 10^{3}$ ($\pm 0.80 \times 10^{3}$)
                      \\& $\rm N(\xi= 300)$ & $-1.30 \times 10^{-7}$ ($\pm 1.21 \times 10^{-6}$)                        & $2.97 \times 10^{-5}$ ($\pm 1.21 \times 10^{-6}$)  & $9.90 \times 10^{3}$ ($\pm 0.81 \times 10^{3}$)
                      \\& $\rm N(\xi= 400)$ & $-2.90 \times 10^{-7}$ ($\pm 1.18 \times 10^{-6}$)                        & $2.98 \times 10^{-5}$ ($\pm 1.18 \times 10^{-6}$)  & $9.93 \times 10^{3}$ ($\pm 0.78 \times 10^{3}$)
                      \\& $\rm N(\xi= 500)$ & $2.99 \times 10^{-6}$ ($\pm 1.20 \times 10^{-6}$)                        & $3.32\times 10^{-5}$ ($\pm 1.20 \times 10^{-6}$)  & $10.0 \times 10^{3}$ ($\pm 0.80 \times 10^{3}$)      \\
\multirow{-7}{*}{11.5}  & $\rm N(\xi= 600)$ & $-7.90 \times 10^{-7}$ ($\pm 1.21 \times 10^{-6}$)                        & $2.92 \times 10^{-5}$ ($\pm 1.21 \times 10^{-6}$)  & $9.75 \times 10^{3}$ ($\pm 0.81 \times 10^{3}$)                   \\ \cline{1-5}
                     & $\rm N(\xi= 000)$ & $3.83 \times 10^{-6}$ ($\pm 5.28 \times 10^{-6}$)                        & $3.29 \times 10^{-5}$ ($\pm 5.29 \times 10^{-6}$)  & $9.69 \times 10^{3}$ ($\pm 3.52 \times 10^{3}$)
                     \\& $\rm N(\xi= 100)$ & $5.62 \times 10^{-6}$ ($\pm 5.49 \times 10^{-6}$)                        & $3.51 \times 10^{-5}$ ($\pm 5.47 \times 10^{-6}$)  & $9.82 \times 10^{3}$ ($\pm 3.65 \times 10^{3}$)
                    \\& $\rm N(\xi= 200)$ & $1.05 \times 10^{-5}$ ($\pm 5.40 \times 10^{-6}$)                        & $4.05 \times 10^{-5}$ ($\pm 5.36 \times 10^{-6}$)  & $10.0 \times 10^{3}$ ($\pm 3.59 \times 10^{3}$)
                     \\& $\rm N(\xi= 300)$ & $-5.70 \times 10^{-7}$ ($\pm 5.28 \times 10^{-6}$)                        & $2.85 \times 10^{-5}$ ($\pm 5.29 \times 10^{-6}$)  & $9.69 \times 10^{3}$ ($\pm 3.52 \times 10^{3}$)
                     \\& $\rm N(\xi= 400)$ & $-1.33 \times 10^{-6}$ ($\pm 5.32 \times 10^{-6}$)                        & $2.89 \times 10^{-5}$ ($\pm 5.32 \times 10^{-6}$)  & $10.0 \times 10^{3}$ ($\pm 3.55 \times 10^{3}$)
                     \\& $\rm N(\xi= 500)$ & $1.36 \times 10^{-5}$ ($\pm 5.53 \times 10^{-6}$)                        & $4.45 \times 10^{-5}$ ($\pm 5.48 \times 10^{-6}$)  & $10.3 \times 10^{3}$ ($\pm 3.66 \times 10^{3}$)      \\
\multirow{-7}{*}{2.8}  & $\rm N(\xi= 600)$ & $-3.5 \times 10^{-6}$ ($\pm 5.40 \times 10^{-6}$)                        & $2.65 \times 10^{-5}$ ($\pm 5.41 \times 10^{-6}$)  & $10.0 \times 10^{3}$ ($\pm 3.60 \times 10^{3}$)                   \\
 \cline{1-5}
                      & $\rm N(\xi= 000)$& $7.63 \times 10^{-6}$ ($\pm 1.13 \times 10^{-5}$)                        & $3.57 \times 10^{-5}$ ($\pm 1.14 \times 10^{-5}$)  & $9.35 \times 10^{3}$ ($\pm 7.53 \times 10^{3}$)
                      \\& $\rm N(\xi= 100)$ & $1.26 \times 10^{-5}$ ($\pm 1.20 \times 10^{-5}$)                        & $4.13 \times 10^{-5}$ ($\pm 1.18 \times 10^{-5}$)  & $9.56 \times 10^{3}$ ($\pm 7.93 \times 10^{3}$)
                      \\& $\rm N(\xi= 200)$ & $2.38 \times 10^{-5}$ ($\pm 1.21 \times 10^{-5}$)                        & $5.37 \times 10^{-5}$ ($\pm 1.19 \times 10^{-5}$)  & $9.96 \times 10^{3}$ ($\pm 7.93 \times 10^{3}$)
                      \\& $\rm N(\xi= 300)$ & $-1.23 \times 10^{-6}$ ($\pm 1.14 \times 10^{-5}$)                        & $2.69 \times 10^{-5}$ ($\pm 1.14 \times 10^{-6}$)  & $9.37 \times 10^{3}$ ($\pm 7.60 \times 10^{3}$)
                     \\& $\rm N(\xi= 400)$ & $-3.21 \times 10^{-6}$ ($\pm 1.20 \times 10^{-5}$)                        & $2.74 \times 10^{-5}$ ($\pm 1.19 \times 10^{-5}$)  & $10.2 \times 10^{3}$ ($\pm 7.93 \times 10^{3}$)
                     \\& $\rm N(\xi= 500)$ & $1.2577 \times 10^{-3}$ ($\pm 7.91 \times 10^{-5}$)                        & $1.2682 \times 10^{-3}$ ($\pm 7.96 \times 10^{-5}$)  & {\color[HTML]{FE0000} $0.35 \times 10^{3}$ ($\pm 52.6 \times 10^{3}$)}   \\
\multirow{-7}{*}{-6.6}  & $\rm N(\xi= 600)$ & $-7.9 \times 10^{-6}$ ($\pm 1.20 \times 10^{-5}$)                        & $2.22 \times 10^{-5}$ ($\pm 1.21 \times 10^{-5}$)  & $10.0 \times 10^{3}$ ($\pm 8.00 \times 10^{3}$)                   \\
\cline{1-5}
                      & $\rm N(\xi= 000)$ & $1.23 \times 10^{-5}$ ($\pm 2.12 \times 10^{-5}$)                        & $3.87 \times 10^{-5}$ ($\pm 2.14 \times 10^{-5}$)  &     {\color[HTML]{FE0000} $8.80 \times 10^{3}$ ($\pm 14.2 \times 10^{3}$)}
                      \\& $\rm N(\xi= 100)$ & $2.52 \times 10^{-5}$ ($\pm 2.28 \times 10^{-5}$)                        & $5.22 \times 10^{-5}$ ($\pm 2.23 \times 10^{-5}$)  & {\color[HTML]{FE0000} $9.82 \times 10^{3}$ ($\pm 15.1 \times 10^{3}$)}
                      \\& $\rm N(\xi= 200)$ & $7.61 \times 10^{-4}$ ($\pm 3.93 \times 10^{-5}$)                        & $7.64 \times 10^{-4}$ ($\pm 3.98 \times 10^{-5}$)  & {\color[HTML]{FE0000} $0.12 \times 10^{3}$ ($\pm 26.4 \times 10^{3}$)}
                      \\& $\rm N(\xi= 300)$ & $-2.40 \times 10^{-6}$ ($\pm 2.14 \times 10^{-5}$)                        & $2.41 \times 10^{-5}$ ($\pm 2.15 \times 10^{-5}$)  & {\color[HTML]{FE0000} $8.83 \times 10^{3}$ ($\pm 14.3 \times 10^{3}$)}
                     \\& $\rm N(\xi= 400)$ & $-6.60 \times 10^{-6}$ ($\pm 2.42 \times 10^{-5}$)                        & $2.41 \times 10^{-5}$ ($\pm 2.42 \times 10^{-5}$)  & {\color[HTML]{FE0000}  $10.2 \times 10^{3}$ ($\pm 16.1 \times 10^{3}$)}             \\& $\rm N(\xi= 500)$ & $1.4261 \times 10^{-3}$ ($\pm 30.0 \times 10^{-5}$)                        & $1.4259 \times 10^{-3}$ ($\pm 29.2 \times 10^{-5}$)  & {\color[HTML]{FE0000}  $-0.06 \times 10^{3}$ ($\pm 190 \times 10^{3}$)}     \\
\multirow{-7}{*}{-13.2}  & $\rm N(\xi= 600)$ & $-1.0995 \times 10^{-3}$ ($\pm 3.75 \times 10^{-5}$)                        & $-1.0961\times 10^{-3}$ ($\pm 3.89 \times 10^{-5}$)  & {\color[HTML]{FE0000}  $1.13 \times 10^{3}$ ($\pm 25.4\times 10^{3}$)}                        \\
 \bottomrule \bottomrule
\end{tabular}
}
\end{table*}

\begin{table*}[htp!]
\setlength{\tabcolsep}{2.5mm}{
\caption{\label{Table:sensitivityOfAWVA} Parameters and some characteristic numerical results of measuring time shift $\tau=3.0 \times 10^{-9}$ under Gaussian noise with different SNR in the AWVA scheme. 1/T (units of MHz) represents sampling frequency of APD.
}
\begin{tabular}{rrrrrrrrr}
\toprule
\toprule
$\rm SNR^{*}$
&  1/T
&  $\rm N(t,\xi_{1})$
&  $\rm N(t,\xi_{2})$
& $\rm \Theta_{0}$
& $\rm \Theta_{\tau}$
& $\Delta {\rm \Theta}$
& $\rm K_{2}^{M}$
& $\rm \overline{K}_{2}(\pm {\rm E}_{2})$           \\ \hline
                     -       &  -       & no-noise               & no-noise    & $1.2442 \times 10^{-9}$                        & $1.1666 \times 10^{-9}$               & $7.76 \times 10^{-11}$ & 0.0258                        \\ \hline
                   & 1  &    $\xi_{1}= 000$                   & $\xi_{2}= 700$ & $1.2462 \times 10^{-9}$                        & $1.1678 \times 10^{-9}$               & $7.84 \times 10^{-11}$ & 0.0261                        \\
                   & 1 &    $\xi_{1}= 100$                   & $\xi_{2}= 710$ &  $1.2427 \times 10^{-9}$                        & $1.1657 \times 10^{-9}$               & $7.70 \times 10^{-11}$ & 0.0256                        \\
                   & 1  &    $\xi_{1}= 200$                   & $\xi_{2}= 720$ &  $1.2452 \times 10^{-9}$                        & $1.1654 \times 10^{-9}$               & $7.98 \times 10^{-11}$ & 0.0266                       \\
                   & 1   & $\xi_{1}= 300$                   & $\xi_{2}= 730$ &  $1.2440 \times 10^{-9}$                        & $1.1674 \times 10^{-9}$               & $7.66 \times 10^{-11}$ & 0.0255                        \\
                   & 1   &    $\xi_{1}= 400$                   & $\xi_{2}= 740$ &  $1.2454 \times 10^{-9}$                        & $1.1680 \times 10^{-9}$               & $7.74 \times 10^{-11}$ & 0.0258                       \\
                   & 1  &    $\xi_{1}= 500$                   & $\xi_{2}= 750$ &  $1.2322 \times 10^{-9}$                        & $1.1545 \times 10^{-9}$               & $7.77 \times 10^{-11}$ & 0.0259                                            \\
\multirow{-7}{*}{7.2} & 1 & $\xi_{1}= 600$& $\xi_{2}= 770 $ & $1.2591 \times 10^{-9}$                        & $1.1817 \times 10^{-9}$               & $7.74 \times 10^{-11}$ & 0.0258 &\multirow{-7}{*}{0.02590($\pm 0.0007$)}                        \\ \cline{1-9}
                   & 1  &    $\xi_{1}= 000$                   & $\xi_{2}= 700$ & $1.1508 \times 10^{-9}$                        & $1.0698 \times 10^{-9}$               & $8.10 \times 10^{-11}$ & 0.0270                        \\
                   & 1  &    $\xi_{1}= 100$                   & $\xi_{2}= 710$ &  $1.3055 \times 10^{-9}$                        & $1.2308 \times 10^{-9}$               & $7.47 \times 10^{-11}$ & 0.0249                        \\
                   & 1  &    $\xi_{1}= 200$                   & $\xi_{2}= 720$ &  $1.1987 \times 10^{-9}$                        & $1.1117 \times 10^{-9}$               & $8.70 \times 10^{-11}$ & 0.0290                       \\
                   & 1  &    $\xi_{1}= 300$                   & $\xi_{2}= 730$ &  $1.2190 \times 10^{-9}$                        & $1.1465 \times 10^{-9}$               & $7.25 \times 10^{-11}$ & 0.0241                        \\
                   & 1  &    $\xi_{1}= 400$                   & $\xi_{2}= 740$ &  $1.2657 \times 10^{-9}$                        & $1.1893 \times 10^{-9}$               & $7.64 \times 10^{-11}$ & 0.0254                       \\
                   & 1  &    $\xi_{1}= 500$                   & $\xi_{2}= 750$ &  $1.1748 \times 10^{-9}$                        & $1.0971 \times 10^{-9}$               & $7.77 \times 10^{-11}$ & 0.0259                                            \\
\multirow{-7}{*}{-3.0} &  1 &$\xi_{1}= 600$& $\xi_{2}= 770 $ & $1.3835 \times 10^{-9}$                        & $1.3072 \times 10^{-9}$               & $7.63 \times 10^{-11}$ & 0.0254 &\multirow{-7}{*}{0.02595($\pm 0.0030$)}                        \\ \cline{1-9}
                     & 1 &    $\xi_{1}= 000$                   & $\xi_{2}= 700$ & $0.6719 \times 10^{-9}$                        & $0.5867 \times 10^{-9}$               & $8.52 \times 10^{-11}$ & 0.0284                        \\
                    & 1 &    $\xi_{1}= 100$                   & $\xi_{2}= 710$ &  $1.6234 \times 10^{-9}$                        & $1.5524 \times 10^{-9}$               & $7.10 \times 10^{-11}$ & 0.0236                        \\
                    & 1 &    $\xi_{1}= 200$                   & $\xi_{2}= 720$ &  $0.9655 \times 10^{-9}$                        & $0.8671 \times 10^{-9}$               & $9.84 \times 10^{-11}$ & 0.0328                       \\
                    & 1 &    $\xi_{1}= 300$                   & $\xi_{2}= 730$ &  $1.1025 \times 10^{-9}$                        & $1.0363 \times 10^{-9}$               & $6.62 \times 10^{-11}$ & 0.0220                        \\
                    & 1 &    $\xi_{1}= 400$                   & $\xi_{2}= 740$ &  $1.3512 \times 10^{-9}$                        & $1.2762 \times 10^{-9}$               & $7.50 \times 10^{-11}$ & 0.0250                       \\
                    & 1 &    $\xi_{1}= 500$                   & $\xi_{2}= 750$ &  $1.0343 \times 10^{-9}$                        & $0.9565 \times 10^{-9}$               & $7.78 \times 10^{-11}$ & 0.0259                                            \\
\multirow{-7}{*}{-12.6}  & 1 & $\xi_{1}= 600$& $\xi_{2}= 770 $ & $1.8158 \times 10^{-9}$                        & $1.7411 \times 10^{-9}$               & $7.47 \times 10^{-11}$ & 0.0249 &\multirow{-7}{*}{0.02608($\pm 0.0067$)}                        \\ \cline{1-9}
                     & 1 &    $\xi_{1}= 000$                   & $\xi_{2}= 700$ & $1.3079 \times 10^{-9}$                        & $1.2151 \times 10^{-9}$               & $9.28 \times 10^{-11}$ & 0.0309                        \\
                    & 1 &    $\xi_{1}= 100$                   & $\xi_{2}= 710$ &  $2.8789 \times 10^{-9}$                        & $2.8145 \times 10^{-9}$               & $6.44 \times 10^{-11}$ & 0.0214                        \\
                     & 1 &    $\xi_{1}= 200$                   & $\xi_{2}= 720$ &  $0.4317 \times 10^{-9}$                        & $0.3892 \times 10^{-9}$               & $4.25 \times 10^{-11}$ & 0.0141                       \\
                    & 1 &    $\xi_{1}= 300$                   & $\xi_{2}= 730$ &  $0.6522 \times 10^{-9}$                        & $0.5974 \times 10^{-9}$               & $5.48 \times 10^{-11}$ & 0.0182                        \\
                     & 1 &    $\xi_{1}= 400$                   & $\xi_{2}= 740$ &  $1.6710 \times 10^{-9}$                        & $1.5987 \times 10^{-9}$               & $7.23 \times 10^{-11}$ & 0.0241                       \\
                    & 1  &    $\xi_{1}= 500$                   & $\xi_{2}= 750$ &  $0.6262 \times 10^{-9}$                        & $0.5484 \times 10^{-9}$               & $7.78 \times 10^{-11}$ & 0.0259                                            \\
\multirow{-7}{*}{-18.6}  & 1 & $\xi_{1}= 600$& $\xi_{2}= 770 $ & $3.3287 \times 10^{-9}$                        & $3.2569 \times 10^{-9}$               & $7.18 \times 10^{-11}$ & 0.0239 &\multirow{-7}{*}{0.02264($\pm 0.0085$)}   \\ \cline{1-9}
                    & 10 &    $\xi_{1}= 000$                   & $\xi_{2}= 700$ & $1.5115 \times 10^{-9}$                        & $1.4299 \times 10^{-9}$               & $8.16 \times 10^{-11}$ & 0.0272                        \\
                    & 10 &    $\xi_{1}= 100$                   & $\xi_{2}= 710$ &  $2.2062 \times 10^{-9}$                        & $2.1381 \times 10^{-9}$               & $6.81 \times 10^{-11}$ & 0.0227                       \\
                     & 10 &    $\xi_{1}= 200$                   & $\xi_{2}= 720$ &  $0.8213 \times 10^{-9}$                        & $0.7503 \times 10^{-9}$               & $7.10 \times 10^{-11}$ & 0.0236                       \\
                    & 10 &    $\xi_{1}= 300$                   & $\xi_{2}= 730$ &  $1.5151 \times 10^{-9}$                        & $1.4387 \times 10^{-9}$               & $7.64 \times 10^{-11}$ & 0.0254                       \\
                     & 10 &    $\xi_{1}= 400$                   & $\xi_{2}= 740$ &  $1.8597 \times 10^{-9}$                        & $1.7841 \times 10^{-9}$               & $7.56 \times 10^{-11}$ & 0.0252                       \\
                    & 10  &    $\xi_{1}= 500$                   & $\xi_{2}= 750$ &  $1.2757 \times 10^{-9}$                        & $1.1975 \times 10^{-9}$               & $7.82 \times 10^{-11}$ & 0.0260                                           \\
\multirow{-7}{*}{-18.6}  & 10 & $\xi_{1}= 600$& $\xi_{2}= 770 $ & $1.8199 \times 10^{-9}$                        & $1.7344 \times 10^{-9}$               & $8.55 \times 10^{-11}$ & 0.0285 &\multirow{-7}{*}{0.02551($\pm 0.0028$)}
\\ \cline{1-9}
                     & 100 &    $\xi_{1}= 000$                   & $\xi_{2}= 700$ & $1.2195 \times 10^{-9}$                        & $1.1449 \times 10^{-9}$               & $7.46 \times 10^{-11}$ & 0.0248                        \\
                    & 100 &    $\xi_{1}= 100$                   & $\xi_{2}= 710$ &  $1.5641 \times 10^{-9}$                        & $1.4852 \times 10^{-9}$               & $7.89 \times 10^{-11}$ & 0.0263                        \\
                     & 100 &    $\xi_{1}= 200$                   & $\xi_{2}= 720$ &  $0.9090 \times 10^{-9}$                        & $0.8319 \times 10^{-9}$               & $7.71 \times 10^{-11}$ & 0.0257                      \\
                    & 100 &    $\xi_{1}= 300$                   & $\xi_{2}= 730$ &  $1.2707 \times 10^{-9}$                        & $1.1929 \times 10^{-9}$               & $7.78 \times 10^{-11}$ & 0.0259                        \\
                     & 100 &    $\xi_{1}= 400$                   & $\xi_{2}= 740$ &  $1.2711 \times 10^{-9}$                        & $1.1931 \times 10^{-9}$               & $7.80 \times 10^{-11}$ & 0.0260                       \\
                    & 100  &    $\xi_{1}= 500$                   & $\xi_{2}= 750$ &  $1.2311 \times 10^{-9}$                        & $1.1559 \times 10^{-9}$               & $7.52 \times 10^{-11}$ & 0.0250                                            \\
\multirow{-7}{*}{-18.6}  & 100 & $\xi_{1}= 600$& $\xi_{2}= 770 $ & $1.4695 \times 10^{-9}$                        & $1.3910 \times 10^{-9}$               & $7.85 \times 10^{-11}$ & 0.0261 &\multirow{-7}{*}{0.02568($\pm 0.0008$)}   \\
 \bottomrule \bottomrule
\end{tabular}
}
\end{table*}
\section{Numerical Results}

After adding the Gaussian white noise with different SNR to the signals, we fit the peak shift  $\delta t$ in the WVA scheme and compute the values of $\rm \Theta$ in the AWVA scheme respectively. In our simulation, the initial temporal probe was chosen as:
\begin{eqnarray}
\label{Eq:initial-pointer-real}
I^{in}_{1}(t;\tau)&=&\left|\left\langle q | \Psi_{i}\right\rangle\right|^{2} \nonumber \\
&=&\frac{1} {(2 \pi \times0.0002^{2})^{1/4}}  e^{- \frac{(t-0.0015)^{2}}{0.0004^{2}} }\,.
\end{eqnarray}
where $I_{0}$ is set as unit, the angle for post-selection was set at $\alpha=0.01$ rad.
{
Note that the sampling frequency 1/T of APDs has a great impact on the auto-correlative intensity $\rm \Theta_{NN}$ in the AWVA scheme, which has been discussed in the previous section. Therefore, the simulation of the final auto-correlative intensity $\rm \Theta_{A+N}$ with 1/T= 10 MHz, 10 MHz and 100 MHz are investigated.
}
The measured results along with various \blue{types of noise} with different sampling frequency 1/T are shown in Fig.~\ref{Fig:HighprobeChangeInTwoScheme} and Fig.~\ref{Fig:ACIschemeDifferentSamplingFrequnecy}. On this basis we compare the sensitivity and the stability (robustness) of the two schemes.
\subsection{The signal-to-noise rate}
\label{Sec:signal-to-noise}
\red{
In information theory, the Fisher information~\cite{PhysRevLett.118.070802,JPHOT.2021.3057671,Zhu2016,PhysRevA.96.052128} is a way of calculating the
fundamental limit of the minimum uncertainty for parameter
estimation, and has been widely used to estimate the SNR in the standard weak measurement.
}
\red{Indeed, the SNR calculation based on the Fisher information is more reasonable and we would investigate the fundamental limit of the AWVA scheme in future.
}
{Herein, we use a more general definition of SNR with voltage magnitudes of signals and noises in the WVA scheme:
}
\begin{eqnarray}
\label{Eq:SNRdefine}
{\rm SNR}= 20 \rm log \frac{Max[V_I(t)]}{Max[V_N(t)]}=
20 \rm log \frac{Max[I_{1}^{out}(t)]}{Max[{ \textbf{N}}(t)]} \, ,
\end{eqnarray}
where $\rm Max[I(t)]$ and $\rm Max[{ \textbf{N}}(t)]$ represent the maximum amplitude of the signal and the noise respectively. {The voltage magnitude $\rm V_I(t)= \mathcal{K} I_{1}^{out}(t)$ is proportional to the amplitude of the signal $I_{1}^{out}(t)$ and the voltage magnitude $\rm V_N(t)= \mathcal{K}\textbf{N}(t)$ is proportional to the amplitude of the noise $\textbf{N}(t)$, where the factor $\mathcal{K}$ represents the coefficient of photoelectric conversion of APD. Note that the SNR may not be well-defined, \red{due} to $\rm {Max[{ \textbf{N}}(t)]}$ increasing without bound as the number of samples is increased.
\red{Note that the left figure in Fig.~\ref{Fig:exzampelnoiseAndSensitivity} indicate that the value of $\rm {Max[{ \textbf{N}}(t)]}$ varies very little in the sampling frequency 1/T range of 1 MHz to 100 MHz. Thus, we neglect this uncertainty of SNR and calculate SNR with the value of $\rm {Max[{ \textbf{N}}(t)]}$ at 1/T=10 MHz.}}
Take the example shown in Fig.~\ref{Fig:exzampelnoiseAndSensitivity}(A), which gives SNR= $20\times \rm log \frac{1.40\times10^{-4}}{3.70  \times10^{-5}} $ = 11.5 dB. The numerical results in the WVA scheme with various \blue{kinds of noise} are presented in Table.~\ref{Table:sensitivityOfWVA}.

For the measurements with the AWVA technique, we define the sensitivity $\rm SNR^{*}$ as an intermediate step in the AWVA scheme and the sensitivity $\rm SNR^{*}$ is obtained by
\begin{eqnarray}
\label{Eq:SNR2define}
{\rm SNR^{*}}=20\times \rm log \frac{Max[I_{21}^{out}(t)]}{Max[{ \textbf{N}}(t)]} \, ,
\end{eqnarray}

In the WVA scheme, the position of \blue{noise} in the optical path has no effect on the calculation of SNR in our simulation, because there is only a single optical path in the WVA scheme. However, in the AWVA scheme, it is worth noting that the 50:50 beam splitter~(shown in Fig.~\ref{Fig:Schemes_model2}) will reduce the strength of the signal, \red{and introduce vacuum noise}. Thus, where the \blue{noise appears} will lead to the different values of $\rm SNR^{*}$. When the noise is assumed to occur only in the optical path before the BS element, the value of $\rm SNR^{*}$ is equal to the value of $\rm SNR$.
\red{Note that the relationship $\rm SNR^{*}$=$\rm SNR^{}$ is true when we neglect the vacuum noise. This is the reason why a bright laser beam must be used - in that case the vacuum noise is small compared to the shot noise. In general, vacuum noise introduced by the BS element will cause $\rm SNR^{*}$ $>$ SNR. }
When the noise is assumed to occur only in the optical path after the BS element, the value of $\rm SNR^{*}$ is reduced and $\rm SNR^{*}=0.5  SNR$. In addition, the value of $\rm SNR^{*}$ will \red{be obtained} at the range $\rm 0.5 SNR < SNR^{*} < SNR$ when the noises occur throughout the optical path in the AWVA scheme. In this paper, in order to highlight the advantages of the scheme, we investigate the lowest $\rm SNR^{*}=0.5  SNR$, because the lower the $\rm SNR^{*}$, the harder it is to detect a useful signal.
{
In addition, comparing the results with larger $\rm SNR^{*}$ to the results with $\rm SNR^{*}=0.5  SNR$ in the AWVA scheme, our simulation can also model the effect of noise before the beamsplitter. The results will be discussed in Sec.~\ref{Sec:EffectsofGaussianSNR}.
}

Note that the relationship ${\rm \Theta}_{A+\textbf{N}}(\tau)$ =${\rm \Theta}_{A}(\tau)$ is only true if the integral time is infinite in Eq.~(\ref{Eq:ACIdefine+noise}). On the other hand, the sampling time (corresponding to the integral time) is finite and the pseudo-random number generator can not generate truly random numbers. Thus, the \blue{various types of noise} will affect the value of $\rm \Theta$ and cause uncertainty in estimating the sensitivity of the scheme with the AWVA technique. In addition, the analysis of the sensitivity to \blue{various types of noise} and the discussion of comparison with these results in the WVA scheme are shown in the next subsection.
\begin{table*}[htp!]
\setlength{\tabcolsep}{1.9mm}{
\caption{\label{Table:sensitivityOfWVAdifferntTau} The numerical results (dimensional quantities in unit of s) for measuring the different $\tau$ under the strong noise at SNR= -18.6 dB in the WVA scheme. The contents ${{\rm E}}_{t0}$ and ${\rm E}_{t \tau}$ in parentheses represent the standard error for estimating the time shifts $\delta t_{0} $ and $\delta t_{\tau} $ respectively. The data in red means the measurement is invalid. $\textbf{N}(\xi\red{=0})$, $\textbf{N}(\xi=100)$ ... $\textbf{N}(\xi=600)$ represent the multiple measurements with the different initial times.}
\begin{tabular}{rrrrr}
\toprule
\toprule
$\tau $            & Noise       & $\delta t_{0} \, \rm (\pm E_{t 0})$                                                    & $\delta t_{\tau} \, \rm (\pm E_{t \tau})$                & $\rm K_{1}\, (\pm {\rm E}_{1})$                                                            \\ \hline
                      & $\rm N(\xi= 000)$ & $1.23 \times 10^{-5}$ ($\pm 2.12 \times 10^{-5}$)                        & $6.57 \times 10^{-5}$ ($\pm 2.16 \times 10^{-5}$)  &     { $8.90 \times 10^{3}$ ($\pm 7.13 \times 10^{3}$)}
                      \\& $\rm N(\xi= 100)$ & $2.52 \times 10^{-5}$ ($\pm 2.28 \times 10^{-5}$)                        & $7.85 \times 10^{-5}$ ($\pm 2.20 \times 10^{-5}$)  & { $8.88 \times 10^{3}$ ($\pm 7.46 \times 10^{3}$)}
                      \\& $\rm N(\xi= 200)$ & $7.6086 \times 10^{-4}$ ($\pm 3.93 \times 10^{-5}$)                        & $7.7576 \times 10^{-4}$ ($\pm 4.28 \times 10^{-5}$)  & {\color[HTML]{FE0000} $2.48 \times 10^{3}$ ($\pm 13.6 \times 10^{3}$)}
                      \\& $\rm N(\xi= 300)$ & $-2.40 \times 10^{-6}$ ($\pm 2.14 \times 10^{-5}$)                        & $5.07 \times 10^{-5}$ ($\pm 2.15 \times 10^{-5}$)  & { $8.85 \times 10^{3}$ ($\pm 7.15 \times 10^{3}$)}
                     \\& $\rm N(\xi= 400)$ & $-6.60 \times 10^{-6}$ ($\pm 2.42 \times 10^{-5}$)                        & $5.49 \times 10^{-5}$ ($\pm 2.41 \times 10^{-5}$)  & { $10.2 \times 10^{3}$ ($\pm 8.05 \times 10^{3}$)}             \\& $\rm N(\xi= 500)$ & $1.4261 \times 10^{-3}$ ($\pm 30.0 \times 10^{-5}$)                        & $1.4251 \times 10^{-3}$ ($\pm 28.2 \times 10^{-5}$)  & {\color[HTML]{FE0000}  $-0.16 \times 10^{3}$ ($\pm 98.6 \times 10^{3}$)}     \\
\multirow{-7}{*}{$6 \times 10^{-9}$}  & $\rm N(\xi= 600)$ & $-1.0995 \times 10^{-3}$ ($\pm 3.75 \times 10^{-5}$)                        & $-1.0920 \times 10^{-3}$ ($\pm 4.07 \times 10^{-5}$)  & {\color[HTML]{FE0000}  $1.25 \times 10^{3}$ ($\pm 13.0\times 10^{3}$)}
\\ \cline{1-5}
                      & $\rm N(\xi= 000)$ & $1.23 \times 10^{-5}$ ($\pm 2.12 \times 10^{-5}$)                        & $9.34 \times 10^{-5}$ ($\pm 2.19 \times 10^{-5}$)  &     { $9.01 \times 10^{3}$ ($\pm 4.84 \times 10^{3}$)}
                      \\& $\rm N(\xi= 100)$ & $2.52 \times 10^{-5}$ ($\pm 2.28 \times 10^{-5}$)                        & $10.4 \times 10^{-5}$ ($\pm 2.18 \times 10^{-5}$)  & { $8.75 \times 10^{3}$ ($\pm 5.22 \times 10^{3}$)}
                      \\& $\rm N(\xi= 200)$ & $7.6086 \times 10^{-4}$ ($\pm 3.93 \times 10^{-5}$)                        & $7.8843 \times 10^{-4}$ ($\pm 4.58 \times 10^{-5}$)  & {\color[HTML]{FE0000} $3.05 \times 10^{3}$ ($\pm 9.45 \times 10^{3}$)}
                      \\& $\rm N(\xi= 300)$ & $-2.40 \times 10^{-6}$ ($\pm 2.14 \times 10^{-5}$)                        & $7.72 \times 10^{-5}$ ($\pm 2.16 \times 10^{-5}$)  & { $8.84 \times 10^{3}$ ($\pm 4.77 \times 10^{3}$)}
                     \\& $\rm N(\xi= 400)$ & $-6.60 \times 10^{-6}$ ($\pm 2.42 \times 10^{-5}$)                        & $8.58 \times 10^{-5}$ ($\pm 2.39 \times 10^{-5}$)  & {  $10.3 \times 10^{3}$ ($\pm 5.34 \times 10^{3}$)}             \\& $\rm N(\xi= 500)$ & $1.4261 \times 10^{-3}$ ($\pm 30.0 \times 10^{-5}$)                        & $1.4238 \times 10^{-3}$ ($\pm 26.8 \times 10^{-5}$)  & {\color[HTML]{FE0000}  $-0.25 \times 10^{3}$ ($\pm 65.5 \times 10^{3}$)}     \\
\multirow{-7}{*}{$9 \times 10^{-9}$} & $\rm N(\xi= 600)$ & $-1.0995 \times 10^{-3}$ ($\pm 3.75 \times 10^{-5}$)                        & $-1.0869\times 10^{-3}$ ($\pm 4.31 \times 10^{-5}$)  & {\color[HTML]{FE0000}  $1.38 \times 10^{3}$ ($\pm 8.95\times 10^{3}$)}
\\
 \cline{1-5}
                      & $\rm N(\xi= 000)$ & $1.23 \times 10^{-5}$ ($\pm 2.12 \times 10^{-5}$)                        & $12.2 \times 10^{-5}$ ($\pm 2.23 \times 10^{-5}$)  &     {$9.12 \times 10^{3}$ ($\pm 3.62 \times 10^{3}$)}
                      \\& $\rm N(\xi= 100)$ & $2.52 \times 10^{-5}$ ($\pm 2.28 \times 10^{-5}$)                        & $13.1 \times 10^{-5}$ ($\pm 2.18  \times 10^{-5}$)  & { $8.77 \times 10^{3}$ ($\pm 3.75 \times 10^{3}$)}
                      \\& $\rm N(\xi= 200)$ & $7.6086\times 10^{-4}$ ($\pm 3.93 \times 10^{-5}$)                        & $8.1703 \times 10^{-4}$ ($\pm 5.26 \times 10^{-5}$)  & {\color[HTML]{FE0000} $4.68 \times 10^{3}$ ($\pm 7.65 \times 10^{3}$)}
                      \\& $\rm N(\xi= 300)$ & $-2.40 \times 10^{-6}$ ($\pm 2.14 \times 10^{-5}$)                        & $10.4 \times 10^{-5}$ ($\pm 2.16 \times 10^{-5}$)  & { $9.21 \times 10^{3}$ ($\pm 2.68 \times 10^{3}$)}
                     \\& $\rm N(\xi= 400)$ & $-6.60 \times 10^{-6}$ ($\pm 2.42 \times 10^{-5}$)                        & $11.7 \times 10^{-5}$ ($\pm 2.37 \times 10^{-5}$)  & {  $10.3 \times 10^{3}$ ($\pm 3.99 \times 10^{3}$)}             \\& $\rm N(\xi= 500)$ & $1.4261 \times 10^{-3}$ ($\pm 30.0 \times 10^{-5}$)                        & $1.4217 \times 10^{-3}$ ($\pm 25.2 \times 10^{-5}$)  & {\color[HTML]{FE0000}  $-0.36 \times 10^{3}$ ($\pm 45.8 \times 10^{3}$)}     \\
\multirow{-7}{*}{$12 \times 10^{-9}$}  & $\rm N(\xi= 600)$ & $-1.0995 \times 10^{-3}$ ($\pm 4.66\times 10^{-5}$)                        & $-1.0797 \times 10^{-3}$ ($\pm 4.63 \times 10^{-5}$)  & {\color[HTML]{FE0000}  $1.65 \times 10^{3}$ ($\pm 7.40\times 10^{3}$)}
\\
\cline{1-5}
                      & $\rm N(\xi= 000)$ & $1.23 \times 10^{-5}$ ($\pm 2.12 \times 10^{-5}$)                        & $15.7 \times 10^{-5}$ ($\pm 2.27 \times 10^{-5}$)  &     { $9.23 \times 10^{3}$ ($\pm 2.92 \times 10^{3}$)}
                      \\& $\rm N(\xi= 100)$ & $2.52 \times 10^{-5}$ ($\pm 2.28 \times 10^{-5}$)                        & $15.6 \times 10^{-5}$ ($\pm 2.18 \times 10^{-5}$)  & { $8.76 \times 10^{3}$ ($\pm 2.96 \times 10^{3}$)}
                      \\& $\rm N(\xi= 200)$ & $7.6086 \times 10^{-4}$ ($\pm 3.93 \times 10^{-5}$)                        & $8.6756 \times 10^{-4}$ ($\pm 6.45 \times 10^{-5}$)  & { $7.11 \times 10^{3}$ ($\pm 6.92 \times 10^{3}$)}
                      \\& $\rm N(\xi= 300)$ & $-2.40 \times 10^{-6}$ ($\pm 2.14 \times 10^{-5}$)                        & $13.1
                      \times 10^{-5}$ ($\pm 2.16 \times 10^{-5}$)  & { $8.86 \times 10^{3}$ ($\pm 2.86 \times 10^{3}$)}
                     \\& $\rm N(\xi= 400)$ & $-6.60 \times 10^{-6}$ ($\pm 2.42 \times 10^{-5}$)                        & $14.7 \times 10^{-5}$ ($\pm 2.34 \times 10^{-5}$)  & {  $10.3 \times 10^{3}$ ($\pm 3.17 \times 10^{3}$)}             \\& $\rm N(\xi= 500)$ & $1.4261 \times 10^{-3}$ ($\pm 30.0 \times 10^{-5}$)                        & $1.4187 \times 10^{-3}$ ($\pm 24.0 \times 10^{-5}$)  & {\color[HTML]{FE0000}  $-0.49 \times 10^{3}$ ($\pm 36.0\times 10^{3}$)}     \\
\multirow{-7}{*}{$15 \times 10^{-9}$} & $\rm N(\xi= 600)$ & $-1.0995 \times 10^{-3}$ ($\pm 3.75 \times 10^{-5}$)                        & $-1.0687 \times 10^{-3}$ ($\pm 5.24 \times 10^{-5}$)  & {\color[HTML]{FE0000}  $2.05 \times 10^{3}$ ($\pm 5.99\times 10^{3}$)}
\\
 \bottomrule \bottomrule
\end{tabular}
}
\end{table*}

\begin{table*}[htp!]
\setlength{\tabcolsep}{2.5mm}{
\caption{\label{Table:sensitivityOfAWVAdifferntTau}
{Parameters and some characteristic numerical results for measuring the different $\tau$ under the strong noises $\textbf{N}(t,\sigma^{2},\xi_{1})$ and $\textbf{N}(t,\sigma^{2},\xi_{2})$ at SNR= -18.6 dB in the AWVA scheme. The sampling frequency is set at 1/T= 10 MHz.
}}
\begin{tabular}{rrrrrrrr}
\toprule
\toprule
$\tau$ (s)
& $\textbf{N}(t,\xi_{1})$
& $\textbf{N}(t,\xi_{2})$
& $\rm \Theta_{0}$
& $\rm \Theta_{\tau}$
& $\Delta {\rm \Theta}$
& $\rm K_{2}^{M}$ &$\rm \overline{K}_{2} \, (\pm {\rm E}_{2})$               \\ \hline
& no noise & no noise
 & $1.2443 \times 10^{-9}$
 & $1.0837 \times 10^{-9}$
 & $16.1 \times 10^{-11}$ & \green{0.0268}          \\ \cline{2-8}
 & $\rm N(\xi_1= 000)$ & $\rm N(\xi_2= 700)$
 & $1.5116 \times 10^{-9}$
 & $1.3418 \times 10^{-9}$
 & $17.0 \times 10^{-11}$ & 0.0283
 \\
  & $\rm N(\xi_1= 100)$ & $\rm N(\xi_2= 710)$
 & $2.2062 \times 10^{-9}$
 & $2.0626 \times 10^{-9}$
 & $14.3 \times 10^{-11}$ & 0.0238
 \\
  & $\rm N(\xi_1= 200)$ & $\rm N(\xi_2= 720)$
 & $0.8214 \times 10^{-9}$
 & $0.6742 \times 10^{-9}$
 & $14.7 \times 10^{-11}$ & 0.0245
 \\
  & $\rm N(\xi_1= 300)$ & $\rm N(\xi_2= 730)$
 & $1.5151 \times 10^{-9}$
 & $1.3575 \times 10^{-9}$
 & $15.7 \times 10^{-11}$ & 0.0261
 \\
  & $\rm N(\xi_1= 400)$ & $\rm N(\xi_2= 740)$
 & $1.8597 \times 10^{-9}$
 & $1.7035 \times 10^{-9}$
 & $15.6 \times 10^{-11}$ & 0.0260
 \\
  & $\rm N(\xi_1= 500)$ & $\rm N(\xi_2= 750)$
 & $1.2757 \times 10^{-9}$
 & $1.1136 \times 10^{-9}$
 & $16.2 \times 10^{-11}$ & 0.0270
 \\
\multirow{-7}{*}{$6 \times 10^{-9}$}
 & $\rm N(\xi_1= 600)$ & $\rm N(\xi_2= 760)$
 & $1.8199 \times 10^{-9}$
 & $1.6441 \times 10^{-9}$
 & $17.6 \times 10^{-11}$ & 0.0293
 &\multirow{-7}{*}{0.02643($\pm 0.00287$)}                \\ \hline
 & no noise & no noise
 & $1.2443 \times 10^{-9}$
 & $0.9973 \times 10^{-9}$
 & $24.7 \times 10^{-11}$ & 0.0274           \\ \cline{2-8}
 & $\rm N(\xi_1= 000)$ & $\rm N(\xi_2= 700)$
 & $1.5116 \times 10^{-9}$
 & $1.2490 \times 10^{-9}$
 & $26.2 \times 10^{-11}$ & 0.0291
 \\
  & $\rm N(\xi_1= 100)$ & $\rm N(\xi_2= 710)$
 & $2.2062 \times 10^{-9}$
 & $1.9815 \times 10^{-9}$
 & $22.4 \times 10^{-11}$ & 0.0249
 \\
  & $\rm N(\xi_1= 200)$ & $\rm N(\xi_2= 720)$
 & $0.8214 \times 10^{-9}$
 & $0.5944 \times 10^{-9}$
 & $22.7 \times 10^{-11}$ & 0.0252
 \\
  & $\rm N(\xi_1= 300)$ & $\rm N(\xi_2= 730)$
 & $1.5151 \times 10^{-9}$
 & $1.2730 \times 10^{-9}$
 & $24.2 \times 10^{-11}$ & 0.0269
 \\
  & $\rm N(\xi_1= 400)$ & $\rm N(\xi_2= 740)$
 & $1.8597 \times 10^{-9}$
 & $1.6194 \times 10^{-9}$
 & $24.0 \times 10^{-11}$ & 0.0266
 \\
  & $\rm N(\xi_1= 500)$ & $\rm N(\xi_2= 750)$
 & $1.2757 \times 10^{-9}$
 & $1.0260 \times 10^{-9}$
 & $25.0 \times 10^{-11}$ & 0.0277
 \\
\multirow{-7}{*}{$9 \times 10^{-9}$}
 & $\rm N(\xi_1= 600)$ & $\rm N(\xi_2= 760)$
 & $1.8199 \times 10^{-9}$
 & $1.5501 \times 10^{-9}$
 & $27.0 \times 10^{-11}$ & 0.0300
 &\multirow{-7}{*}{0.02720($\pm 0.00280$)}                \\ \hline
  & no noise & no noise
 & $1.2443 \times 10^{-9}$
 & $0.9091 \times 10^{-9}$
 & $33.5 \times 10^{-11}$ & 0.0279           \\ \cline{2-8}
 & $\rm N(\xi_1= 000)$ & $\rm N(\xi_2= 700)$
 & $1.5116 \times 10^{-9}$
 & $1.1535 \times 10^{-9}$
 & $35.8 \times 10^{-11}$ & 0.0298
 \\
  & $\rm N(\xi_1= 100)$ & $\rm N(\xi_2= 710)$
 & $2.2062 \times 10^{-9}$
 & $1.8966 \times 10^{-9}$
 & $30.9 \times 10^{-11}$ & 0.0257
 \\
  & $\rm N(\xi_1= 200)$ & $\rm N(\xi_2= 720)$
 & $0.8214 \times 10^{-9}$
 & $0.5128 \times 10^{-9}$
 & $30.8 \times 10^{-11}$ & 0.0256
 \\
  & $\rm N(\xi_1= 300)$ & $\rm N(\xi_2= 730)$
 & $1.5151 \times 10^{-9}$
 & $1.1870 \times 10^{-9}$
 & $32.8 \times 10^{-11}$ & 0.0273
 \\
  & $\rm N(\xi_1= 400)$ & $\rm N(\xi_2= 740)$
 & $1.8597 \times 10^{-9}$
 & $1.5336 \times 10^{-9}$
 & $32.6 \times 10^{-11}$ & 0.0271
 \\
  & $\rm N(\xi_1= 500)$ & $\rm N(\xi_2= 750)$
 & $1.2757 \times 10^{-9}$
 & $0.9364 \times 10^{-9}$
 & $33.9 \times 10^{-11}$ & 0.0282
 \\
\multirow{-7}{*}{$12 \times 10^{-9}$}
 & $\rm N(\xi_1= 600)$ & $\rm N(\xi_2= 760)$
 & $1.8199 \times 10^{-9}$
 & $1.4566 \times 10^{-9}$
 & $36.3 \times 10^{-11}$ & 0.0302
 &\multirow{-7}{*}{0.02770($\pm 0.00250$)}                \\ \hline
   & no noise & no noise
 & $1.2443 \times 10^{-9}$
 & $0.8208 \times 10^{-9}$
 & $43.4 \times 10^{-11}$ & 0.0282           \\ \cline{2-8}
 & $\rm N(\xi_1= 000)$ & $\rm N(\xi_2= 700)$
 & $1.5116 \times 10^{-9}$
 & $1.0575 \times 10^{-9}$
 & $45.4 \times 10^{-11}$ & 0.0303
 \\
  & $\rm N(\xi_1= 100)$ & $\rm N(\xi_2= 710)$
 & $2.2062 \times 10^{-9}$
 & $1.8100 \times 10^{-9}$
 & $39.6 \times 10^{-11}$ & 0.0264
 \\
  & $\rm N(\xi_1= 200)$ & $\rm N(\xi_2= 720)$
 & $0.8214 \times 10^{-9}$
 & $0.4308 \times 10^{-9}$
 & $39.1 \times 10^{-11}$ & 0.0261
 \\
  & $\rm N(\xi_1= 300)$ & $\rm N(\xi_2= 730)$
 & $1.5151 \times 10^{-9}$
 & $1.1010 \times 10^{-9}$
 & $41.4 \times 10^{-11}$ & 0.0276
 \\
  & $\rm N(\xi_1= 400)$ & $\rm N(\xi_2= 740)$
 & $1.8597 \times 10^{-9}$
 & $1.4477 \times 10^{-9}$
 & $41.2 \times 10^{-11}$ & 0.0274
 \\
  & $\rm N(\xi_1= 500)$ & $\rm N(\xi_2= 750)$
 & $1.2757 \times 10^{-9}$
 & $0.8466 \times 10^{-9}$
 & $42.9 \times 10^{-11}$ & 0.0286
 \\
\multirow{-7}{*}{$15 \times 10^{-9}$}
 & $\rm N(\xi_1= 600)$ & $\rm N(\xi_2= 760)$
 & $1.8199 \times 10^{-9}$
 & $1.3628 \times 10^{-9}$
 & $45.7 \times 10^{-11}$ & 0.0304
 &\multirow{-7}{*}{0.02811($\pm \green{0.00229}$)}                \\
 \bottomrule \bottomrule
\end{tabular}
}
\end{table*}
\begin{table*}[htp!]
\setlength{\tabcolsep}{2.5mm}{
\caption{\label{Table:sensitivityOfAWVAdifferntTau_F100MHz}
{Parameters and some characteristic numerical results for measuring the different $\tau$ under the strong noises $\textbf{N}(t,\sigma^{2},\xi_{1})$ and $\textbf{N}(t,\sigma^{2},\xi_{2})$ at SNR= -18.6 dB in the AWVA scheme. The sampling frequency is set at 1/T= 100 MHz.
}}
\begin{tabular}{rrrrrrrr}
\toprule
\toprule
$\tau$ (s)
& $\textbf{N}(t,\xi_{1})$
& $\textbf{N}(t,\xi_{2})$
& $\rm \Theta_{0}$
& $\rm \Theta_{\tau}$
& $\Delta {\rm \Theta}$
& $\rm K_{2}^{M}$ &$\rm \overline{K}_{2} \, (\pm {\rm E}_{2})$               \\ \hline
& no noise & no noise
 & $1.2443 \times 10^{-9}$
 & $1.0837 \times 10^{-9}$
 & $16.1 \times 10^{-11}$ & 0.0267           \\ \cline{2-8}
 & $\rm N(\xi_1= 000)$ & $\rm N(\xi_2= 700)$
 & $1.2195 \times 10^{-9}$
 & $1.0651 \times 10^{-9}$
 & $15.4 \times 10^{-11}$ & 0.0257
 \\
  & $\rm N(\xi_1= 100)$ & $\rm N(\xi_2= 710)$
 & $1.5641 \times 10^{-9}$
 & $1.4010 \times 10^{-9}$
 & $16.3 \times 10^{-11}$ & 0.0272
 \\
  & $\rm N(\xi_1= 200)$ & $\rm N(\xi_2= 720)$
 & $0.9090 \times 10^{-9}$
 & $0.7495 \times 10^{-9}$
 & $15.9 \times 10^{-11}$ & 0.0265
 \\
  & $\rm N(\xi_1= 300)$ & $\rm N(\xi_2= 730)$
 & $1.2707 \times 10^{-9}$
 & $1.1104 \times 10^{-9}$
 & $16.0 \times 10^{-11}$ & 0.0266
 \\
  & $\rm N(\xi_1= 400)$ & $\rm N(\xi_2= 740)$
 & $1.2711 \times 10^{-9}$
 & $1.1095 \times 10^{-9}$
 & $16.1 \times 10^{-11}$ & 0.0268
 \\
  & $\rm N(\xi_1= 500)$ & $\rm N(\xi_2= 750)$
 & $1.2311 \times 10^{-9}$
 & $1.0754 \times 10^{-9}$
 & $15.6 \times 10^{-11}$ & 0.0260
 \\
\multirow{-7}{*}{$6 \times 10^{-9}$}
 & $\rm N(\xi_1= 600)$ & $\rm N(\xi_2= 760)$
 & $1.4695 \times 10^{-9}$
 & $1.3071 \times 10^{-9}$
 & $16.2 \times 10^{-11}$ & 0.0270
 &\multirow{-7}{*}{0.02654($\pm 0.00084$)}                \\ \hline
  & no noise & no noise
 & $1.2443 \times 10^{-9}$
 & $0.9973 \times 10^{-9}$
 & $24.7 \times 10^{-11}$ & 0.0274           \\ \cline{2-8}
 & $\rm N(\xi_1= 000)$ & $\rm N(\xi_2= 700)$
 & $1.2195 \times 10^{-9}$
 & $0.9818 \times 10^{-9}$
 & $23.8 \times 10^{-11}$ & 0.0264
 \\
  & $\rm N(\xi_1= 100)$ & $\rm N(\xi_2= 710)$
 & $1.5641 \times 10^{-9}$
 & $1.3132 \times 10^{-9}$
 & $25.1 \times 10^{-11}$ & 0.0279
 \\
  & $\rm N(\xi_1= 200)$ & $\rm N(\xi_2= 720)$
 & $0.9090 \times 10^{-9}$
 & $0.6636 \times 10^{-9}$
 & $24.5 \times 10^{-11}$ & 0.0272
 \\
  & $\rm N(\xi_1= 300)$ & $\rm N(\xi_2= 730)$
 & $1.2707 \times 10^{-9}$
 & $1.0246 \times 10^{-9}$
 & $24.6 \times 10^{-11}$ & 0.0273
 \\
  & $\rm N(\xi_1= 400)$ & $\rm N(\xi_2= 740)$
 & $1.2711 \times 10^{-9}$
 & $1.0223 \times 10^{-9}$
 & $24.8 \times 10^{-11}$ & 0.0276
 \\
  & $\rm N(\xi_1= 500)$ & $\rm N(\xi_2= 750)$
 & $1.2311 \times 10^{-9}$
 & $0.9914 \times 10^{-9}$
 & $24.0 \times 10^{-11}$ & 0.0267
 \\
\multirow{-7}{*}{$9 \times 10^{-9}$}
 & $\rm N(\xi_1= 600)$ & $\rm N(\xi_2= 760)$
 & $1.4695 \times 10^{-9}$                     & $1.2196 \times 10^{-9}$
 & $25.0 \times 10^{-11}$ & 0.0278
 &\multirow{-7}{*}{0.02727($\pm 0.00087$)}                \\ \hline
   & no noise & no noise
 & $1.2443 \times 10^{-9}$
 & $0.9091 \times 10^{-9}$
 & $33.5 \times 10^{-11}$ & 0.0279           \\ \cline{2-8}
 & $\rm N(\xi_1= 000)$ & $\rm N(\xi_2= 700)$
 & $1.2195 \times 10^{-9}$
 & $0.8967 \times 10^{-9}$
 & $32.3 \times 10^{-11}$ & 0.0269
 \\
  & $\rm N(\xi_1= 100)$ & $\rm N(\xi_2= 710)$
 & $1.5641 \times 10^{-9}$
 & $1.2236 \times 10^{-9}$
 & $34.1 \times 10^{-11}$ & 0.0284
 \\
  & $\rm N(\xi_1= 200)$ & $\rm N(\xi_2= 720)$
 & $0.9090 \times 10^{-9}$
 & $0.5760 \times 10^{-9}$
 & $33.3 \times 10^{-11}$ & 0.0277
 \\
  & $\rm N(\xi_1= 300)$ & $\rm N(\xi_2= 730)$
 & $1.2707 \times 10^{-9}$
 & $0.9375 \times 10^{-9}$
 & $33.3 \times 10^{-11}$ & 0.0277
 \\
  & $\rm N(\xi_1= 400)$ & $\rm N(\xi_2= 740)$
 & $1.2711 \times 10^{-9}$
 & $0.9332 \times 10^{-9}$
 & $33.8 \times 10^{-11}$ & 0.0282
 \\
  & $\rm N(\xi_1= 500)$ & $\rm N(\xi_2= 750)$
 & $1.2311 \times 10^{-9}$
 & $0.9056 \times 10^{-9}$
 & $32.5 \times 10^{-11}$ & 0.0271
 \\
\multirow{-7}{*}{$12 \times 10^{-9}$}
 & $\rm N(\xi_1= 600)$ & $\rm N(\xi_2= 760)$
 & $1.4695 \times 10^{-9}$
 & $1.1301 \times 10^{-9}$
 & $33.9 \times 10^{-11}$ & 0.0282
 &\multirow{-7}{*}{0.02774($\pm 0.00084$)}                \\ \hline
    & no noise & no noise
 & $1.2443 \times 10^{-9}$
 & $0.8208 \times 10^{-9}$
 & $43.4 \times 10^{-11}$ & 0.0282           \\ \cline{2-8}
 & $\rm N(\xi_1= 000)$ & $\rm N(\xi_2= 700)$
 & $1.2195 \times 10^{-9}$
 & $0.8115 \times 10^{-9}$
 & $40.8 \times 10^{-11}$ & 0.0272
 \\
  & $\rm N(\xi_1= 100)$ & $\rm N(\xi_2= 710)$
 & $1.5641 \times 10^{-9}$
 & $1.1339 \times 10^{-9}$
 & $43.0 \times 10^{-11}$ & 0.0287
 \\
  & $\rm N(\xi_1= 200)$ & $\rm N(\xi_2= 720)$
 & $0.9090 \times 10^{-9}$
 & $0.4883 \times 10^{-9}$
 & $42.1 \times 10^{-11}$ & 0.0281
 \\
  & $\rm N(\xi_1= 300)$ & $\rm N(\xi_2= 730)$
 & $1.2707 \times 10^{-9}$
 & $0.8505 \times 10^{-9}$
 & $42.0 \times 10^{-11}$ & 0.0280
 \\
  & $\rm N(\xi_1= 400)$ & $\rm N(\xi_2= 740)$
 & $1.2711 \times 10^{-9}$
 & $0.8438 \times 10^{-9}$
 & $42.7 \times 10^{-11}$ & 0.0285
 \\
  & $\rm N(\xi_1= 500)$ & $\rm N(\xi_2= 750)$
 & $1.2311 \times 10^{-9}$
 & $0.8195 \times 10^{-9}$
 & $41.2 \times 10^{-11}$ & 0.0275
 \\
\multirow{-7}{*}{$15 \times 10^{-9}$}
 & $\rm N(\xi_1= 600)$ & $\rm N(\xi_2= 760)$
 & $1.4695 \times 10^{-9}$
 & $1.0405 \times 10^{-9}$
 & $42.9 \times 10^{-11}$ & 0.0286
 &\multirow{-7}{*}{0.02808($\pm 0.00088$)}
 \\
 \bottomrule \bottomrule
\end{tabular}
}
\end{table*}
\begin{figure*}[t]
	\centering
	\subfigure
{
	\vspace{-0.4cm}
	\begin{minipage}{0.315\linewidth}
	\centering
	\centerline{\includegraphics[scale=0.14,angle=0]{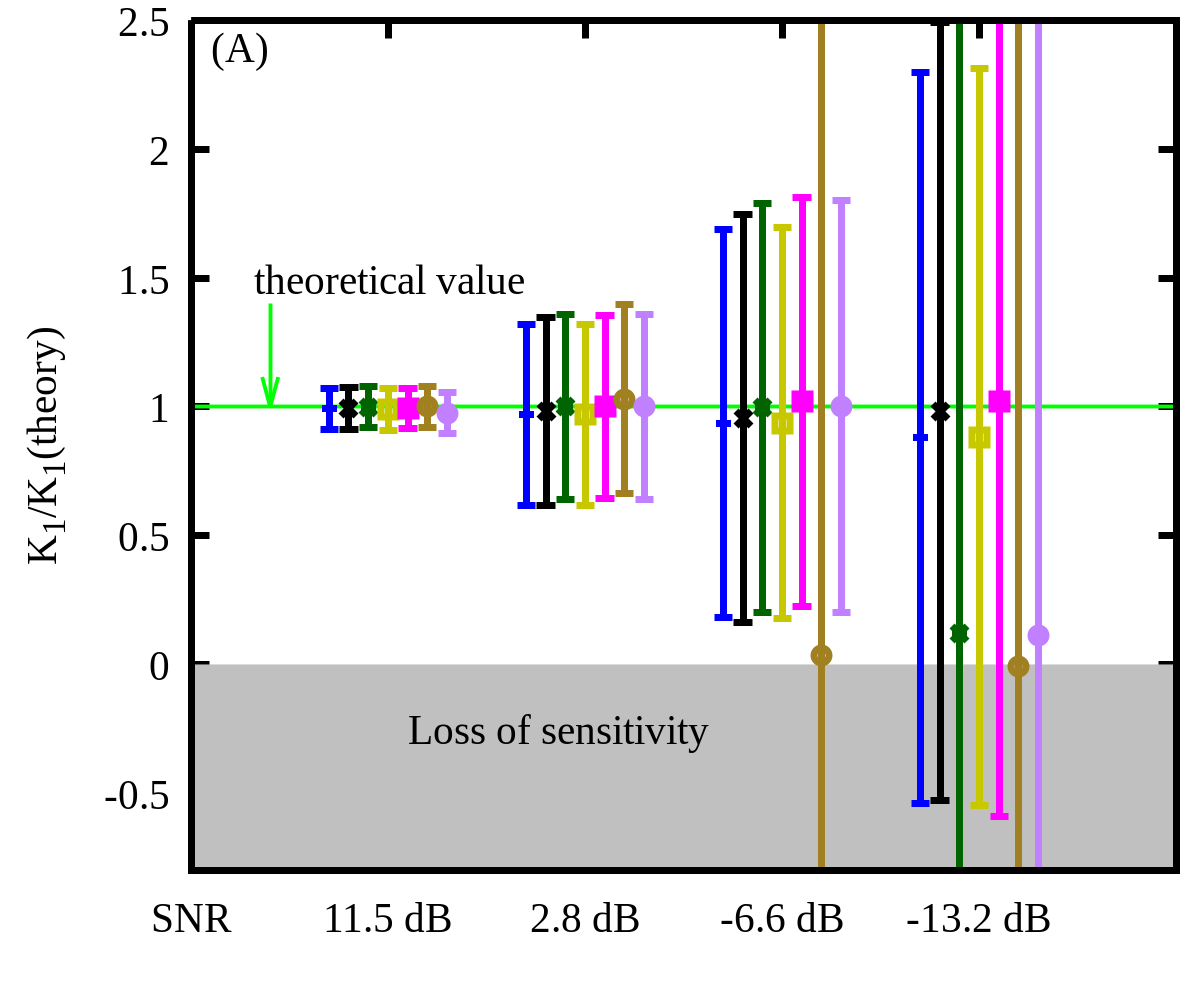}}
	\end{minipage}
}
\subfigure
{
	\vspace{-0.4cm}
	\begin{minipage}{0.315\linewidth}
	\centering
	\centerline{\includegraphics[scale=0.14,angle=0]{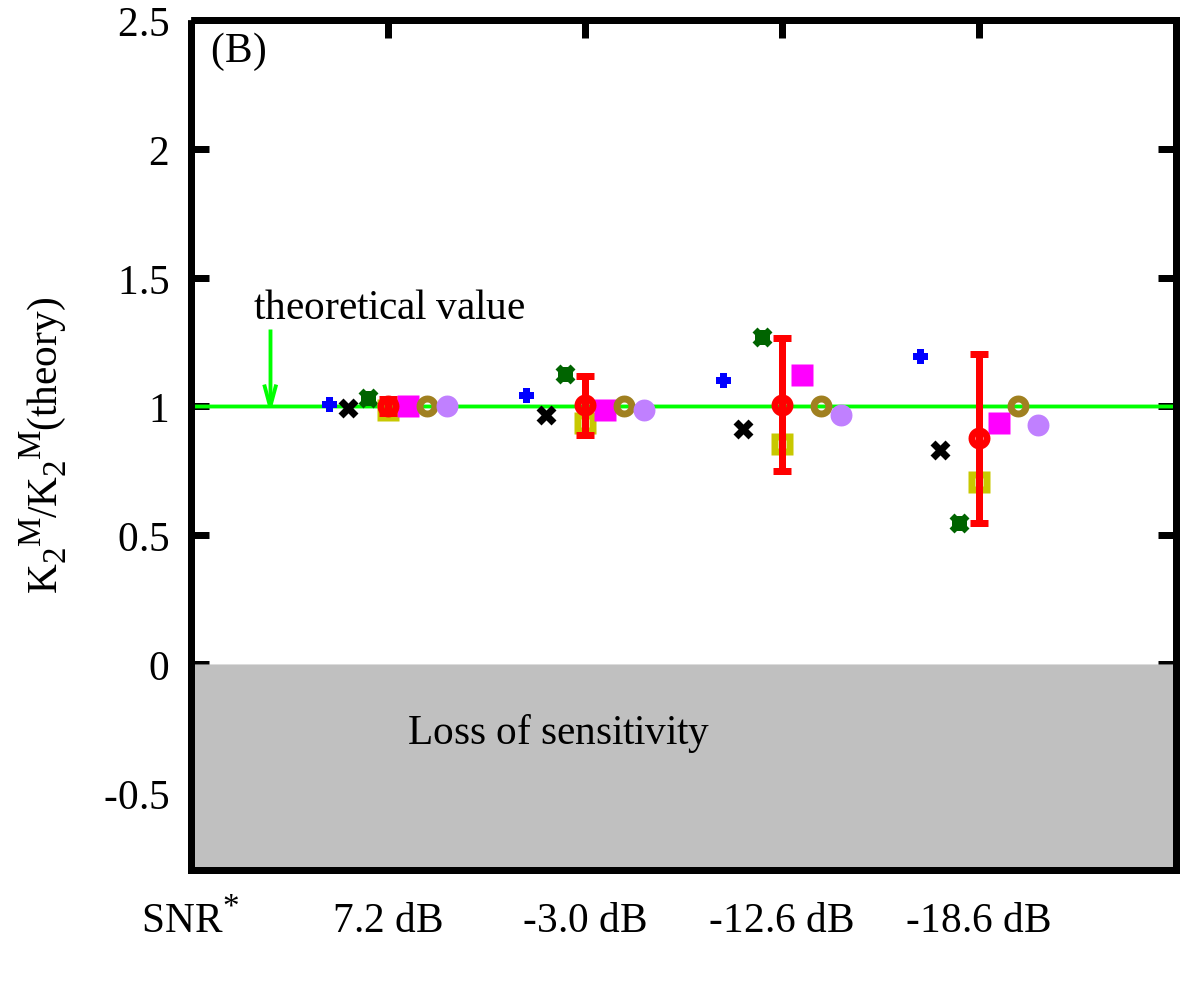}}
	\end{minipage}
}
\subfigure
{
	\vspace{-0.4cm}
	\begin{minipage}{0.315\linewidth}
	\centering
	\centerline{\includegraphics[scale=0.14,angle=0]{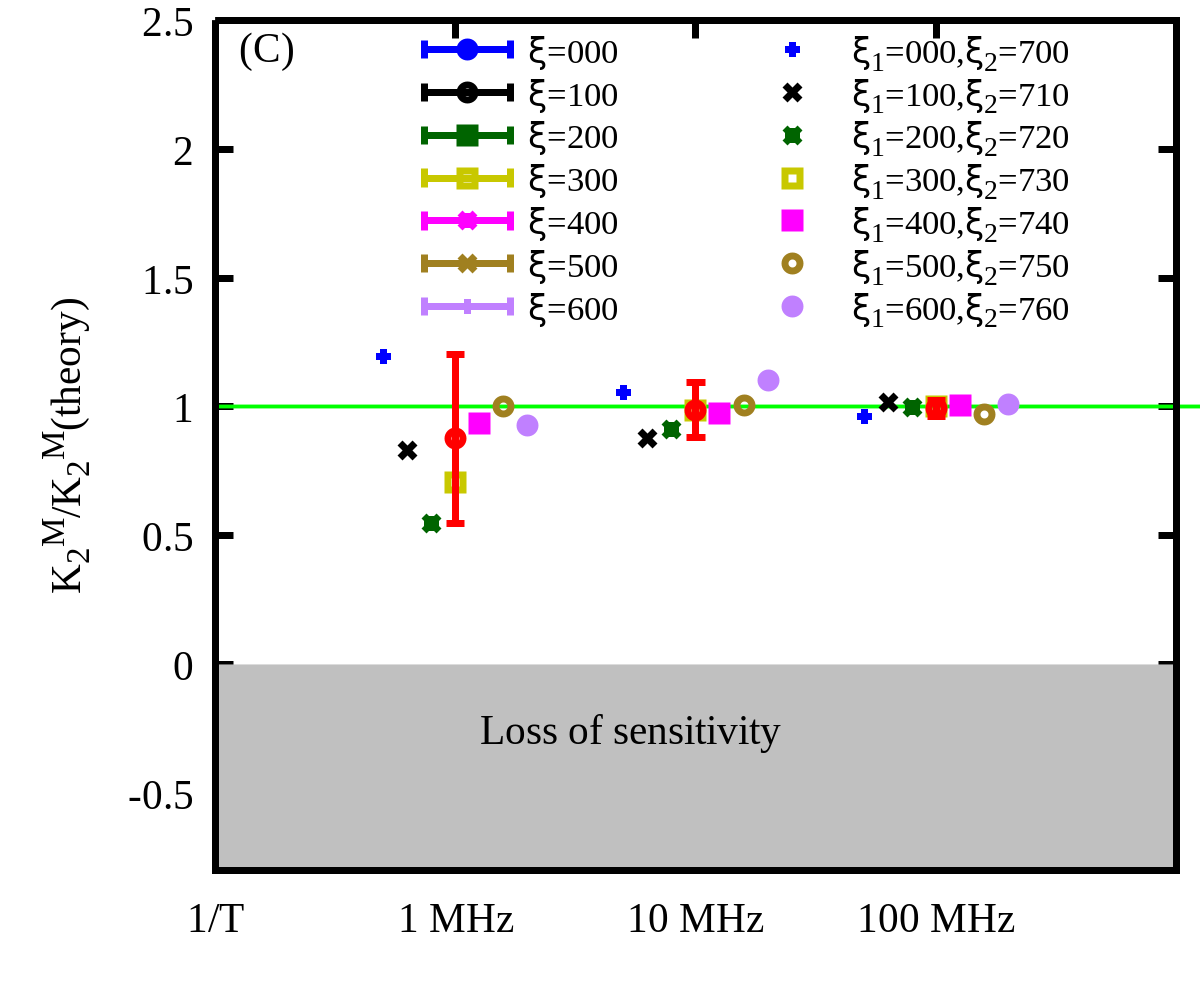}}
	\end{minipage}
}
	\vspace{-0.4cm}
\vspace*{0mm} \caption{\label{Fig:scheme12sensitivity}
{
The normalized sensitivity in the WVA scheme and in the AWVA scheme under the Gaussian noises with different SNR and sampling frequency: (A). the normalized sensitivity in the WVA scheme under  noises  $\textbf{N}(t,\sigma^{2},\xi)$ with different SNR; (B). the normalized sensitivity in the AWVA scheme under noises  $\textbf{N}(t,\sigma^{2},\xi_{1})$ and $\textbf{N}(t,\sigma^{2},\xi_{2})$ at sampling frequency 1/T= 1 MHz with different SNR; (C). the normalized sensitivity in the AWVA scheme under noises  $\textbf{N}(t,\sigma^{2},\xi_{1})$ and $\textbf{N}(t,\sigma^{2},\xi_{2})$ at SNR= -18.6 dB with different sampling frequency 1/T.
The gray band represents the measurements failing to effectively detect the final signal.  The {``}red" data with the error bar represent the final results of seven statistical averages in the AWVA scheme. }
}
\end{figure*}
\subsection{The sensitivity with the standard error}
In order to calculate the sensitivity in the two schemes, we did the simulation with both schemes with the time shift $\tau$.  In the WVA scheme, the sensitivity is defined as
\begin{eqnarray}
\label{Eq:sensitydefinescheme1}
{\rm K_{1}}=\frac{\Delta(\delta t)}{\Delta(\tau)}=
\frac{\delta t_{\tau}-\delta t_{0}}{\tau},
\end{eqnarray}
where peak value shift $\delta t_{0} $ of the temporal pointer and its standard error ${\rm E}_{t0}$ represent the results with the time shift $\tau$= 0s. The peak value shift $\delta t_{\tau} $ of the temporal pointer and its standard error ${\rm E}_{t \tau}$ are obtained by fitting the Gaussian profile of the signal $I_1^{out}(t,\tau)$ with least square method. Then, the statistical error ${\rm E}_1$ for estimating the value of $\rm K_1$ can be calculated from Eq.~(\ref{Eq:sensitydefinescheme1}) with the law of error propagation: ${\rm E}_1=(|{\rm E}_{t \tau}|+|{\rm E}_{t 0}|)/\tau$. Finally, the sensitivity $\rm K_1$ with its the statistical error ${\rm E}_1$ under different simulation conditions are shown in Table~\ref{Table:sensitivityOfWVA} and Table~\ref{Table:sensitivityOfWVAdifferntTau}.

Note that the quantities measured in the AWVA scheme are the values of $\rm \Theta$, therefore, the sensitivity in the AWVA scheme is defined as:
\begin{eqnarray}
\label{Eq:sensitydefinescheme2}
{\rm K_{2}(t)} &=&\frac{\Delta[\rm \Theta(t;\tau)]}{\Delta(\tau)}= \frac{\Theta_{0}-\Theta_{\tau} }{\tau}  \, ,
\end{eqnarray}
where $\Theta_{\tau}$ represents the result $\Theta_{A+N}(t;\tau)$ calculated from Eq.~(\ref{Eq:ACIdefine+noise2}), $\Theta_{0}$ represents the initial value of the measurement without the time shift $\tau$ and can be calculated from the {data} $I_{22+N}^{out}$ on APD2 in Fig.~\ref{Fig:Schemes_model2}:
\begin{eqnarray}
\label{Eq:ACIdefine+noiseInitial}
\Theta_{0}&=&\int_{0}^{t}  I_{22+\textbf{N}}^{out}(t^{\prime}) \times  I_{22+\textbf{N}}^{out}(t^{\prime}) dt^{\prime}
\end{eqnarray}
Note that the sensitivity ${\rm K_{2}(t)}$ depends on the scope of time integral $t$. The dependence of the sensitivity on $t$ with \blue{various types of noise} is shown in Fig.~\ref{Fig:exzampelnoiseAndSensitivity}(B) and Fig.~\ref{Fig:exzampelnoiseAndSensitivity}(C), and one can find  that the maximum sensitivity ${\rm K_{2}^{M}}$ is achieved when \red{integrated} to $t= 1.5 $ ms. In addition, we neglect the standard error of ${\rm K_{2}^{M}}$, since {$\Theta$} can be estimated by using a high vertical resolution oscilloscope.
We show the maximum sensitivity ${\rm K_{2}^{M}}$ without standard errors in Table.~\ref{Table:sensitivityOfAWVA} and Table.~\ref{Table:sensitivityOfAWVAdifferntTau}. In this paper, we define the central value ${\rm \overline{K}_{2}}$ with its statistical error ${\rm E}_{2}$ in the AWVA scheme, which is created by calculating the seven statistical averages of ${\rm K_{2}^{M}}$ with different measurements:
\begin{eqnarray}
\label{Eq:scheme2-sensitivityError}
\rm \overline{K}_{2}= \sum \limits_{i=1}^{n} {\rm K_{2}^{M}(i)}/n , \, \, {\rm E}_{2}=\rm Max \{ |\overline{K}_{2} - K_{2}^{M}(i)| \} \,,
\end{eqnarray}
where the ${\rm K_{2}^{M}(i)}$ represents the result of the ith measurement with the different value of $\xi$, n represents the total number of measurements and n=7.  For different simulation conditions, the central value ${\rm \overline{K}_{2}}$ with its statistical error ${\rm E}_{2}$ in the AWVA scheme are displayed in Table~\ref{Table:sensitivityOfAWVA} and Table~\ref{Table:sensitivityOfAWVAdifferntTau}.

To compare the sensitivity in the two schemes, we further normalized them with their corresponding theoretical results in Fig.~\ref{Fig:scheme12sensitivity} and Fig.~\ref{Fig:scheme12sensitivitywithTau}.  Note that when the value of the sensitivity $\rm K$ $<$ 0, corresponding to the \blue{``}Loss of sensitivity" area, the measurement is invalid.
The central value ${\rm \overline{K}_{2}}$ with its statistical error ${\rm E}_{2}$ in the AWVA scheme are displayed as the \blue{``}red" {data} with error bar in Fig.~\ref{Fig:scheme12sensitivity} and Fig.~\ref{Fig:scheme12sensitivitywithTau}.

\subsection{Effects of Gaussian noise with different SNR}
\label{Sec:EffectsofGaussianSNR}
Fig.~\ref{Fig:HighprobeChangeInTwoScheme} displays the results in the two schemes under the Gaussian white noise  with different SNRs. In addition, the corresponding results of the sensitivity are shown in Table.~\ref{Table:sensitivityOfWVA} and Table.~\ref{Table:sensitivityOfAWVA}. Combining the results of the normalized sensitivity in Fig.~\ref{Fig:scheme12sensitivity}, we come to the following summaries of the effects of noises in the two schemes:

1). In general, the AWVA scheme outperforms the WVA scheme, because the statistical error of the normalized sensitivity in the AWVA scheme is much smaller than that in the WVA scheme at the same level of SNR.
Note that our results may be obtained by assuming that the time resolution and the vertical resolution of the oscilloscope can meet the requirements of our scheme, where the vertical resolution of the oscilloscope limits the accuracy of the value of $\rm \Theta$. On the other hand, the signal processing modules (Fig.~\ref{Fig:Schemes_model3}) do not need to be implemented in hardware if the weak measurement does not meet the real-time measurement and this process may generate other kinds of noise. Thus, the results of $\rm \Theta$ can be calculated mathematically after the measurement from the collected data on APD1 and APD2 of the scheme in Fig.~\ref{Fig:Schemes_model2}.

2). There is no doubt that the intensity of noise significantly affects the results in both schemes. As shown in Fig.~\ref{Fig:scheme12sensitivity}, the statistical error of $\rm K_{1}$ and $\rm \overline{K}_{2}$ increase, when the SNR is smaller. However, when the SNR is smaller than -6.6 dB, the central values of the measurement under noise $\rm \textbf{N}(SNR=\,-6.6\,dB,\xi=\, 500)$, $\rm N(SNR=\,-13.2\,dB,\xi= \,300)$, $\rm \textbf{N}(SNR=\,-13.2\,dB,\xi=\, 500)$ and $\rm \textbf{N}(SNR=\,-13.2\,dB,\xi= 600)$ deviates greatly from the theoretical value in WVA scheme. In addition, the measurements under noise at SNR= -13.2 dB in the WVA scheme are invalid due to the error bars extending into the {``}Loss of sensitivity" area. Meanwhile, under the strong noise with negative-dB SNR, the scheme with the AWVA technique gives more accurate results with smaller statistical errors than the scheme with the WVA technique.

{
3). Note that we obtain $\rm SNR^{*}=0.5  SNR$ in the AWVA scheme by assuming that noise occurs after the beamsplitter. When the noises occur throughout the optical path, the value of $\rm SNR^{*}$ will \red{be obtained} in the range $\rm 0.5 SNR < SNR^{*} < SNR$. As shown in Fig.~\ref{Fig:scheme12sensitivity}, the results corresponding to $\rm 0.5 SNR < SNR^{*} < SNR$ are equivalent to shifting the curve to the right and lead to a smaller error bar compering to the results with $\rm SNR^{*}=0.5  SNR$. Therefore, the AWVA scheme performs the WVA scheme no matter where the noise appears.
}

In conclusion, the advantage of the scheme with the AWVA technique is more obvious when the SNR is lower. Note that the statistical errors in the AWVA scheme get \red{larger} when SNR is lower. Therefore, there is also a lower limit (corresponding to the minimum SNR) to how effective the AWVA scheme can be measured.

\subsection{Effects of the sampling frequency}

{
In this paper, we only investigate the influence of the sampling frequency 1/T on estimating the auto-correlative intensity $\rm \Theta_{}$ in the AWVA scheme due to the estimation of $\rm \Theta_{}$ strongly depending on  the sampling frequency.
In Sec.~\ref{Sec:WVAandAWVAunderNoises}, the influence of the different sampling frequency 1/T  on the auto-correlative intensity $\rm \Theta_{NN}$ of noises has been discussed. And the right figure of Fig.~\ref{Fig:ACIschemeDifferentSamplingFrequnecy} indicates that increasing sampling frequency gets a lower $\rm \Theta_{NN}$. The effects of the different sampling frequencies on the auto-correlative intensity $\rm \Theta_{A+N}$ of noises and the sensitivity in the AWVA scheme are shown in Table.~\ref{Table:sensitivityOfAWVA} and  Fig.~\ref{Fig:scheme12sensitivity}(C). Predictably, increasing sampling frequency 1/T can dramatically enhance the AWVA scheme's robustness.
}

{
It is worth noting that the upper limit of the sampling frequency 1/T is set at 100 MHz. There are two reasons for the choice in our simulation. One is that the results with 1/T=100 MHz are obviously better than the results with 1/T=10 MHz. And our simulation with 1/T=1000 MHz on Simulink \red{takes a prohibitively large amount of} time. Therefore the simulation with a larger 1/T=100 MHz is unnecessary. Another reason is that, in order to achieve sampling accuracy, the realistic sampling frequency normally needs to reach 3-10 times of the theoretical sampling frequency according to the Nyquist theorem~\cite{6674179}. Furthermore, \red{it is difficult to guarantee that} the time resolution (1/T) and the vertical resolution of the oscilloscope meet the requirements at the same time since the vertical resolution must be high enough to detect the shift of $\rm \Theta$.
}

\subsection{Effects of the coupling strength}
In the two schemes for measuring the time shifts $\tau$ induced by a birefringent crystal, the time shift $\tau$ serves as the coupling strength in the weak measurements. Next, we will show the results with the two schemes at different coupling strengths $\tau$. Note that this discussion is necessary \red{because} the two schemes can be transformed to measure other physical quantities. In Fig.~\ref{Fig:scheme12sensitivitywithTau}, we show the results of $\delta t$ in the WVA scheme and the shift of $\rm \Theta$(t=1.5 ms) in the AWVA scheme, as well as their corresponding sensitivities, under the noise $\textbf{N}(t,\sigma^{2}=4.0 \times 10^{-7})$ with SNR= -13.2 dB. Note that the sensitivity $\rm K_{1}$ and $\rm \overline{K}_{2}$ have been normalized to \red{highlight} the deviation from the theoretical values. Fig.~\ref{Fig:scheme12sensitivitywithTau} shows the deviations of the results in the two schemes from the theoretical value. We come to the following summaries of the effects of the coupling strength in the two schemes:
\begin{figure*}[t]
	\centering
	\subfigure
{
	\vspace{-0.4cm}
	\begin{minipage}{0.315\linewidth}
	\centering
	\centerline{\includegraphics[scale=0.135,angle=0]{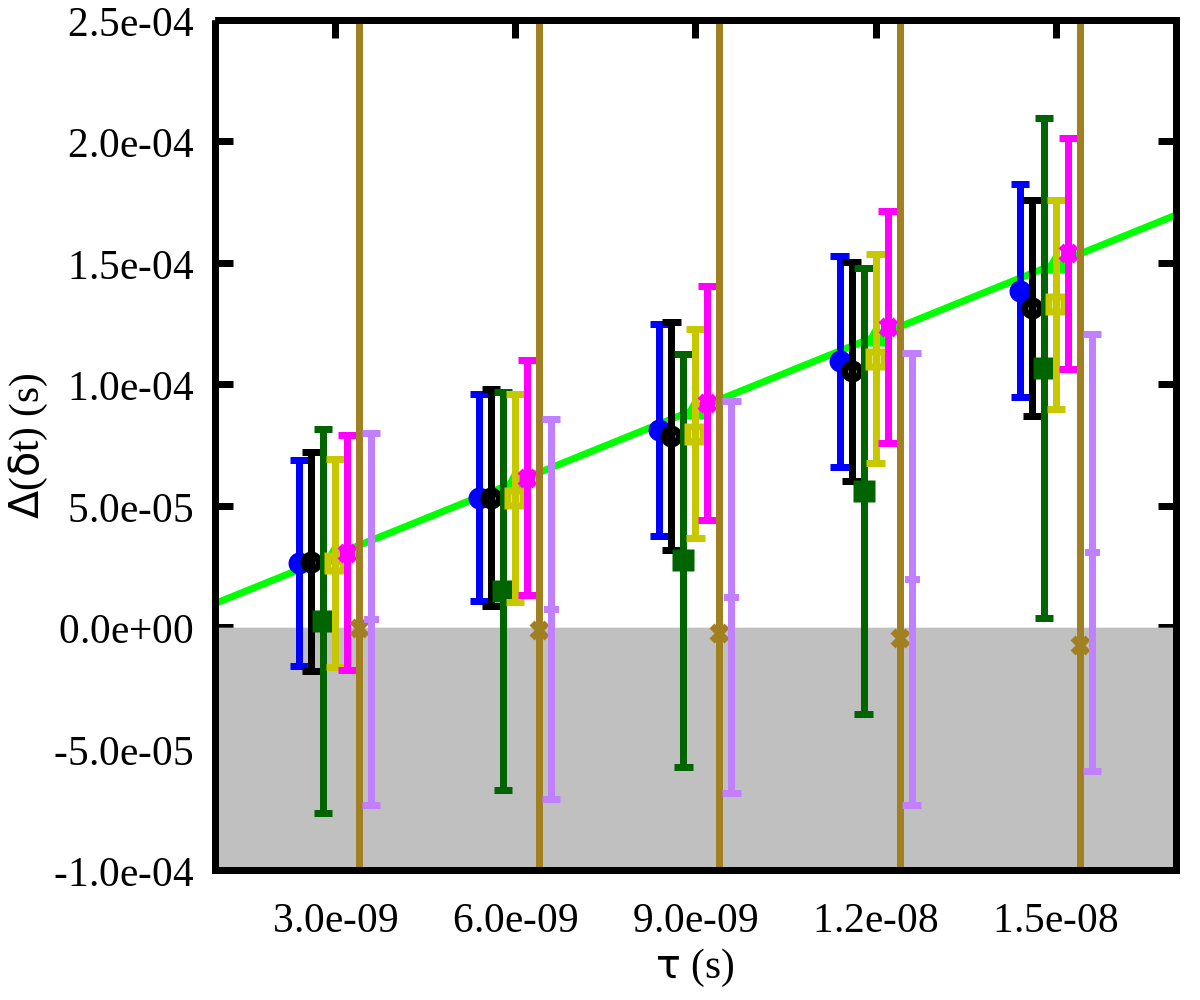}}
	\end{minipage}
}
\subfigure
{
	\vspace{-0.4cm}
	\begin{minipage}{0.315\linewidth}
	\centering
	\centerline{\includegraphics[scale=0.135,angle=0]{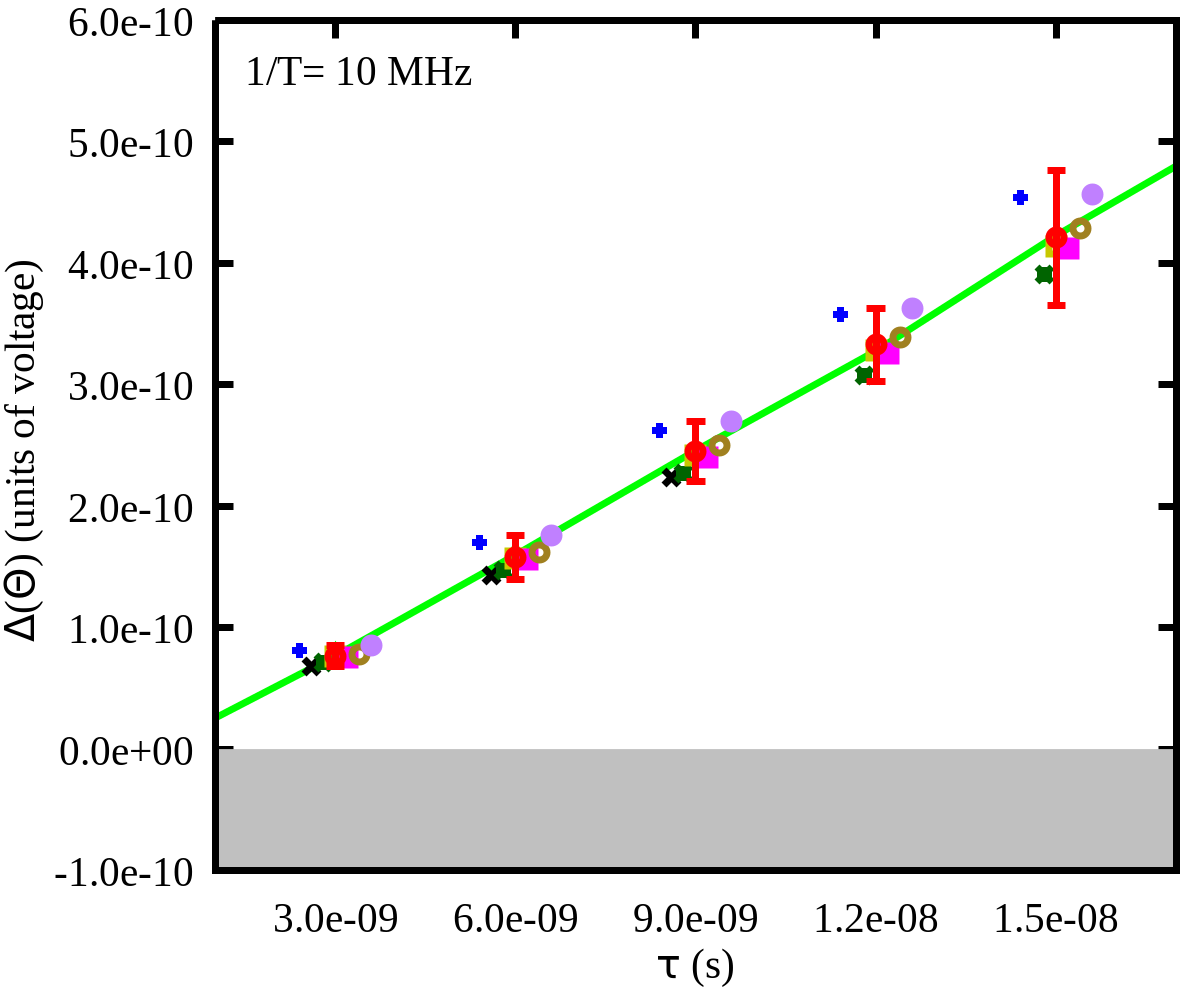}}
	\end{minipage}
}
\subfigure
{
	\vspace{-0.4cm}
	\begin{minipage}{0.315\linewidth}
	\centering
	\centerline{\includegraphics[scale=0.135,angle=0]{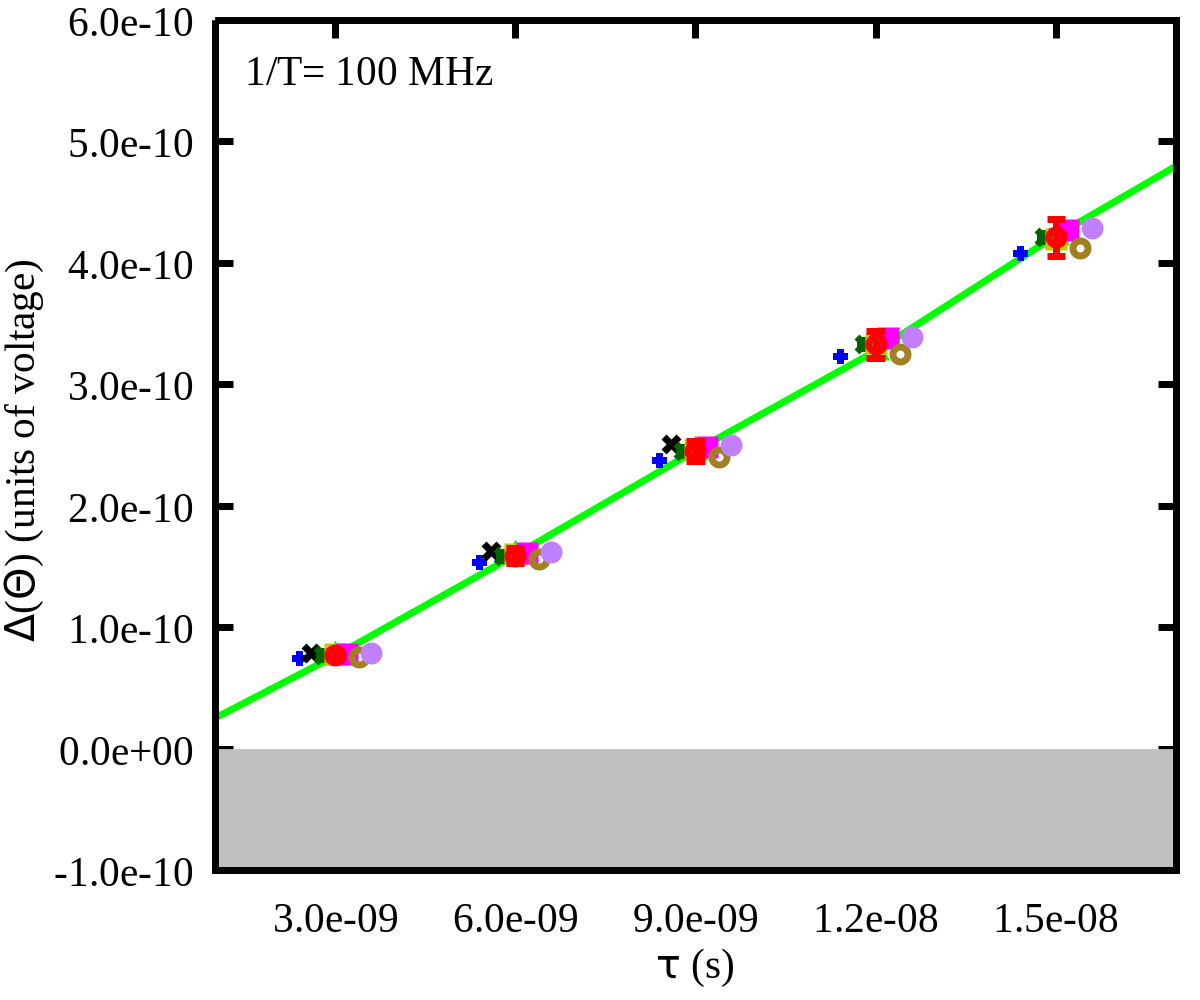}}
	\end{minipage}
}

	\vspace{-0.4cm}
	\subfigure
{
	\vspace{-0.4cm}
	\begin{minipage}{0.315\linewidth}
	\centering
	\centerline{\includegraphics[scale=0.135,angle=0]{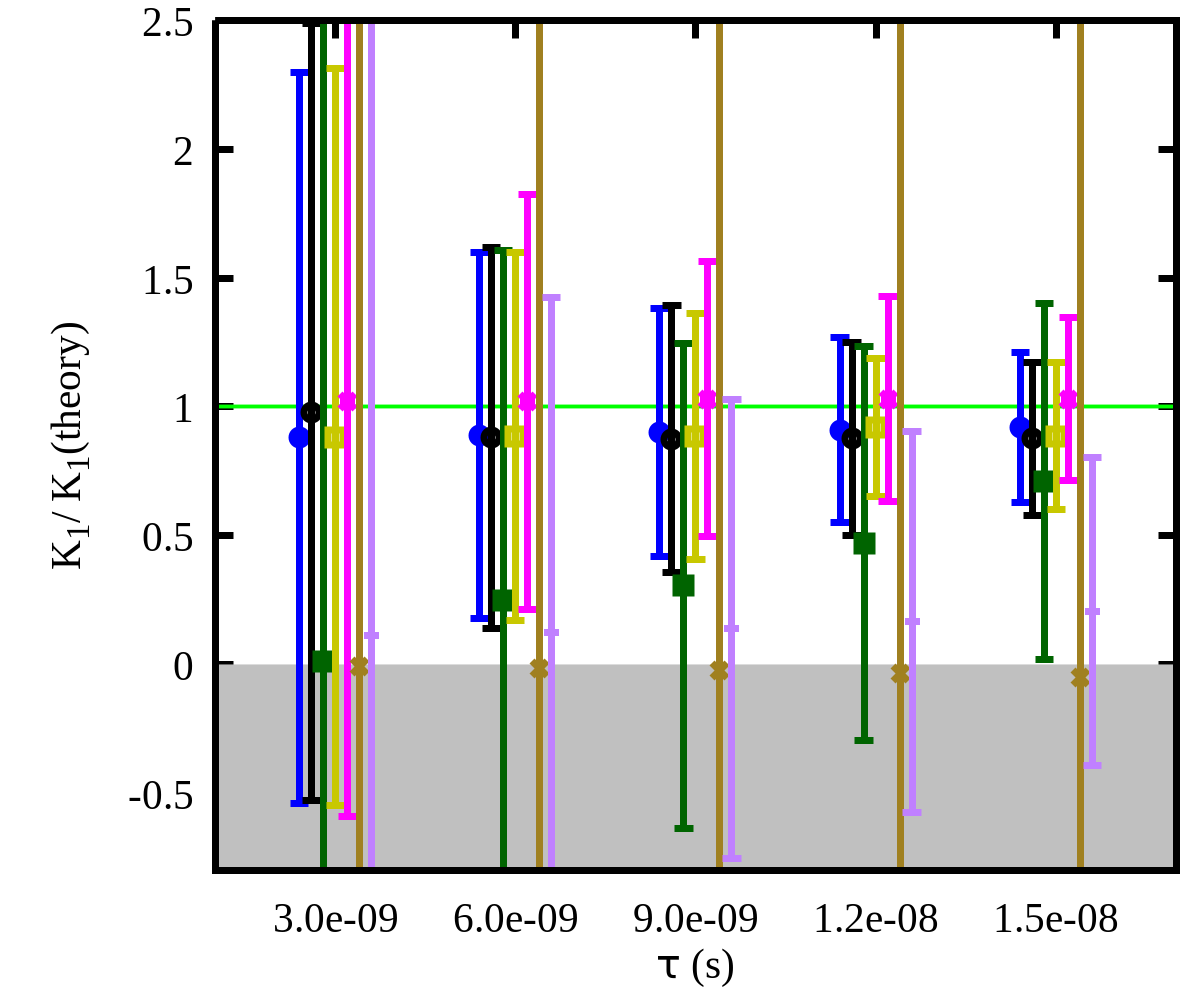}}
	\end{minipage}
}
\subfigure
{
	\vspace{-0.4cm}
	\begin{minipage}{0.315\linewidth}
	\centering
	\centerline{\includegraphics[scale=0.135,angle=0]{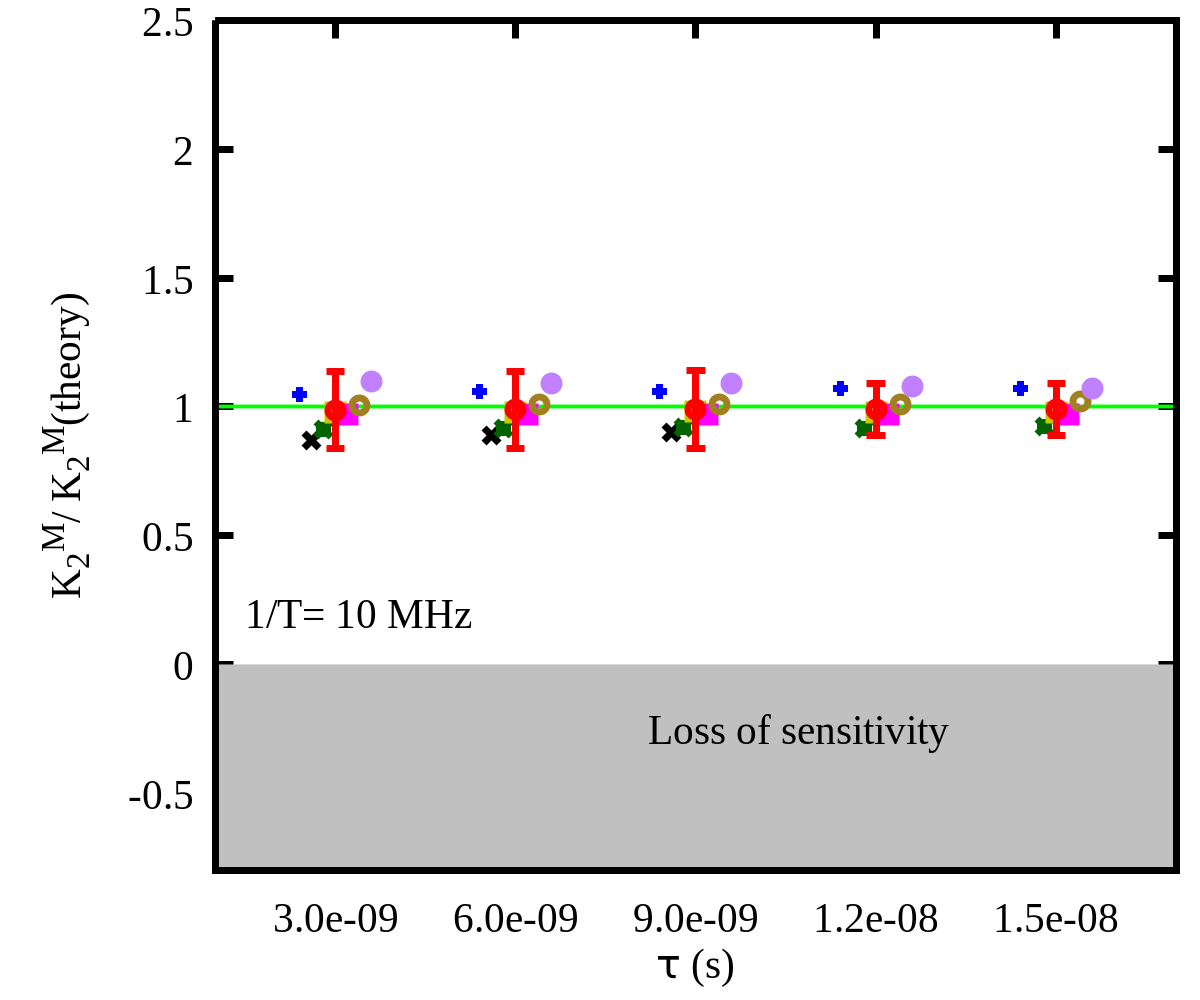}}
	\end{minipage}
}
\subfigure
{
	\vspace{-0.4cm}
	\begin{minipage}{0.315\linewidth}
	\centering
	\centerline{\includegraphics[scale=0.135,angle=0]{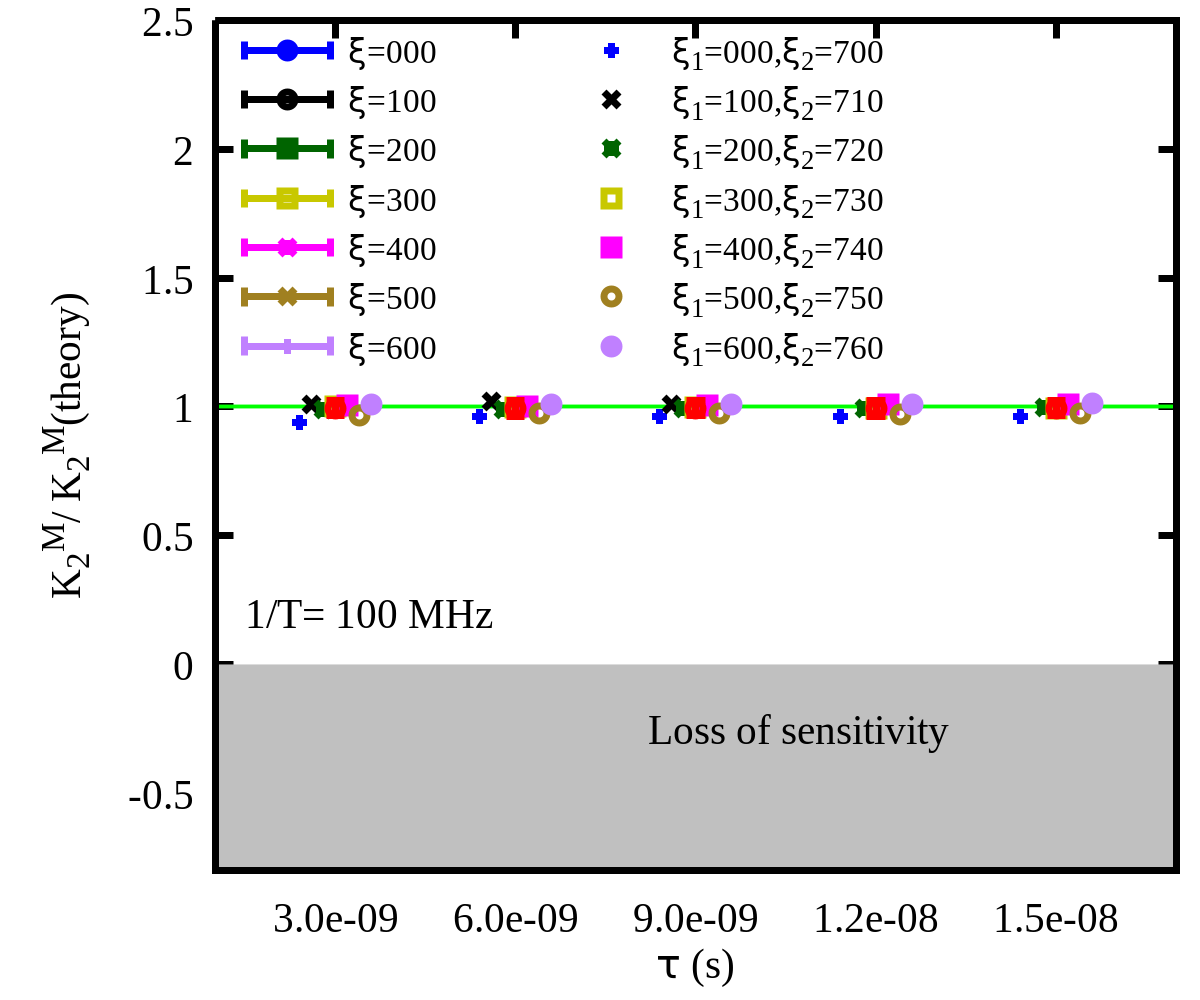}}
	\end{minipage}
}
	\vspace{-0.4cm}
\vspace*{0mm} \caption{\label{Fig:scheme12sensitivitywithTau} The shifts $\Delta (\delta t)$  as well as the shifts  $\Delta $($\rm \Theta$) (upper panels), and the corresponding sensitivity (lower panels) with respect to the coupling strength $\tau$ under the Gaussian noise with SNR= -12.6 dB.
{
The left panel shows the results under noises $\textbf{N}(t,\sigma^{2},\xi)$ with different seeds $\xi$ in the WVA scheme,
the middle panel shows the results at sampling frequency 1/T= 10 MHz under noises  $\textbf{N}(t,\sigma^{2},\xi_{1})$ and $\textbf{N}(t,\sigma^{2},\xi_{2})$ with different seeds $\xi_{1}$ and $\xi_{2}$ in the AWVA scheme,
and the right panel shows the results at sampling frequency 1/T= 100 MHz under noises  $\textbf{N}(t,\sigma^{2},\xi_{1})$ and $\textbf{N}(t,\sigma^{2},\xi_{2})$ with different seeds $\xi_{1}$ and $\xi_{2}$ in the AWVA scheme.
The {``}red" data with the error bar represent the final results of seven statistical averages in the AWVA scheme.
}
}
\end{figure*}

(1). The upper panels in Fig.~\ref{Fig:scheme12sensitivitywithTau} display the shifts $\Delta (\delta t)$ in the WVA scheme and the shifts $\Delta $($\rm \Theta$) in the AWVA scheme dependence of the coupling strength $\tau$. Under the same noise with SNR= -13.2 dB, the measurements with their error bars in the WVA scheme extending into the {``}Loss of sensitivity" area are invalid, while all the measurements in the AWVA scheme give valid results with statistical errors. Besides, the central values of measuring $\Delta $($\rm \Theta$) in the AWVA scheme agree with the theoretical results in absence of {noise}. Nevertheless, the central values of measuring $\Delta (\delta t)$ in the WVA scheme deviated greatly from the theoretical values. In addition, the magnitudes of the statistical errors in the WVA scheme are independent of the coupling strength $\tau$, and it is clear to verify that when discarding the measurements with the noise $\rm \textbf{N}(\sigma =10^{-5}, \xi=500)$ and $\rm \textbf{N}(\sigma =10^{-5}, \xi=600)$. While the magnitudes of the error bars in the AWVA scheme get bigger when the coupling strength $\tau$ increases.

(2). The lower panels in Fig.~\ref{Fig:scheme12sensitivitywithTau} display the normalized sensitivity $\rm {K}_{1}$ and $\rm \overline{K}_{2}$ dependence of the coupling strength $\tau$.
The results in the WVA scheme indicate that the statistical errors of estimating $\rm {K}_{1}$ get smaller when the coupling strength $\tau$ increases. While results in the AWVA scheme indicate that the statistical errors of estimating $\rm \overline{K}_{2}$ are independent of the coupling strength $\tau$. Furthermore, if the experimental conditions permit multiple measurements, the measurements, where the central value deviates greatly from the theoretical value, can be eliminated to obtain the results with smaller error bars.

(3). The middle panels (1/T= 10 MHz) and right panels (1/T= 100 MHz) show the influence of the sampling frequency on the results in the AWVA scheme. There is little deviation between the central value and the theoretical value with sampling frequency 1/T= 100 MHz. Therefore, increasing the sampling frequency can enhance the robustness of detecting the change of the coupling strengths $\tau$.

In conclusion, the scheme with the WVA technique is superior to the scheme with the AWVA technique when the coupling strength $\tau >$  $1.5 \times 10^{-8}$ s. On the other hand, the advantage of the scheme with the AWVA technique is more obvious when the coupling strength is lower
{and the sampling frequency is higher.}

\subsection{The measurements with different random seed}
\label{Sec:different_random_seed}
So far, the previous results and discussion have been based on the assumption that the thermal noise and shot noise of the detection may not cause the different series of \blue{noise} detected on APD1 and APD2 in the AVWA scheme. Therefore, the shift $\Delta {\rm \Theta}$ as well as sensitivity $\rm K_{2}$ were calculated from the {data for}  $\rm \Theta (\tau=0~s)$ and $\rm \Theta (\tau)$ with the same $\xi$. In other words, the noise is the same when you measure $\rm \Theta (\tau=0~s)$ and $\rm \Theta (\tau)$. Meanwhile, we calculated  $\Delta (\delta t)$ from the {data for}  $\delta t_{0} $ and $\delta t_{\tau} $ with the same $\xi$. Note that these measurements can only be completed under special experimental conditions. Further, we use the data with the different $\xi$ to estimate the shift $\Delta (\delta t)$ and $\Delta {\rm \Theta}$. In particular, the results of the measurements with different random seed were calculated and were shown in Table.~\ref{Table:manyMeasurements}. Where the average results $ \delta \overline{t}_{0}(\pm \overline{{\rm E}}_{t0})$, $ \delta \overline{t}_{\tau}(\pm \overline{{\rm E}}_{t\tau})$, $\overline{\Theta}_{0}(\pm \overline{{\rm E}}_{c0})$ and $  \overline{\Theta}_{\tau}(\pm \overline{{\rm E}}_{c\tau})$ of multiple measurements are redefined:
\begin{eqnarray}
\label{Eq:manyMeasurements}
 \delta \overline{t}_{0}&=& \sum \limits_{i=1}^{n} { \delta {t}_{0} (i)}/n , \, \, \overline{{\rm E}}_{t0}={\rm Max} \{ |\delta \overline{t}_{0} - {\delta {t}_{0} (i)}| \} \,,  \\
 \delta \overline{t}_{\tau}&=&  \sum \limits_{i=1}^{n} { \delta {t}_{\tau} (i)}/n , \, \, \overline{{\rm E}}_{t\tau}={\rm Max }\{ |\delta \overline{t}_{0} - {\delta {t}_{\tau} (i)}| \} \,,
  \end{eqnarray}
 \begin{eqnarray}
\label{Eq:manyMeasurements2}
 \overline{\Theta}_{0}&=& \sum \limits_{i=1}^{n} { {\Theta}_{0} (i)}/n , \, \,\overline{{\rm E}}_{c0}={\rm Max} \{ | \overline{\Theta}_{0} - {{\rm \Theta}_{0} (i)}| \} \,,  \\
  \overline{{\rm \Theta}}_{\tau}&=& \sum \limits_{i=1}^{n} {{\rm  \Theta}_{\tau} (i)}/n , \, \, \overline{{\rm E}}_{c\tau}={\rm Max }\{ | \overline{{\rm \Theta}}_{\tau} - {{\rm \Theta}_{\tau} (i)}| \} \,,
\end{eqnarray}
where the (i) represents the result of the ith measurement with the different value of $\xi$, n represents the total number of measurements and n=7. Then, the corresponding shifts $\Delta (\delta \overline{t})$ and $\Delta \overline{\Theta}$ are obtained by the relationships $\Delta (\delta \overline{t})=\delta \overline{t}_{\tau}-\delta \overline{t}_{0} $, $\Delta \overline{\Theta}=\overline{\Theta}_{0}- \overline{\Theta}_{\tau}$. Furthermore, the sensitivity $\rm \overline{K}_{1}(\pm \overline{E}_{1})$ in scheme WVA and the sensitivity $\rm \overline{\overline{K}}_{2}(\pm \overline{E}_{2})$ can be calculated by
\begin{eqnarray}
\label{Eq:manyMeasurementsSensitivity}
 \overline{{\rm K}}_{1}= \delta \overline{t}/ \tau ,\,\,\,
\overline{{\rm E}}_{1}=(\overline{{\rm E}}_{t0}+\overline{{\rm E}}_{t\tau})/ \tau,
\\
 \overline{\overline{{\rm K}}}_{2}= \delta \overline{t}/ \tau ,\,\,\,
\overline{{\rm E}}_{2}=(\overline{{\rm E}}_{c0}+\overline{{\rm E}}_{c\tau})/ \tau.
\end{eqnarray}
\begin{table*}[htp!]
\setlength{\tabcolsep}{1.1mm}{
\caption{\label{Table:manyMeasurements} The numerical average results (dimensional quantities in unit of s) of multiple measurement time shift $\tau=3.0 \times 10^{-9}s$ under Gaussian noise with different SNR.  }
\begin{tabular}{rrrrrrr}
\toprule
\toprule
$\rm SNR$
& 1/T
& $ \delta \overline{t}_{0}(\pm \overline{{\rm E}}_{t0})$
& $ \delta \overline{t}_{\tau}(\pm \overline{{\rm E}}_{t\tau})$
& $\Delta (\delta \overline{t})$
& $\overline{{\rm K}} _{1}\, (\pm \overline{{\rm E}}_{1}) /1.0 \times10^{3}$
 \\ \hline
11.5   & 1 &  $8.97 \times 10^{-7}$ ($\pm 2.09 \times 10^{-6}$) &  $3.09 \times 10^{-5}$ ($\pm 2.30 \times 10^{-6}$)&  $3.00 \times 10^{-5}$ ($\pm 4.39 \times 10^{-6}$) &
1.000($\pm$0.146)\\
2.8   & 1 &  $4.02 \times 10^{-6}$ ($\pm 9.58 \times 10^{-6}$) &  $3.38 \times 10^{-5}$ ($\pm 1.07 \times 10^{-5}$)&  $2.98 \times 10^{-5}$ ($\pm 2.03 \times 10^{-5}$) &
0.993($\pm$0.676)\\
-6.6  & 1 & $1.84 \times 10^{-4}$ ($\pm 1.07 \times 10^{-3}$) &  $2.11 \times 10^{-4}$ ($\pm 1.06 \times 10^{-3}$)&  $2.70 \times 10^{-5}$ ($\pm 2.13 \times 10^{-3}$) &
0.900($\pm$71.00)\\
-13.2   & 1 &  $1.59 \times 10^{-4}$ ($\pm 1.27 \times 10^{-3}$) &  $1.76 \times 10^{-4}$ ($\pm 1.27 \times 10^{-3}$)&  $1.70 \times 10^{-5}$ ($\pm 2.54 \times 10^{-3}$) &
0.566($\pm$84.66)\\     \hline
SNR
& 1/T
& $\rm  \overline{\Theta}_{0}(\pm \rm \overline{E}_{c0})$
& $\rm  \overline{\Theta}_{\tau}(\pm \rm \overline{E}_{c\tau})$
& $\Delta {\rm \overline{\Theta}}$
&$\rm \overline{\overline{K}}_{2} \, (\pm \rm \overline{E}_{2})/0.0258$
\\ \hline
11.5   & 1 &  $1.2450 \times 10^{-9}$ ($\pm 1.41 \times 10^{-11}$) &  $1.1672 \times 10^{-9}$ ($\pm 1.45 \times 10^{-11}$)&  $7.78 \times 10^{-11}$ ($\pm 2.86 \times 10^{-11}$) &
1.005($\pm$0.369)\\
2.8   & 1 &  $1.2426 \times 10^{-9}$ ($\pm 1.41 \times 10^{-10}$) &  $1.1646 \times 10^{-9}$ ($\pm 1.43 \times 10^{-10}$)&  $7.80 \times 10^{-11}$ ($\pm 2.84 \times 10^{-10}$) &
1.008($\pm$3.669)\\
-6.6   & 1 &  $1.2235 \times 10^{-9}$ ($\pm 5.92 \times 10^{-10}$) &  $1.1452 \times 10^{-9}$ ($\pm 5.96 \times 10^{-10}$)&  $7.83 \times 10^{-11}$ ($\pm 1.19 \times 10^{-09}$) &
1.012($\pm$15.37)\\
-13.2   & 1 &  $1.5567 \times 10^{-9}$ ($\pm 2.17 \times 10^{-09}$) &  $1.4886 \times 10^{-9}$ ($\pm 1.77 \times 10^{-09}$)&  $6.81 \times 10^{-11}$ ($\pm 3.94 \times 10^{-09}$) &
0.880($\pm$50.90)\\
-13.2   & 10 &  $1.5728 \times 10^{-9}$ ($\pm 7.52 \times 10^{-10}$) &  $1.4961 \times 10^{-9}$ ($\pm 7.46 \times 10^{-10}$)&  $7.67 \times 10^{-11}$ ($\pm 1.50 \times 10^{-09}$) &
0.991($\pm$19.38)\\
-13.2   & 100 &  $1.2764 \times 10^{-9}$ ($\pm 3.67 \times 10^{-10}$) &  $1.1993 \times 10^{-9}$ ($\pm 3.67 \times 10^{-10}$)&  $7.71 \times 10^{-11}$ ($\pm 7.34 \times 10^{-10}$) &
0.996($\pm$9.483)\\
 \bottomrule \bottomrule
\end{tabular}
}
\end{table*}
Finally, we display these average results of the multiple measurements with different random seeds in Table.~\ref{Table:manyMeasurements}. The simulation results show that the AWVA scheme with larger error bars has no advantage over the WVA scheme in the environment of weak noises with SNR= 11.5 dB and SNR= 2.8 dB {when the sampling frequency is set 1/T= 1 MHz; however, the error bars can diminish when \red{a larger sampling frequency is employed}.
When multiple measurements are completed under strong noises with SNR= -6.6 dB and SNR= -13,2 dB, the central value $\rm \overline{\overline{K}}_{2}/0.0258$ (0.0258 corresponding to the theoretical value without noise) is closer to the theoretical value 1, and the values of $\rm \overline{E}_{2}/0.0258$ are much smaller than that of statistical errors $\rm \overline{E}_{1}/1.0\times 10^{3}$.
In addition, the results of multiple measurements with a lager sampling frequency 1/T have smaller deviation of central value $\rm \overline{\overline{K}}_{2}/0.0258$ from the theoretical value and statistical errors $\rm \overline{E}_{2}$.
}
In general, when multiple measurements are completed under strong noises with negative-dB SNR, the AWVA scheme may have the potential to outperform the WVA scheme with a smaller deviation from the theoretical value and statistical errors.
{The AWVA technique is a kind of easily realized and \red{novel} scheme that can adapt to strong noise for real-time estimation of unknown small evolution parameters.}

\section{Summary and discussions}
We have performed a new scheme with auto-correlative weak-value amplification for precision phase estimation. A \blue{simulated} experiment for estimating the time shift with the standard weak-value amplification (WVA) technique and the auto-correlative weak-value amplification (AWVA) technique has been derived under Gaussian white noise with different signal-noise-ratio. In addition, a new quality (pointer), namely the auto-correlative intensity $\rm \Theta$, is defined in our scheme with the AWVA technique to estimate the small signal. Compared to fitting the shift of the Gaussian pointer with the standard WVA technique, measuring the shift of $\rm \Theta$ has the advantage of suppressing noise, and the advantage of the scheme with the AWVA technique is more obvious when the SNR and the coupling strength are lower. Therefore,
our results have demonstrated that the AWVA technique outperforms the standard WVA technique in the time domain even when the signal \blue{is submerged} in noise (SNR $<$ 0). The robustness suggests that the AWVA technique can be applied for a vast range of weak measurements in the time domain.

Note that we assume that the contribution of all noise is Gaussian white noise in this paper. However, the real \blue{technical noise} and environmental noise are far more complex than Gaussian white noise. In addition, \blue{in} many areas of physics \blue{Gaussian noise has} been replaced by colored noise (non-Gaussian)~\cite{PhysRevA.38.5938,PhysRevD.45.2843,WU2020124253,PhysRevE.101.052205}. Therefore, the auto-correlative weak-value amplification technique under the colored noises or the real noises will be investigated in our future work. {Finally, we must admit that the precision and robustness of AWVA technique depends on both the sampling frequency (1/T) and the vertical resolution of the oscilloscope. Thus, our scheme needs further experimental verification.}

After completion of this work, we found that the AWVA technique can also be achieved on the weak measurement in the frequency domain. The quantity $\rm \Theta$ can also be obtained \red{mathematically} by the integral of momentum, and \red{we will explore how to realize this scheme} in our future study.

\begin{acknowledgments}
This study is financially supported by the National Science Foundation of China (No. 41630317), MOST Special Fund from the State Key Laboratory of Geological Processes and Mineral Resources, China University of Geosciences (No. MSFGPMR01-4), the National Key Research and Development Program of China (No. 2018YFC1503705), and the Fundamental Research Funds for National Universities, China University of Geosciences (Wuhan). \red{ A.C.D. acknowledges support from the EPSRC, Impact Acceleration Account (EP/R511705/1).}
\end{acknowledgments}
\bibliography{reference}
\end{document}